\documentclass[longauth]{aa}  
\usepackage[english]{babel}
\usepackage[utf8]{inputenc}
\usepackage[T1]{fontenc}
\usepackage{graphicx}
\usepackage{amsmath,amssymb}
\usepackage{txfonts} % se commenti usa altro formato testo A&A
\usepackage{hyperref}
\usepackage{etoolbox}
\usepackage{wasysym}
\usepackage{ccaption}
\usepackage{threeparttable}
\usepackage{natbib}

\defcitealias{2020ApJ...895...31B}{B20}
\newcommand{\Msun}{$M_{\odot}$}
\newcommand{\Rsun}{$R_{\odot}$}
\newcommand{\Ni}{$^{56}$Ni}
\newcommand{\kms}{km s$^{-1}$}

\newcommand{\Ha} {\mbox{H$\alpha$}\,}

\begin{document} 

\title{The fast rise of the unusual Type IIL/IIb SN~2018ivc}

\titlerunning{The fast rise of SN~2018ivc}

\author{A. Reguitti\inst{1,2}\fnmsep\thanks{E-mail: andrea.reguitti@inaf.it},
R. Dastidar\inst{3,4},
G. Pignata\inst{5},
K. Maeda\inst{6},
T. J. Moriya\inst{7,8},
H. Kuncarayakti\inst{9,10}, 
{\'O}. Rodr{\'\i}guez\inst{4}, \\
M. Bersten\inst{11,12,13},
J. P. Anderson\inst{14,4},
P. Charalampopoulos\inst{9}, 
M. Fraser\inst{15},
M. Gromadzki\inst{16}, 
D. R. Young\inst{17}, \\
S. Benetti\inst{2}, 
Y.-Z. Cai\inst{18,19,20},
N. Elias-Rosa\inst{2,21}, 
P. Lundqvist\inst{22},
%C. Inserra$^{d}$, %M. Pursiainen, %A. Razza, %F. Ragosta$^{e}$,%F. Olivares E., %P. Wiseman, %S. J. Brennan$^{f}$, %G. Leloudas, %J. Antilen,%T.-W. Chen$^{h}$, 
R. Carini\inst{23}, 
S. P. Cosentino\inst{24,25}, 
L. Galbany\inst{21,26}, \\
M. Gonzalez-Bañuelos\inst{21},
C. P. Guti\'errez\inst{26,21},
M. Kopsacheili\inst{21}, 
J. A. Pineda G.\inst{3,4}, 
M. Ramirez\inst{3,4} 
}

\authorrunning{Reguitti et al.} 

\institute{
INAF – Osservatorio Astronomico di Brera, Via E. Bianchi 46, I-23807 Merate (LC), Italy
\and
INAF – Osservatorio Astronomico di Padova, Vicolo dell'Osservatorio 5, I-35122 Padova, Italy
\and
Instituto de Astrofìsica, Universidad Andres Bello, Fernandez Concha 700, Las Condes, 8320000 Santiago RM, Chile
\and
Millennium Institute of Astrophysics (MAS), Nuncio Monse\~{n}or S\'{o}tero Sanz 100, Providencia, 8320000 Santiago RM, Chile
\and
Instituto de Alta Investigación, Universidad de Tarapacá, Casilla 7D, Arica, Chile
\and
Department of Astronomy, Kyoto University, Kitashirakawa-Oiwake-cho, Sakyo-ku, Kyoto, 606-8502, Japan
\and
National Astronomical Observatory of Japan, National Institutes of Natural Sciences, and Graduate Institute for Advanced Studies, SOKENDAI, 2-21-1 Osawa, Mitaka, Tokyo 181-8588, Japan
\and
School of Physics and Astronomy, Monash University, Clayton, Victoria 3800, Australia
\and
Department of Physics and Astronomy, University of Turku, FI-20014 Turku, Finland
\and
Finnish Centre for Astronomy with ESO (FINCA), FI-20014 University of Turku, Finland
\and
Instituto de Astrofísica de La Plata (IALP), CCT-CONICET-UNLP. Paseo del Bosque S/N, B1900FWA, La Plata, Argentina
\and
Facultad de Ciencias Astronómicas y Geofísicas, Universidad Nacional de La Plata, Paseo del Bosque S/N, B1900FWA, La Plata, Argentina
\and
Kavli Institute for the Physics and Mathematics of the Universe (WPI), The University of Tokyo, 5-1-5 Kashiwanoha, Kashiwa, Chiba, 277-8583, Japan
\and
European Southern Observatory, Alonso de C\'ordova 3107, Casilla 19, 8320000 Santiago, Chile
\and
School of Physics, O’Brien Centre for Science North, University
College Dublin, Belfield, Dublin 4, Ireland
\and
Astronomical Observatory, University of Warsaw, Al. Ujazdowskie 4, 00-478 Warszawa, Poland
\and
Astrophysics Research Centre, School of Mathematics and Physics, Queen’s University Belfast, Belfast BT7 1NN, UK
\and
Yunnan Observatories, Chinese Academy of Sciences, Kunming 650216, P.R. China
\and
Key Laboratory for the Structure and Evolution of Celestial Objects, Chinese Academy of Sciences, Kunming 650216, P.R. China
\and
International Centre of Supernovae, Yunnan Key Laboratory, Kunming 650216, P.R. China
\and
Institute of Space Sciences (ICE-CSIC), Campus UAB, Carrer de Can Magrans, s/n, E-08193 Barcelona, Spain
\and
The Oskar Klein Centre, Department of Astronomy, Stockholm University, AlbaNova, SE-10691 Stockholm, Sweden
\and
INAF - Osservatorio Astronomioco di Roma, Via Frascati 33, I-00078, Monte Porzio Catone (RM), Italy
\and
Università degli Studi di Catania, Dip. di Fisica e Astronomia "Ettore Majorana", Via S. Sofia 64, I-95123 Catania, Italy
\and
INAF - Osservatorio Astrofisico di Catania, Via S. Sofia 78, I-95123 Catania, Italy
\and
Institut d'Estudis Espacials de Catalunya (IEEC), 08860 Castelldefels (Barcelona), Spain
%$^{d}$Cardiff %$^{h}$Taiwan
}

\date{Accepted XXX. Received 2024; in original form 2018}
 
\abstract{
We present an analysis of the photometric and spectroscopic dataset of the Type II supernova (SN) 2018ivc in the nearby (10 Mpc) galaxy Messier~77. Thanks to the high cadence of the CHASE survey, we observed the SN rising very rapidly by nearly three magnitudes in five hours (or 18 mag d$^{-1}$). The $r$-band light curve presents four distinct phases: the maximum light is reached in just one day, then a first, rapid linear decline precedes a short-duration plateau. Finally, a long, slower linear decline lasted for one year. Following a radio rebrightening, we detected SN~2018ivc four years after the explosion.
The early spectra show a blue, nearly featureless continuum, but the spectra evolve rapidly: after about 10 days a prominent H$\alpha$ line starts to emerge, with a peculiar profile, but the spectra are heavily contaminated by emission lines from the host galaxy.
He I lines, namely $\lambda\lambda$5876,7065, are also strong. On top of the former, a strong absorption from the \ion{Na}{I} doublet is visible, indicative of a non-negligible internal reddening. From its equivalent width, we derive a lower limit on the host reddening of $A_V\simeq1.5$ mag, while from the Balmer decrement and a match of the $B-V$ colour curve of SN~2018ivc to that of the comparison objects, a host reddening of $A_V\simeq3.0$ mag is obtained.
The spectra are similar to those of SNe II, but with strong He lines. Given the peculiar light curve and spectral features, we suggest SN~2018ivc could be a transitional object between the Type IIL and Type IIb SNe classes.
In addition, we found signs of interaction with circumstellar medium in the light curve, making SN~2018ivc also an interacting event.
Finally, we modelled the early multi-band light curves and photospheric velocity of SN~2018ivc to estimate the explosion and CSM physical parameters.
}


\maketitle
%-------------------------------------------------------------------

\section{Introduction} \label{introduction}

Modern fast astronomical surveys, with a cadence of one day or less, can detect very young transients a mere hours after their explosions, including supernovae (SNe), the final fate of the most massive stars. Some examples of these surveys are the Zwicky Transient Factory (ZTF; \citealt{2019PASP..131a8002B}) and the Asteroid Terrestrial-impact Last Alert System (ATLAS; \citealt{2018PASP..130f4505T, SmithK2020PASP..132h5002S}). 
Several surveys are specifically built to discover new fast-evolving objects by selecting a limited number of sky fields around nearby galaxies and reducing the time cadence to obtain multiple exposures in a single night. The CHilean Automatic Supernova sEarch survey (CHASE, PI Pignata, \citealt{2009AIPC.1111..551P, Hamuy2012MmSAI..83..388H}) project is designed as such.

With a time interval of a few hours between each image, it is possible to follow the photometric evolution of the elusive first phases after the explosion.
%A recent example of the application of this strategy, which allowed also to constrain the explosion epoch to within a mere five hours, is offered by SN 2023ixf \citep{2023arXiv230606097H}.
The few events detected so early have revealed fast rises, brightening by one order of magnitude or more in a few hours due to the rapid increase in the radius of the emitting photosphere. Core-collapse SNe (CCSNe) are believed to show a sharp increase in luminosity towards an ephemeral peak and then a decline in the first hours after the explosion, a feature called `Shock Break-Out' \citep[SBO,][]{1977ApJS...33..515F, 2010ApJ...725..904N, 2017hsn..book..967W}.

Type II SNe by definition show emission lines from hydrogen, i.e. from the Balmer series \citep{1997ARA&A..35..309F}, and have been subdivided into different categories according to the properties of their light curves or features in the spectra.
Based on the photometrical evolution, Type II SNe were historically divided into Type IIP and Type IIL \citep{1979A&A....72..287B} depending on the declining slope after the maximum light. The former are characterised by a nearly constant luminosity lasting for 3-4 months, a `plateau' (hence the `P' label), while the latter continue to fade linearly, with a mean slope of more than 1~mag/100 days \citep{Patat1994A&A...282..731P, 2016MNRAS.459.3939V}.
Detailed studies have revealed that the difference in the light curves between the two typologies is reflected in some physical properties, with SNe IIL being on average more luminous than SNe IIP, and having hotter progenitors (Yellow and Red Supergiant stars, respectively, e.g. \citealt{2009ARA&A..47...63S, VanDyk2017hsn..book..693V} and references therein).
Nowadays, with catalogues of hundreds of SNe, the gap between the two classes has been filled up, exhibiting a continuum in the distribution of slopes \citep[e.g.][]{2014ApJ...786...67A, 2015ApJ...799..208S, Galbany2016AJ....151...33G, 2016MNRAS.459.3939V}.

If in the spectra of a Type II SN helium lines begin to appear and then dominate at late phases (1-2 months), the SN is labelled as a Type~IIb \citep{1987ApJ...318..664W, Filippenko1993ApJ...415L.103F}. In this type of objects, Balmer lines may disappear from the spectra taken months after maximum, with He lines becoming the dominant features. Concerning the light curves, a sub-sample of SNe IIb showed a secondary peak about three weeks after the first maximum, noticeable in all filters \citep[see for instance][]{Richmond1994AJ....107.1022R, Kumar2013MNRAS.431..308K}, due to the cooling following the shock breakout.
Finally, a sub-class of SNe II shows narrow Balmer emission lines, hence they are called Type IIn SNe \citep{Schlegel1990MNRAS.244..269S, Filippenko1997ARA&A..35..309F, Fraser2020RSOS....700467F}. These narrow lines are generated within a slow-moving ($\sim10^2$ \kms) circumstellar medium (CSM) surrounding the progenitor star; when the fast SN ejecta collide with the dense CSM they start to interact: the kinetic energy of the ejecta is efficiently converted into radiation, providing an additional power source to the light curve and making the SN brighter.

It is rare to observe the very first light coming from a new SN event, because the cadence of a survey might not be high enough, or the object evolves too rapidly.
In this paper, we present the observations of the fast rise of SN~2018ivc, caught just after the explosion thanks to the observing strategy of the CHASE program, and its subsequent evolution, which shows spectroscopic and photometric features common to both Type~IIL and Type IIb classes of SNe.
We perform also an independent analysis to the one previously conducted by \cite{2020ApJ...895...31B} (hereafter \citetalias{2020ApJ...895...31B}).

The structure of the paper is the following: in Sect. \ref{discovery} we report how the object was discovered, and the characteristics of the host galaxy. 
In Sect. \ref{photometry} we present the photometric data of SN~2018ivc, along with the evolution of the light curves and the comparison of the absolute light curve with similar fast-rising objects. 
The spectroscopic data are analysed in Sect. \ref{spectroscopy}, together with a brief description of the spectral evolution, the estimation of the internal reddening and a comparison with reference objects. 
In Sect. \ref{discussion} we discuss the possibility that SN~2018ivc may be a transitional object between the two SNe types IIL and IIb, based on the presence of common features of both classes, and even an interacting SN. We also present our modelling of the early light curve with \texttt{SNEC} and the explosion parameters we derived.
Finally, in Sect. \ref{conclusion} we summarise our conclusions. % observational facilities used, the data reduction techniques, the 

\section{Discovery and host galaxy} \label{discovery}

SN~2018ivc (a.k.a. DLT18aq, ZTF18acrcogn, ATLAS18zot, PS19aht) was officially discovered by the Distance Less Than 40 Mpc (DLT40; \citealt{2018AAS...23124511S}) Supernova search on 2018 November 24.076 (UT) in the galaxy NGC~1068 \citep{2018TNSTR1816....1V}, at celestial coordinates $\alpha$ = 02:42:41.29, $\delta = -$00:00:31.71 (J2000). The transient was quite bright, at 14.65 AB mag in the Clear filter. Their last non-detection, down to 19.35 mag, dates back to 2018 November 19 (five days earlier).
%The DLT40 survey utilises the 0.4-meter Panchromatic Robotic Optical Monitoring and Polarimetry Telescopes (PROMPT) at Cerro Tololo Inter-American Observatory (CTIO). %The discovery was achieved with one of them, the PROMPT5. 

\begin{figure}
    \includegraphics[width=\columnwidth]{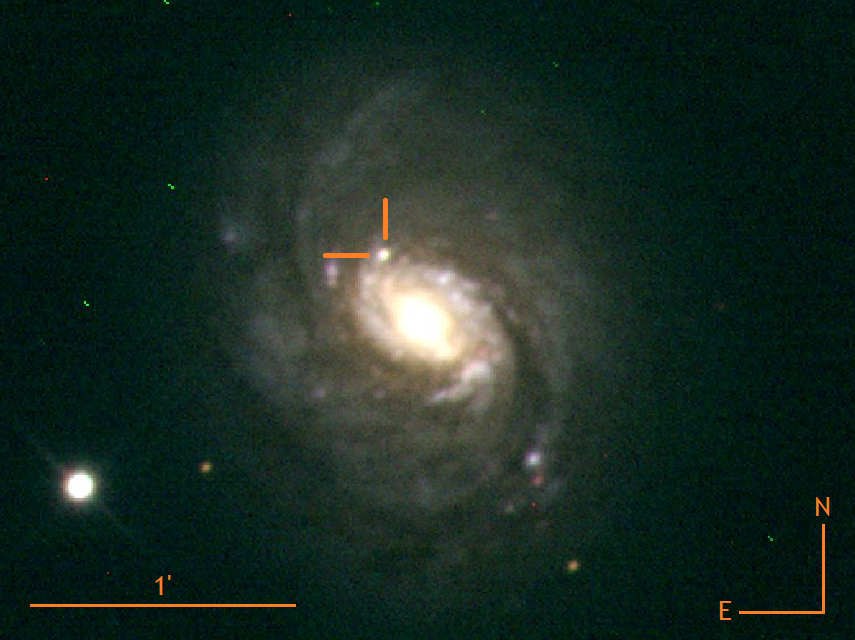}
    \caption{Colour image of the host galaxy M77 and of SN~2018ivc taken on 2018 November 27, four days after the explosion, with the PROMPT6 telescope. The image is a combination of the frames obtained with the $B$, $V$ and $R$ filters using the $RGB$ technique. The orientation, scale of the image and SN location are marked.}
    \label{Fig1:colorful}
    \end{figure}

The spectroscopic classification of SN~2018ivc was performed at the Asiago Astrophysical Observatory with the 1.22-meter \textit{Galileo} Telescope. The transient was classified as a young Type II SN, because of a blue featureless continuum \citep{2018ATel12239....1O}, as confirmed by \cite{2018ATel12240....1Z} and \cite{2018TNSCR1818....1Y}.
A colour image of SN~2018ivc within its host galaxy taken four days after discovery is shown in Fig. \ref{Fig1:colorful}.

\subsection{Host galaxy}
NGC 1068 (or Messier 77), is a spiral galaxy of type (R)SA(rs)b according to \cite{deVaucouleurs1991rc3..book.....D}. The galaxy hosts a well known active galactic nucleus, classified also as a Seyfert 2 galaxy \citep{1993ApJ...414..552O}.
The Milky Way reddening in the direction of NGC 1068 is $A_{V,MW}=0.091$ mag \citep{2011ApJ...737..103S}; however, the internal reddening is much larger and uncertain, as discussed in Sect. \ref{Sect:internal_extinction}.

The NASA/IPAC Extragalactic Database (NED\footnote{https://ned.ipac.caltech.edu}) reports 11 measurements of distance to NGC 1068. We adopt the most recent estimation, which corresponds to a distance modulus ($\mu$) of 30.02$\pm$0.39~mag \citep{Nasonova2011A&A...532A.104N} obtained through the Tully-Fisher method. This translates into a distance of 10.1$^{+2.0}_{-1.7}$ Mpc, which is the same distance assumed by \citetalias{2020ApJ...895...31B}.
The redshift of NGC 1068 $z=0.003793\pm0.000010$ \citep{1999ApJS..121..287H} corresponds to an heliocentric radial velocity of 1137$\pm$3 \kms.

\section{Photometry} \label{photometry}
%The CHASE survey is a project devoted to the discovery and study of young SNe in nearby galaxies. This survey also uses the eight PROMPT telescopes at CTIO \citep{2005NCimC..28..767R}. The survey has discovered more than 130 SNe \citep{2012MmSAI..83..388H}. A clear filter is used to improve the efficiency, allowing to spot fainter transients. The strategy of the survey is taking three images of the same field every night, with the aim to constrain the explosion epoch of a new transient within a few hours. The region around M77 was added to the catalogue of fields of the search due to the equatorial position and because of the proximity, that allows to discover intrinsically fainter events.

\subsection{Observational facilities}
The photometric follow-up of SN~2018ivc was performed with a plethora of instruments and telescopes around the world, available to our collaborations, whose characteristics are reported in Table \ref{Tab:instruments}. Observations were done in the optical with Open, Johnson-Cousins (JC) $BVRI$, Sloan $ugriz$, ATLAS $cyan,orange$ ($c,o$) filters and in the near-infrared (NIR) with $JHKs$ filters.

\begin{table}\begin{small}
\caption{Observational facilities and instrumentation used in the photometric follow-up of SN~2018ivc. For each instrument we report the relative telescope with the diameter, the geographical location and the filters used.}
\label{Tab:instruments}
\begin{tabular}{llll}
\hline
Telescope & Location & Instrument & Filters \\
\hline
ATLAS (0.4m) & Hawaii & ACAM1 & $c,o$ \\
PROMPT5 (0.4m) & CTIO & Apogee & $BVgriz$ \\
PROMPT6 (0.4m) & CTIO & Apogee & $BVR$ \\
PROMPT1 (0.6m) & CTIO & Apogee & Open ($r$) \\ %Real discover
PROMPT8 (0.6m) & CTIO & Apogee & $BVRI$ \\
TRAPPIST (0.6m) & La Silla & Fairchild & $BVRI$ \\
Schmidt (0.67m) & Asiago & Moravian & $uBVgri$ \\
Oschin (1.22m) & Mt. Palomar & ZTF & $g$ \\
SMARTS (1.3m) & La Silla & ANDICAM & $BVRI$ \\
SMARTS (1.3m) & La Silla & ANDICAM-IR & $JH$ \\
LT (2.0m) & La Palma & IO:O & $r$ \\
MPG/ESO (2.2m) & La Silla & WFI & $R$ \\
NOT (2.56m) & La Palma & ALFOSC & $uBVgriz$ \\
CFHT (3.58m) & Mauna Kea & MegaPrime & $i$ \\
NTT (3.58m) & La Silla & EFOSC & $BVRIri$ \\
NTT (3.58m) & La Silla & SOFI & $JHKs$ \\
SUBARU (8.2m) & Mauna Kea & FOCAS & $VR$ \\
VLT (8.2m) & Paranal & FORS2 & $r$ \\
\hline
\end{tabular}\end{small}
\end{table}

\subsection{Data reduction}
For the photometric data reduction, we used a dedicated pipeline called \texttt{SNOoPY}\footnote{SNOoPy is a package for SN~photometry using PSF fitting and/or template subtraction developed by E. Cappellaro at the Padova Astronomical Observatory. A package description can be found at http://sngroup.oapd.inaf.it/snoopy.html.}. The instrumental magnitudes were determined through the PSF-fit method. For Sloan filter images, the photometric zero points and colour terms were computed through a sequence of reference stars from the \textsl{SDSS} survey in the SN~field. For JC $BV$ filters, the magnitudes of the reference stars were taken from the APASS DR10 catalogue\footnote{https://www.aavso.org/apass}. To calibrate the frames in JC $RI$, we converted the $gri$ magnitudes of the \textsl{SDSS} standard stars in the field into $BVRI$ magnitudes, adopting the conversion formulas of \cite{2002AJ....123.2121S} (see their Table 7).
Finally, for NIR images, the magnitudes were calibrated with the 2MASS catalogue \citep{2006AJ....131.1163S}.
Photometric errors were estimated through artificial star experiments, also accounting for uncertainties in the PSF-fitting procedure.

Because of the complex environment around the SN, we used the template-subtraction technique\footnote{The technique consists in the alignment, transformation to the same scale, and PSF matching of the template to the science images, in order to be able to perform the subtraction.} to remove the background underlying the object. For the JC $I$ and Sloan filters, the archive images of the \textsl{SDSS} DR12\footnote{https://dr12.sdss.org/fields} survey were used as templates, while for $BVR$ filters we chose the FORS2 images taken at the ESO Very Large Telescope (VLT) on 13 September 2016.
Open filter magnitudes were calibrated as $SDSS$ Sloan-$r$ ones, because the quantum efficiency of the detector peaks at a wavelength near the maximum sensitivity of that filter.

We caution the reader that we found a discrepancy between our Sloan $g$-band photometric measurements and those published by \citetalias{2020ApJ...895...31B}; the details are discussed in Appendix \ref{g-discrepancy}.

\subsection{Light curve evolution}
Because of the treatment of the Open filter images as Sloan-$r$, the light curve in this filter is the best sampled, especially in the earliest phases, and it is used as a reference to describe the evolution of the light curves of SN~2018ivc.

\begin{figure}
    \includegraphics[width=1.0\columnwidth]{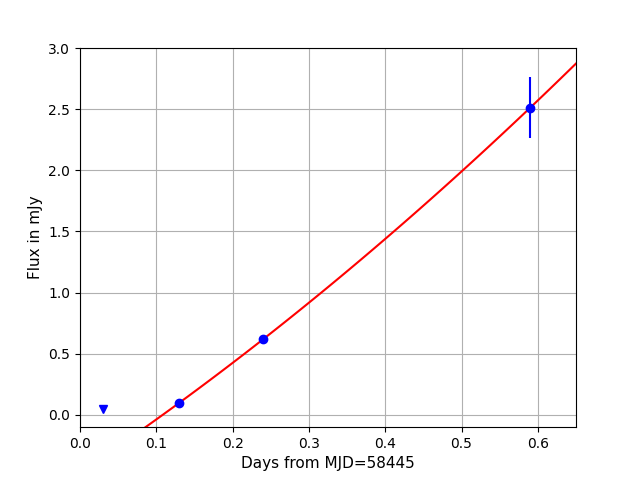}
    \caption{
    A parabolic fit on the fluxes of the three detections of November 23. The explosion is predicted to occur when the curve crosses the zero flux level, at MJD 58445.11. The last upper limit is shown as a downwards triangle.
    }
    \label{Fig:rise_fit}
    \end{figure}

The CHASE survey, using the PROMPT1 telescope, reports an upper limit on 22 November 2018. The non-detection is not stringent, at only 18.3 mag. At the beginning of the following Chilean night (November 23.03), we registered the last non-detection, this time a much deeper one ($>19.7$ mag, with a threshold at 2.5$\sigma$).
However, 2.5 hours later (Nov 23.13), we observed the first detection of a new source at mag 18.9$\pm$0.2. Another 2.5 hours later (Nov 23.24), the transient had reached mag 16.9$\pm$0.05. Consequently, in about five hours SN~2018ivc increased in brightness by at least 2.8 magnitudes, which is equivalent to $>$13.3 mag day$^{-1}$, but between the first and second detection (2nd and 3rd PROMPT observation) the rise was even faster, with an increase of 2 mag in just 2.5 hours, or a rate of $18.2\pm2.1$ mag day$^{-1}$.
The Japanese amateur astronomer K. Itagaki detected the object eight hours after our third image (Nov~23.59) at mag 15.4\footnote{https://www.rochesterastronomy.org/sn2018/sn2018ivc.html}. Thus, SN 2018ivc rose by more than 4 magnitudes in less than 14 hours.
We performed a parabolic fit (i.e. a ‘fireball model') of the fluxes of the first three detections of November 23 (the first two from CHASE plus the Itagaki's measurement, Fig. \ref{Fig:rise_fit}) to better constrain the explosion epoch as the instant the flux is equal to 0.
Interestingly, the zero flux is found at MJD 58445.108, only half an hour before the first PROMPT detection, consistent with our assumed interval for the explosion epoch (see below).

\begin{figure}
    \includegraphics[width=1\columnwidth]{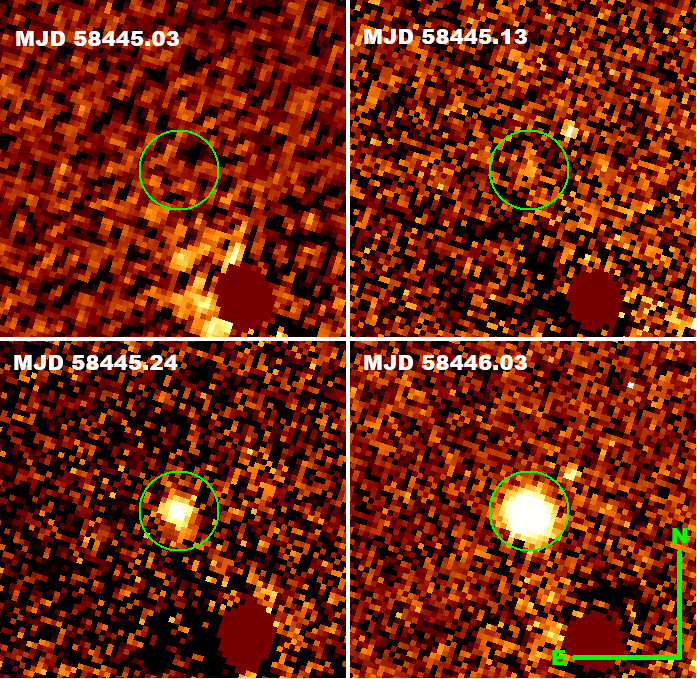}
    \caption{
    Four stamps of the template-subtracted images from the PROMPT telescopes of 23 November 2018, showing the fast rise of SN~2018ivc after the explosion.
    Top-left: The last non-detection with a threshold of 2.5$\sigma$. Top-right: the first detection. Bottom-left: the second, much brighter detection of the night. Bottom-right: the first PROMPT image from the following night, with the source close to maximum light.
    The colour scales are roughly the same on all the stamps.
    All the green circles are centred on the position of the transient calculated by the PSF fitting and have a radius of 5".}
    \label{Fig:discovery}
    \end{figure}

While our last non-detection is not much fainter than the first detection, it is nonetheless the most stringent upper limit available from TNS or from the literature.
%usa modello nakar sari bsg, ma andrebbe usato curvefit sopra dati lasciando texpl come param libero
%Given the very short time interval between the last non-detection and the first detection, we are able to constrain the explosion epoch of SN~2018ivc in a time window of just 2.5~hours.
We cannot exclude a scenario with a slower rise just after the explosion and before our observations.
Finally, we also made use of our hydrodynamical modelling of the light curve, presented in Sect. \ref{Sect:modelling}, to provide an independent constrain of the explosion epoch, as described in the Appendix \ref{Appendix_B}.
For all of the above reasons, we assume the explosion epoch as the middle point between the last upper limit and the first detection, on MJD 58445.08$^{+0.06}_{-0.9}$ (Nov 23.08).
In Fig. \ref{Fig:discovery} we show four cut-outs of the template-subtracted images from the PROMPT telescopes on the night of November 23, and at maximum light the night after, showing how the object has increased in brightness in just five hours and in the first 24 hours.

\begin{figure*}
    \includegraphics[width=1.1\columnwidth]{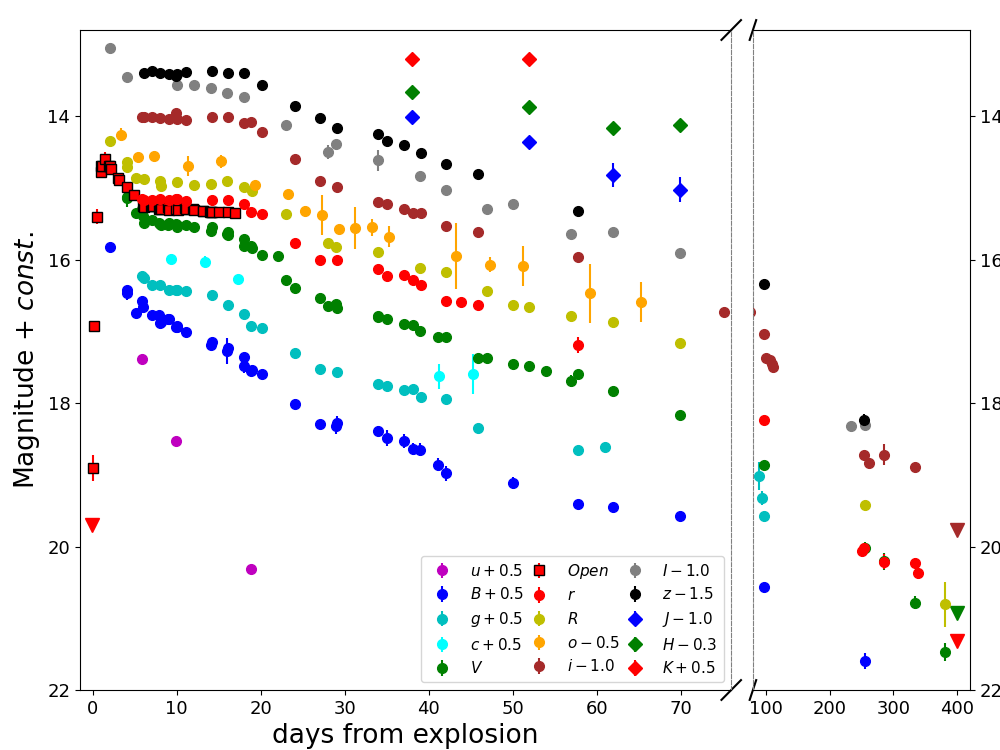} \includegraphics[width=0.96\columnwidth]{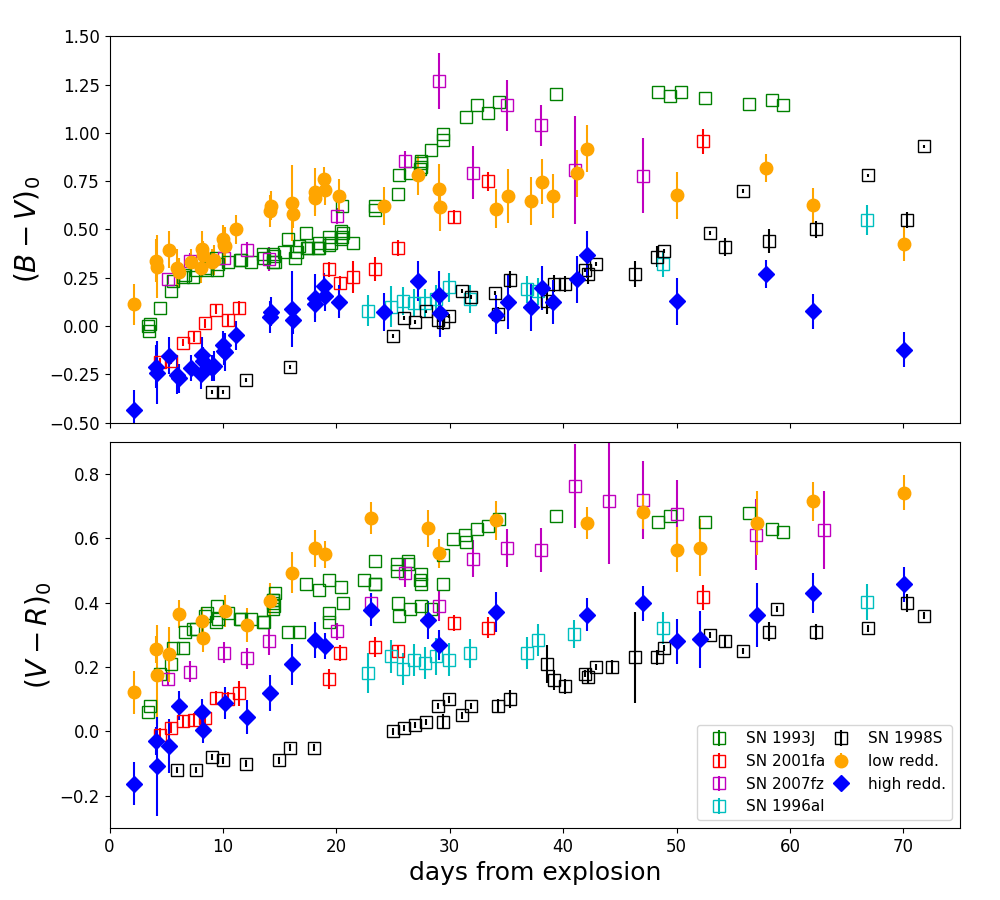}
    \caption{Left: Open, JC $BVRI$, Sloan $ugriz$, ATLAS $c,o$ and NIR $JHK$ light curves over the 1-year follow-up of SN~2018ivc. The graph has been broken into two windows (before and after +3 months) for clarity. Downward triangles mark upper limits.
    Right: $B-V$ and $V-R$ reddening-corrected colour curves of SN~2018ivc (filled points) and of the comparison objects: SNe 1993J, 2001fa, 2007fz, 1996al and 1998S. For SN~2018ivc, we consider both the low (orange circles) and high (blue diamonds) reddening scenarios (see Sect. \ref{Sect:internal_extinction}).}
    \label{Fig:LC+CLC}
    \end{figure*}

The next day, the DLT40 survey officially discovered the transient. At that epoch, the object was already near maximum light, at Clear mag 14.65. We then started the multi-band follow-up campaign. 
The complete photometric evolution of SN~2018ivc over one year of follow-up is plotted in Fig. \ref{Fig:LC+CLC}, left panel. The observed optical Sloan, Johnson, clear, NIR and ATLAS magnitudes are tabulated in Tables \ref{tab:sloan} to \ref{tab:atlas}.

The Open+$r$-band light curve can be divided into four distinct phases: a very fast rise to maximum light, that was reached just 1.5 days after the first detection, a first, steep linear decline, a short-duration plateau and a long, shallower linear decline.
The timescale to maximum light described above is considerably short with respect to the median value of a large sample of SNe II \citep[7.5$\pm$0.3~days;][]{santiago2015MNRAS.451.2212G}, though it has been observed in at least one other object \citep[the sub-luminous Type~IIP SN~2005cs,][]{2006A&A...460..769T, 2009MNRAS.394.2266P}. % (7.5$\pm$0.3~days for \citealt{santiago2015MNRAS.451.2212G}; 9.8$\pm$1.8~days for \citealt{Bruch2023ApJ...952..119B})
    %\begin{figure}
    %\includegraphics[width=1.1\columnwidth]{Images/broken_axis}
    %\caption{The Clear and $r$-band light curves over one year of follow-up. \textit{The graph has been broken in two windows for clarity.} The light curve can be clearly divided into four stages: a fast rise to maximum light, reached in about 1.5 days, a first, steep linear decline, a short-duration plateau and a long, shallower linear decline. In both graphs the phase is relative to the assumed explosion time.}
    %\end{figure}
After the maximum, SN~2018ivc started to fade linearly, with a slope of $13.5\pm0.3$ mag (100~d)$^{-1}$. This fast linear decline only lasts 4.5 days and is followed by a short-duration phase at nearly constant luminosity, resembling a plateau lasting for about 10 days.
In the redder bands a short plateau is observed, in the $V$ and $g$ filters the light curves show a change in the slope, becoming almost flat, while in the $B$ filter the flattening feature is absent.
Later on, the object started again to fade with a second linear decline of $3.7\pm0.2$ mag (100 d)$^{-1}$, %faster than the slope expected from the radioactive decay of $^{56}$Co (0.98 mag (100 d)$^{-1}$), 
which is observed up to 3~months after the explosion, when the object went behind the Sun. However, this decline seems not to be monotonic, for example at around +30 days it slightly slows down in the $r$-band. 
\citetalias{2020ApJ...895...31B} also noted the short plateau and the second flattening in their light curve at the same phases and interpreted those as a signature of interaction between the SN~ejecta and the CSM around the progenitor.
A 2-week plateau, between +5 and +18 days, is observed in the $i$- and $z$-band light curves. Unfortunately, the NIR campaign started too late to verify if this short plateau was also visible in the $JHK$ filters. 
%The $B$-band light curve, instead, is monotonic, with a linear decline all the way down to 100 days. The rate of decline in the first 50 days is $6.4\pm0.2$ mag (100~d)$^{-1}$.
%A hint of a flattening at +30 days is visible also in the $B$-band light curve, while in the $V$ filter the decline is not strictly linear, showing signs of oscillations. Again, \citetalias{2020ApJ...895...31B} suggest those oscillations are evidences of ejecta-CSM interaction.

Observations conducted between August and October 2019, after the Solar conjunction, show a linear decline still ongoing, but at a shallower rate: in the $V$-band the rate of decline between +250 and +380 days is $1.2\pm0.1$ mag (100~d)$^{-1}$, similar to other SNe~II (see the $s_3$ parameter values in \citealt{2014ApJ...786...67A}). %a bit faster than the rate expected from the $^{56}$Co decay.
In contrast, the $r$- and $i$-band light curves around +300 days are more flat, probably due to an intensification of the H$\alpha$ and Ca~II NIR emission lines, respectively, likely caused by stronger interaction. %, but because of the lack of a nebular spectrum (see next Section) this cannot be proven. %B20 mostrano uno spettro nebulare.
%The $r$-band light curve in this phase is nearly constant, showing a small bump or at least a flattening (Fig. \ref{Fig:LC+CLC}, right panel), Finally, the $i$-band light curve also is flat during the nebular stage. 
The object was still detected in Hubble Space Telescope (\textit{HST}) images taken on 2 December 2020 (+740 d), at $F555W=22.71\pm0.02$ and $F555W=21.97\pm0.03$ Vegamag \citep{Baer-Way2024arXiv240112185B}.

In September 2022, one month after the ALMA observations of \cite{Maeda2023ApJ...945L...3M}, in which they observed a radio rebrightening of SN 2018ivc, we imaged the field of the object with the 8.2-m VLT+FORS2, under favourable sky conditions (seeing $\sim$0.7-0.8"), in the $r$-band. Remarkably, we detect a source at the SN position at $r=$21.1 mag four years after the explosion.
%To convince the reader of the authenticity of our detection, in Fig. \ref{Fig:FORS} we show a cut-out of the template-subtracted image from VLT+FORS2 on the night of 23 September 2022, centred at the position of SN 2018ivc.\\
SN 2001ig is another SN IIb that showed modulations and bumps in its radio light curve \citep{Ryder2004MNRAS.349.1093R} and was detected in the optical three years after the explosion \citep{Ryder2006MNRAS.369L..32R}. In that case, the radio rebrightenings were interpreted as due to interaction with CSM shells produced in a binary system, with the optical detection being the survived massive companion star.

 %   \begin{figure}\centering
 %   \includegraphics[width=0.75\columnwidth]{Images/FORS2_late_subtracted.png}
 %   \caption{
 %   Stamp of the template-subtracted VLT+FORS2 image of 23 September 2022, in which a source is detected at the position of SN 2018ivc at $R=$21.1 mag. A 2016 VLT+FORS2 $R$-band archival image was used as the template.
 %   The green circle is centred on the position of SN 2018ivc and has a radius of 1".}
 %   \label{Fig:FORS}
 %   \end{figure}
    
\subsubsection{Comparison objects}
Based on the peculiarity of the early light curve, with a linear decline but with a short-duration plateau, and the strong He I lines in the spectra (see Sect. \ref{spectroscopy}), we constructed a small sample of comparison objects, selecting the transitional Type IIL SNe~2001fa and 2007fz \citep{2014MNRAS.445..554F}, the Type IIb SNe~1993J \citep[e.g.][]{2000AJ....120.1487M, 2000AJ....120.1499M}, 2011fu \citep{2015MNRAS.454...95M} and 2013df \citep{2014MNRAS.445.1647M}, and the Type IIn SNe~1996al \citep{2016MNRAS.456.3296B} and 1998S\footnote{We mention that nowadays SN 1998S is considered as a Type IIL SN with flash spectroscopy features \citep{Gal-Yam2014Natur.509..471G} in the early spectra.} \citep{2000ApJ...536..239L, 2001MNRAS.325..907F, 2004MNRAS.352..457P}.

\subsection{Colour curves}
We calculated the extinction-corrected $B-V$ and $V-R$ colour curves of SN~2018ivc, assuming both host reddening scenarios (discussed in Sect. \ref{Sect:internal_extinction}), and compared them with the comparison objects. The colour curves are plotted in Fig. \ref{Fig:LC+CLC}, right panel.
Both the $B-V$ and $V-R$ colours become rapidly redder from the first observation up to $\sim$1 month after the explosion, then the increase is slower, and from +50 days onward the $B-V$ colour becomes bluer again.

In the low internal reddening scenario, the colours of SN~2018ivc are more similar to those of SNe 1993J and 2007fz, while in the high reddening scenario they are more akin to SN~2001fa and the SNe~IIn 1996al and 1998S (the latter always remains much bluer, but the $(B-V)_0$ colours are the same between +30 and +40 days).
Type IIn SNe are known to show bluer colours due to the contribution of the ejecta-CSM interaction, that makes the continuum bluer with respect to other sub-types of SNe II \citep{Hillier2019A&A...631A...8H, osmar2020MNRAS.494.5882R}.

\subsection{Absolute light curve and rise slope comparison}
\begin{figure}
    \includegraphics[width=1.02\columnwidth]{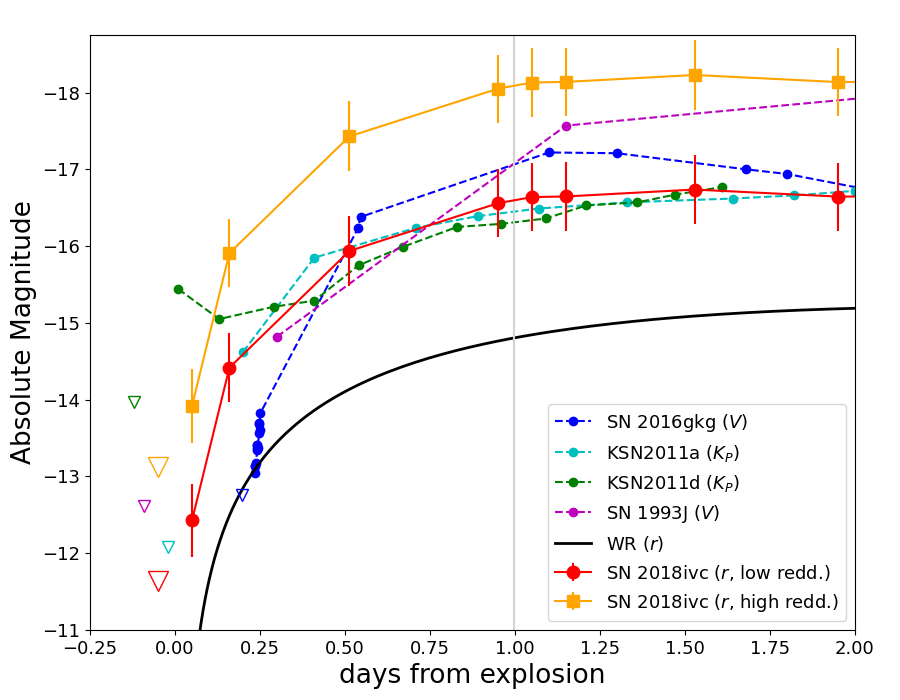}
    \caption{Comparison of the absolute light curves of SNe~2018ivc, 1993J, 2016gkg (adopting the explosion epoch from \citealt{Arcavi2017ApJ...837L...2A}), KSN2011a and KSN2011d in the first two days after the explosion. For SN~2018ivc, the Clear (calibrated as $r$-band) magnitudes are plotted, for SNe 1993J and 2016gkg the $V$-band and for the two \textsl{Kepler}'s SNe the $K_P$-filter ones. For SN~2018ivc, the error bars are also plotted, summing in quadrature the errors from the photometric measurements, the $\mu$ and the internal reddening. Both the low and high host reddening scenarios are considered. 
    The official discovery date of SN~2018ivc is marked with a vertical line. The downward triangles mark the last non-detection of each SN.
    The absolute $r$-band light curve following the SBO from a typical WR model ($M=$5~\Msun, $R=$10~\Rsun, $E_{expl}=$1~foe) of \citet{2010ApJ...725..904N} is also over-plotted (black line). }
    \label{fig:absolute+nakar&sari}
    \end{figure}

We constructed the absolute $r$-band light curve of SN~2018ivc, with the adopted $\mu$ and the low host reddening case of $A_{V,host}=1.5\pm$0.2 mag, as discussed in Sect. \ref{Sect:internal_extinction}. %This corresponds to $A_{r,host}=1.27\pm$0.17 mag.
Given the very fast rise the first day after the explosion (with a rate of 18 mag in 1 day), we choose to compare it with the rise of objects for which there is a claim of SBO detection, i.e. the Type IIb SN~2016gkg \citep{2018Natur.554..497B} and the two Type~IIP KSN~2011a and KSN~2011d (\citealt{2016ApJ...820...23G}, but see \citealt{2017ApJ...848....8R}), the latter two being discovered by the \textsl{Kepler} spacecraft. The prototypical Type~IIb SN~1993J \citep{1993ApJ...417L..71W, Richmond1994AJ....107.1022R, woosley1994ApJ...429..300W} is also used as a reference. The comparison of the absolute light curves of these four objects is shown in Fig.~\ref{fig:absolute+nakar&sari}.
%These comparison SNe provide us another way to estimate the explosion epoch of SN 2018ivc. We performed the same parabolic fit to the observations of the 4 SNe (1993J, 2016gkg, KSN 2011a and 2011d) during the first day since the discovery (three days for SN 1993J to increase the number of data-points, and removing the first two points of KSN 2011d). The best fits are shown in the bottom panels of Fig. 3.
%These comparison SNe provide us another way to estimate the explosion epoch of SN 2018ivc using the rise time to the first peak: the explosion date of SNe 1993J and 2016gkg are well established (\citealt{Wheeler1993SSRv...66..425W} and \citealt{Arcavi2017ApJ...837L...2A}, respectively), and their rise times are 2.2 d \citep{Richmond1994AJ....107.1022R} and 1.2 d \citep{Bersten2012ApJ...757...31B}, respectively.
%Therefore, their average rise time is $1.7\pm0.5$ d.
%Given that SN 2018ivc reached its maximum brightness at $r\simeq14.6$ mag on MJD 58446.6, we can estimate that its explosion occurred on MJD 58444.9$\pm$0.5 (4.3 hours before our assumed time), though the error bar is evidently overestimated, as already on MJD 58445.13 we have our first detection.

SN 2016gkg was serendipitously observed to double in flux in just 20 minutes, which is equivalent to a rising slope of $43\pm6$ mag~d$^{-1}$ \citep{2018Natur.554..497B}; the fastest SN~rise known to date. The hydrodynamic simulation of \cite{2018Natur.554..497B} reveals that this behaviour is consistent with the rise during the SBO phase. Compared with the rise we observed in SN~2018ivc ($18\pm2$ mag~d$^{-1}$), the rise of SN~2016gkg is much faster. However, because the SBO rise and cooling has a typical duration of much less than 2.5 hours (see the hydrodynamic models from \citealt{2018Natur.554..497B} and \citealt{2011ApJS..193...20T}), it could also have occurred in SN~2018ivc between the 1st and 2nd CHASE observation.
In this case, SN~2018ivc could have risen even faster than observed, and a higher temporal cadence of the survey may have spotted the SBO signature.
Another possibility is that the CSM might have created an extended pseudo-photosphere, which would cause the SBO to last for days, and hence to blend with the rising light curve.
However, another possibility is that we actually missed the SBO because it occurred before our first observation. In fact, our last pre-discovery upper limit is not stringent enough to rule out an explosion epoch earlier than the one we assumed.

The \textsl{Kepler} mission claimed to have observed the SBO rise of two Type~IIP SNe, thanks to the fast cadence of its instrument. KSN~2011a rises faster than the model prediction by \cite{2011ApJ...728...63R}; this was interpreted as being due to the SN~shock propagating into the CSM, that allowed the conversion of more kinetic energy into luminosity and dilute the SBO signal \citep{2010ApJ...724.1396O, 2011ApJ...729L...6C, 2016ApJ...820...23G}. Between the first two detections, KSN~2011a has a rise slope of 5.9 mag d$^{-1}$: taking into account the last non-detection, a lower limit for the rise slope of 11.5~mag d$^{-1}$ is obtained, still slower than that observed in SN~2018ivc.
In KSN~2011d, the cooling after the SBO was indeed observed (i.e. the light curve was declining just after discovery), but the rise was not. From the last non-detection, we can derive a lower limit of the slope of the rising as 11.3 mag d$^{-1}$.

The selected objects show heterogeneous behaviours (especially before 0.5 d) probably due to different physical conditions of the exploding stars and their close environments.

\section{Spectroscopy} \label{spectroscopy}
Together with the photometric follow-up campaign of SN~2018ivc, we started a spectroscopic one, that lasted two months and during which we collected nine optical spectra, whose log is reported in Table~\ref{tab:spectra}. The classification spectrum is the one obtained at the Asiago Observatory, even if an earlier spectrum from the Kanata Observatory is available\footnote{https://wis-tns.weizmann.ac.il/object/2018ivc}. All the spectra will be publicly released on the \textsc{WISeREP} interface \citep{2012PASP..124..668Y}.

\begin{table*}
\caption{Log of the spectroscopic observations of SN~2018ivc. For each spectrum the date, the observed spectral range, the resolution, the exposure time and the telescope used are reported. The phases reported are relative to the assumed explosion time (MJD 58445.08).}
\label{tab:spectra}
\begin{tabular}{lllllll}
\hline
Date & MJD & Phase & Coverage & Resolution & Exposure & Telescope + Instrument + Grism \\
 & & (d) & (\AA) & (\AA) & (s) & \\
\hline
2018-11-24 & 58446.60 & 1.5  & 3770-9650 & 7  & 2700 & Kanata 1.5m + HOWPol \\
2018-11-24 & 58446.99 & 1.9  & 3600-7880 & 11 & 1200 & Galileo 1.22m + B\&C + 300 tr \\
2018-12-02 & 58454.99 & 9.9 & 3600-9680 & 12 & 800  & NOT 2.56m + ALFOSC + gr4 \\
2018-12-09 & 58461.15 & 16.1 & 3330-9920 & 14 & 900  & NTT 3.58m + EFOSC2 + gr11/16\\
2018-12-11 & 58463.89 & 18.8 & 3530-9600 & 14 & 800  & NOT 2.56m + ALFOSC + gr4 \\
2018-12-15 & 58467.13 & 22.0 & 3330-9920 & 9  & 900  & NTT 3.58m + EFOSC2 + gr11/16\\
2019-01-07 & 58490.91 & 45.8 & 3750-9600 & 18 & 2000 & NOT 2.56m + ALFOSC + gr4 \\
2019-01-15 & 58498.04 & 52.9 & 3640-9230 & 9  & 1800 & NTT 3.58m + EFOSC2 + gr13\\
2019-01-19 & 58502.87 & 57.8 & 3600-9600 & 13 & 1200 & NOT 2.56m + ALFOSC + gr4 \\
\hline
\end{tabular}\end{table*}

\subsection{Data reduction}
The spectra were pre-reduced and calibrated using standard IRAF routines, including correction for bias and flat-field, extraction of the 1-D spectrum, removal of background and cosmic rays, and wavelength and flux calibrations, using arc lamps and spectrophotometric standard stars. The spectra were also corrected for the strongest telluric absorption bands.
The spectra from the NOT telescope were reduced using the \texttt{ALFOSCGUI}\footnote{https://sngroup.oapd.inaf.it/foscgui.html} pipeline, designed specifically for the quick reduction of photometric and spectroscopic data taken with the \textsc{ALFOSC} instrument, in the frame of the NOT Unbiased Transients Survey (NUTS) project \citep{2016ATel.8992....1M}.
The spectra from the NTT telescope were collected through the extended European Southern Observatory Spectroscopic Survey of Transient Objects (ePESSTO; \citealt{2015A&A...579A..40S}) collaboration, and reduced with a \texttt{pyraf}-based pipeline (\texttt{PESSTO}), optimised for the \textsc{EFOSC2} instrument.

The subtraction of the underlying background was particularly difficult because the object is located at the edge of the brightest part of the galaxy disk (as can be seen in Fig. \ref{Fig1:colorful}). Hence, the contamination from the host, in the form of forbidden and narrow Balmer emission lines, is strong and spatially variable in the neighbourhood of the SN (they are stronger towards the centre of the galaxy, and fainter outwards).

\begin{figure*}
    \includegraphics[width=2.05\columnwidth]{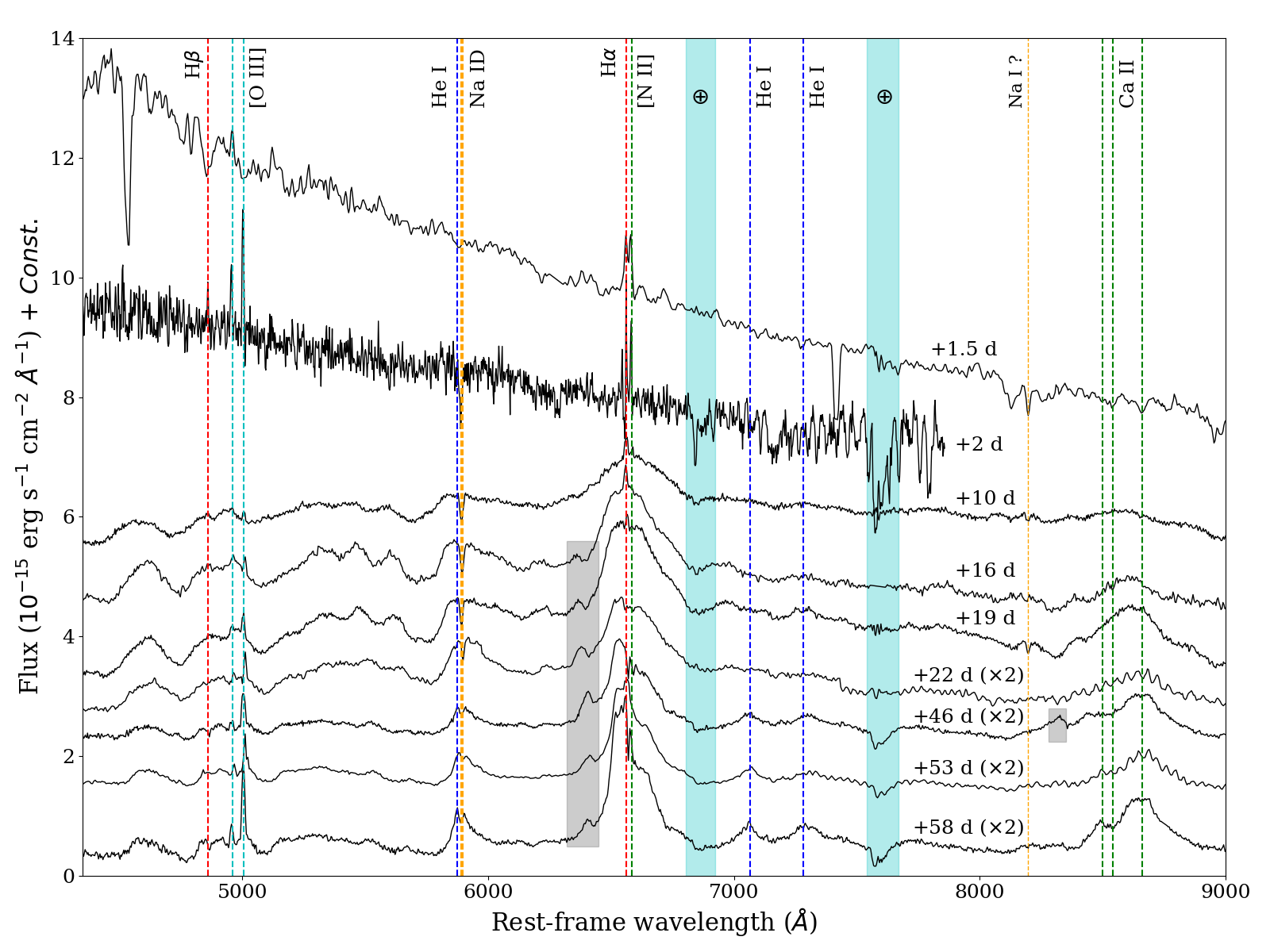}
    \caption{Spectral sequence of SN~2018ivc, spanning two months of evolution after the discovery. The phases relative to the explosion are reported. The spectra have been corrected for redshift and Milky Way extinction only, as the host galaxy reddening is highly uncertain. The principal identified transitions are highlighted with vertical dashed lines. The possible high velocity feature discussed in Sect. \ref{Sec:interaction} is also marked in gray.
    Each spectrum is shifted by a constant for graphical purposes. The regions most affected by the telluric absorption bands are marked in cyan.}
    \label{fig:spectra}
    \end{figure*}

Finally, the spectra are corrected for the redshift of the host galaxy and the Galactic reddening, but not for the internal one, because of its uncertainties (Sect. \ref{Sect:internal_extinction}). The sequence of spectra is shown in Fig. \ref{fig:spectra}.

\subsection{Line identification and spectral evolution}\label{subsec:line_identification}
In the earliest spectra, over a blue featureless continuum, narrow emissions from the host galaxy are detected: those are narrow H$\alpha$ and H$\beta$, [O III] $\lambda\lambda$4959,5007 and [N II] $\lambda$6584. This latter line is present in all the spectra, and its contamination deteriorates the profile of the H$\alpha$ line from the SN. An absorption line from the unresolved Na ID doublet is already visible, suggesting the presence of a conspicuous internal extinction along the line of sight.
We also tentatively identify another absorption from the Na I $\lambda$8195.

Starting from the +10 days spectrum, a broad emission from the Ca II NIR triplet appears, each line (after a deblending) with a $v_{FWHM}$ between $\sim$6500 and $\sim$8000 km s$^{-1}$, and it increases in strength over time. H$\alpha$ now presents a broad component, with a FHWM of 12000 km~s$^{-1}$, from the fast SN~expanding ejecta. To the blue side of H$\alpha$, a broad bump (with a FWHM close to the one of H$\alpha$ line) from He I $\lambda$5876 is visible, apparently with a broad P Cygni profile, and the strong absorption from Na I on top of the line. The blue part of the spectrum is characterised by some emission features from broad H$\beta$ and blends of metal lines, likely due to Fe II. 
The principal features are located at 5000-5500 \AA~and 4400-4700 \AA. Later on, as the continuum fades, these features become more prominent. 

In the following ten days, the velocities derived from the width of the lines slightly decrease, with the FWHM of the now boxy shaped H$\alpha$ going down to $\sim$11000~km~s$^{-1}$, while the FWHM of He I $\lambda$5876 and H$\beta$ diminish to 7000 and 8000 km s$^{-1}$, respectively. The broad absorption of He I $\lambda$5876 to its blue side is also somewhat boxy in shape.

\begin{figure}
    \includegraphics[width=1.02\columnwidth]{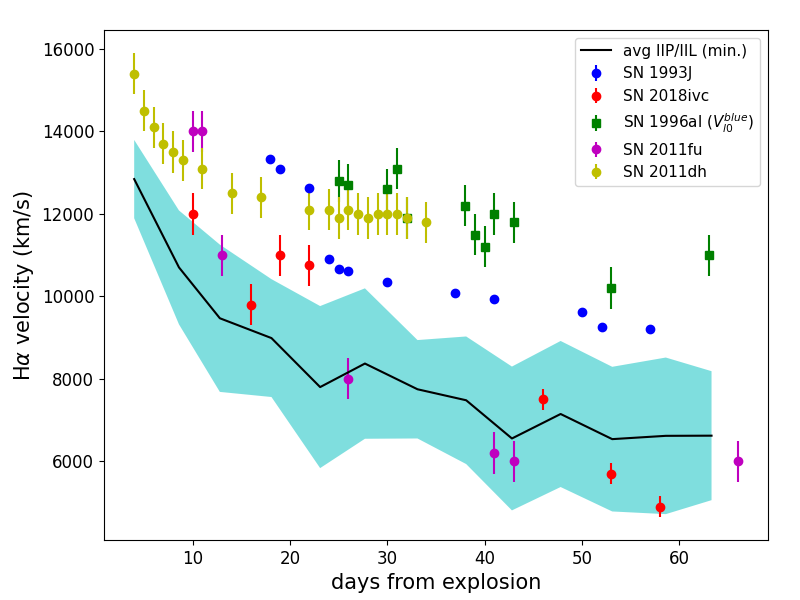}
    \caption{Comparison of velocities from H$\alpha$ in spectra of SNe 1993J, 1996al (considering the blue component of H$\alpha$), 2011fu \citep{Kumar2013MNRAS.431..308K}, 2011dh \citep{Marion2014ApJ...781...69M} 2018ivc (this work), and a sample of SNe IIP/IIL \citep[][using the minimum of the H$\alpha$ absorption]{Gutierrez2017ApJ...850...89G}. For SN 2018ivc, the velocity is derived from the position of the minimum of the absorption.}
    \label{fig:halfa_vel}
    \end{figure}

We collected three spectra at later phases (between +46 and +58 d) that are quite similar to each other. The expansion velocity from the width of the H$\alpha$ line has diminished to about 8000 km s$^{-1}$. Overall, the velocity evolution of H features is consistent with that found by \citetalias{2020ApJ...895...31B}, where spectra at similar phases are available.
The velocity evolution of the H$\alpha$ line in SN 2018ivc is shown in Fig. \ref{fig:halfa_vel}, where is compared to the sample of SNe IIP/IIL from \cite{Gutierrez2017ApJ...850...89G} and to some notable SNe IIb.
%The ejecta velocities of SN~2018ivc from H$\alpha$ at early phases are higher than the average velocity of the sample of SNe IIP/IIL from \cite{Gutierrez2017ApJ...850...89G}, and more similar to SNe IIb rather than SNe IIL, but become lower after one month (see Fig. \ref{fig:halfa_vel} for a comparison).

In the red part of the late spectra, two other emission lines from He~I are detected, namely the $\lambda$7065 and $\lambda$7281, with a $v_{FWHM}\sim4000$~\kms. 
The detection of strong He I lines is typical of Type~IIb SNe, but in those objects the Balmer lines fade away at late phases, while in the spectra of SN~2018ivc H$\alpha$ is always the strongest line. However, the spectroscopic follow-up campaign did not last sufficiently long to check if the Balmer lines disappeared in the months after the explosion.
Because the three late spectra were taken closely in time, and appear very similar, we average-combined them in a single, higher signal-to-noise spectrum, that is used for the comparisons in Sect.~\ref{discussion}.

\begin{figure}
    \includegraphics[width=1.01\columnwidth]{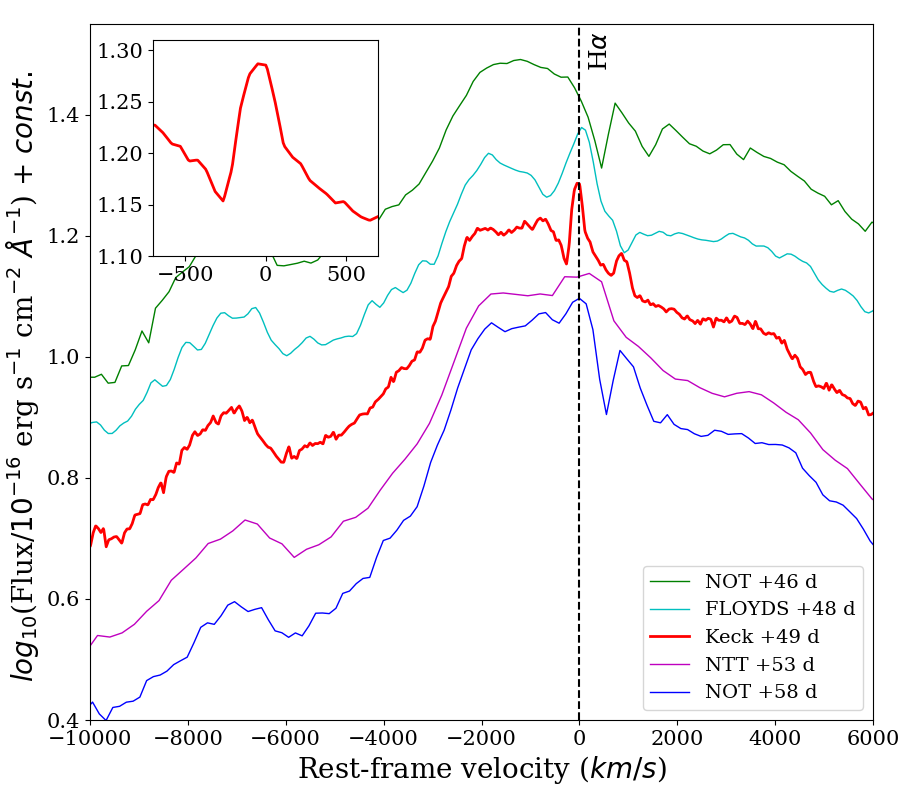}
    \caption{H$\alpha$ line of the +49 d Keck-I/LRIS spectrum of SN~2018ivc published by \citetalias{2020ApJ...895...31B}. The rest-frame H$\alpha$ wavelength is marked and coincides with the narrow H$\alpha$ emission with a narrow P Cygni profile. The HV feature at 6400 \AA~is visible on the left side (see Sect. \ref{Sec:interaction}). A 2m Faulkes/FLOYDS spectrum taken the day before the Keck-I one, also reported in \citetalias{2020ApJ...895...31B}, already shows the narrow absorption feature.
    Our lower-resolution spectra taken near that phase are also shown, in which a `small' emission feature is visible at rest-frame zero velocity.
    A zoom on the P Cygni feature in the Keck spectrum is plotted in the blow-up box.} %'suggesting' that the aforementioned narrow emission line may be real.
    \label{fig:halfa_zoom}
    \end{figure}

In the +49 d Keck-I/LRIS spectrum of SN~2018ivc published by \citetalias{2020ApJ...895...31B}, a narrow H$\alpha$ emission line with a P Cygni profile is visible on top of the broad H$\alpha$ emission (Fig. \ref{fig:halfa_zoom}). This line seems to be present also in their FLOYDS +48 d spectrum, and more marginally in our +53 and +58 d spectra. Although the narrow emission can be contamination from the host galaxy, as the [N~II] $\lambda$6584 line is still visible, there is the possibility of a wind or an expanding shell in front of the ejecta, producing the narrow P Cygni absorption.
This would indicate the presence of a CSM with a complex structure around the progenitor \citep{Dessart2022A&A...660L...9D}.
The expansion velocity of this putative shell, derived from the position of the minimum of the P Cygni absorption, is 250 km s$^{-1}$. The same CSM velocity was measured in SN~1996al \citep{2016MNRAS.456.3296B}. According to \cite{Smith2014ARA&A..52..487S}, this wind velocity is definitively fast for a RSG, and compatible with a more compact progenitor, such as a blue supergiant or a He star.

\citetalias{2020ApJ...895...31B} noted an emission feature to the blue side of H$\alpha$ at $\sim$6400~\AA, which is present also in our spectra of SN~2018ivc (see Figures \ref{fig:spectra} and \ref{fig:halfa_zoom}). The feature is present in all the spectra between phase +15 and +60 days and is strongest in the +22 and +46 d spectra, when the $r$-band light curve presents a change in the declining slope. 
They interpreted it as a high velocity (HV) blue-shifted emission feature from the H$\alpha$ line, generated in a fast-moving ($\sim10^4$ km~s$^{-1}$) clump, approaching the observer.
The identification of this feature as [O I] $\lambda\lambda$6300,6364 is not ruled out; however the rest-frame central wavelength of this feature changes from 6360 to 6415 \AA~between phase $+$15 d and $+$46 d. If this feature would be from the [O I] $\lambda$6300 transition, it would be redshifted by a velocity increasing with time from $+$2850 \kms to $+$5400 \kms.
Therefore, it is easier to explain this feature as a slowing-down H$\alpha$ HV feature instead.
Also, our spectra were taken too early for the SN to have entered in the nebular phase, when [O I] lines are typically observed.

Looking again at Fig. \ref{fig:halfa_zoom}, we note the emergence of a broad shoulder to the red at $\sim+$4000 \kms\,over time, while the peak of the broad H$\alpha$ is offset by about $-$2000 \kms\,to the blue \citep[see also][]{Anderson2014MNRAS.441..671A}.
This, together with the presence of the HV feature, suggests that the ejecta are strongly asymmetric.
% typically one sees a offset at early times that then goes away. You see a less pronounced blueshift at early times, but then the peak in your last spectrum is offset, which is the opposite behaviour to normal.

\subsection{Comparison with various objects}
\begin{figure}
    \includegraphics[width=1.03\columnwidth]{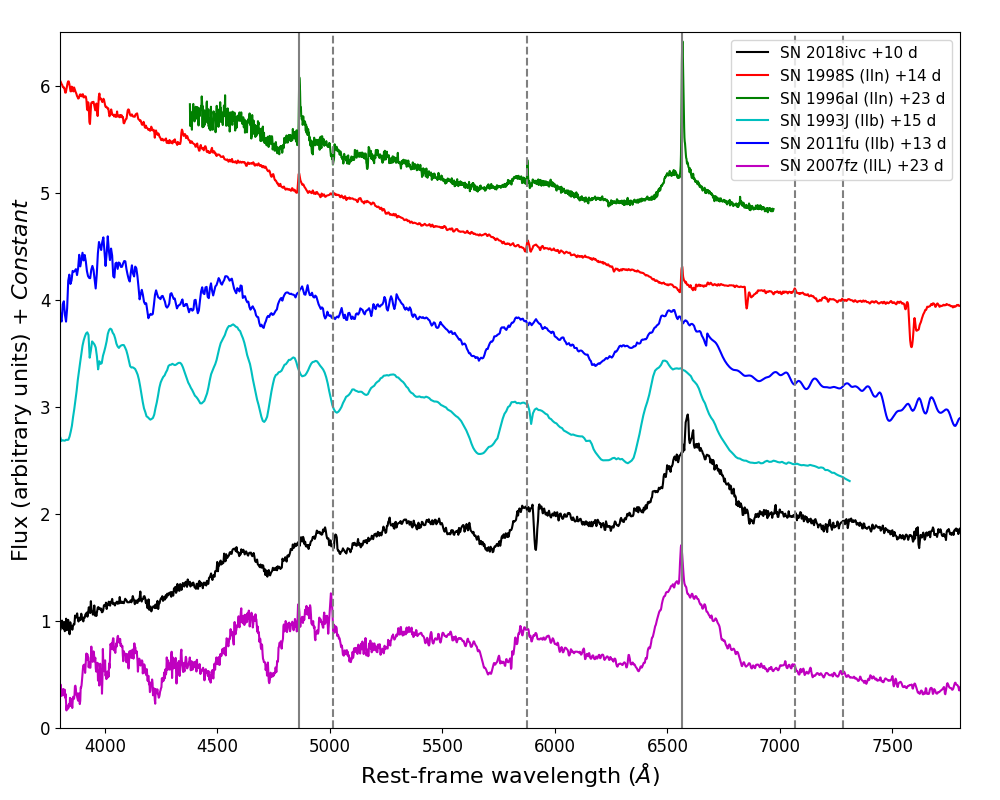}
    \caption{
    Comparison of an early spectrum of SN~2018ivc (+10~d) with those at similar epochs of SNe IIn 1996al and 1998S, SNe~IIb 1993J and 2011fu, SN IIL 2007fz.
    The true spectral appearance of SN 2018ivc in the first days is hidden by the effects of interaction.
    H and He transitions are marked by the continuous and dashed vertical lines, respectively.
    }
    \label{fig:early_spectra}
    \end{figure}

We used the tool \texttt{GELATO} \citep{2008A&A...488..383H} to search for objects with spectra similar to those of SN~2018ivc, both at the early and late phases. Interestingly, at $+1.5/+2$ d the software finds a match with two well-known Type IIn SNe, 1996al and 1998S. \citetalias{2020ApJ...895...31B} interpreted the deviations from the linear decline in the light curve as evidence of interaction between the fast SN~ejecta and a CSM, which would make SN~2018ivc an interacting SN.
However, the spectra of SN 2018ivc never show narrow H emission lines, usual indication of the presence of a CSM surrounding the progenitor star.
Therefore, SN 2018ivc cannot be considered as a Type IIn SN.
%Nonetheless, SN~2018ivc is not a Type IIn SN, neither at late phases, because of the lack of narrow emissions in the spectra from a pre-existing CSM around the progenitor of the SN. 
%Moreover, a CSM would have diluted the photons from the SBO, resulting in a slower rise after the explosion \citep{2018MNRAS.476.2840M}, that is contrary to the early observations.
Running on the early spectra, \texttt{GELATO} gives two Type IIb events, SNe~1993J and 2011fu as comparison objects. This is in agreement with the presence of intense He I emission lines.
The early spectrum of SN 2018ivc and the spectra of the comparison objects are plotted in Fig. \ref{fig:early_spectra}.

\subsection{Internal extinction}\label{Sect:internal_extinction}
The deep absorption from the Na {\sc i} D $\lambda\lambda$5890,5896 doublet indicates the presence of a non-negligible internal extinction. We measured the equivalent width (EW) of the unresolved doublet in all the available spectra, and calculated a mean value of 3.0$\pm$0.4~\AA.
This value is not useful for accurately estimating the dust extinction, as the relations by \cite{2012MNRAS.426.1465P} and \cite{1997A&A...318..269M} between the EW of the Na I doublet and the extinction from the host galaxy saturate for EW $>$0.8 and $>$0.6~\AA, respectively. The saturation of the Na I lines precludes a precise measure of $A_{V,host}$, but it is likely high.
The relation by \cite{2003fthp.conf..200T}, despite being likely saturated, would provide an additional reddening from the host of $A_{V,host}$=1.5$\pm$0.2 mag. Finally, we use the relation of \cite{Rodriguez2023ApJ...955...71R} valid for SNe II to estimate a lower limit on the host galaxy extinction of $A_{V,host}\geq1.4\pm0.3$ mag.
These values are similar to that assumed by \citetalias{2020ApJ...895...31B} who, from a $HST$ pre-explosion image of the field of SN~2018ivc, and from measurement of the Balmer decrement in a VLT+MUSE spectrum of the SN site years before the explosion, estimated a colour excess $E(B-V)\approx0.5$~mag. %Conversely, the relation of \cite{2018A&A...609A.135S} gives $A_{V,host}$=2.34$\pm$0.55 mag.
%We remark that the internal extinction calculated previously is only an estimate, affected by an uncertainty likely much larger, thus we do not corrected the spectra for the host reddening.

As the Type IIn SN~1996al is a comparison object to SN~2018ivc, given its similar light curve evolution and early spectra, we tried to correct the $B-V$ colour curve of SN~2018ivc to match that of SN~1996al at early phases.
%By looking at the $B-V$ colour curve of SN~2018ivc corrected for an internal extinction of $A_{V,host}$=1.5 mag (Fig. \ref{Fig:LC+CLC}, right panel), we see that it is still too red for an interacting SN, being redder at early phases than SN~1993J, for which no evidences of interaction were found. It is also much redder than the type IIn SN~1996al, which is a comparison object to SN~2018ivc given its similar light curve evolution and early spectra.
We interpolated the $(B-V)_0$ colour curve of SN~1996al between the discovery and +50 d phase and calculated the median difference of $B-V$ colour between the two objects, to get the same average colour evolution. We found a colour excess of $E(B-V)=1.04\pm0.04$ mag ($A_V=3.22\pm0.09$~mag, of which 0.09 mag from the MW). % we get the same average colour evolution.
By matching in the same manner the $V-R$ colours between +20 and +70 d we derived a total colour excess $E(V-R)=0.63\pm0.04$ mag (equivalent to $E(B-V)=1.21\pm0.08$ mag).
%Doing the same for the $V-R$ colours (between +20 and +70 d) gives a total colour excess $E(V-R)=0.63\pm0.04$ mag (equivalent to $E(B-V)=1.21\pm0.08$ mag).
%Specifically, by adding an additional $E(B-V)$ of 0.017 mag (corresponding to $A_V=0.053$ mag) to the extinction amount assumed for the high reddening scenario ($A_V=3.0$ mag, for a total of $A_V=3.05$ mag) we get a match of the $(B-V)_0$ colours of SN~1996al and SN~2018ivc between +20 and +50 d.
%The necessary internal extinction to obtain a match between the colour curves of SN~1996al and SN~2018ivc is much higher than that adopted by \citetalias{2020ApJ...895...31B}, but it is also similar to the one we derived from the Balmer decrement of nearby H II regions close to the SN~explosion site.

\begin{figure}
    \includegraphics[width=1\columnwidth]{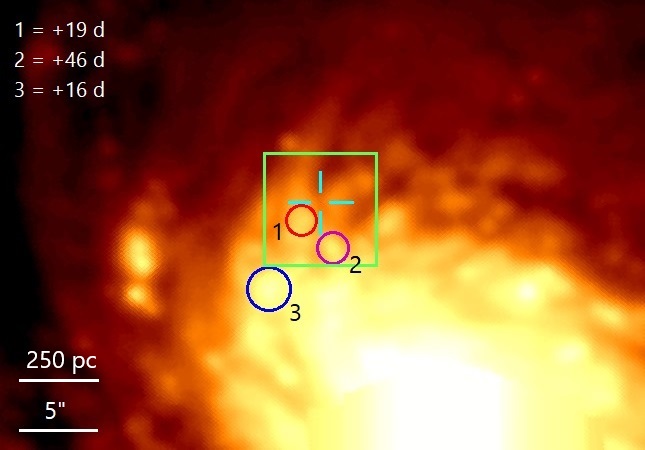} %0.96
    \caption{Archival $r$-band image from the Dark Energy Survey DR1 of the field of SN~2018ivc. The cross marks the SN~position. The field-of-view of VLT+MUSE in narrow-field mode is shown by the square. The H II regions from which we extracted the spectra to evaluate the Balmer decrement are highlighted by the circles. The spectrum of the first H II region is extracted from the spectrum of 2018-12-11 (+19 d), the second from the 2019-01-07 (+46 d) spectrum and the third from the 2018-12-09 (+16 d) spectrum.}
    \label{fig:hii_reg}
    \end{figure}

From our long-slit spectroscopy of SN~2018ivc, we extracted the spectra of three different H~II regions adjacent or close to the SN~position, that fell within the slit when our spectra were taken. The H II regions selected are highlighted in Fig. \ref{fig:hii_reg}. In particular, we measured the flux of H$\alpha$ and H$\beta$ emission lines to evaluate the Balmer decrement. % (Fig.~\ref{fig:hii_reg}, right)
From three different spectra, we obtained an average H$\alpha$/H$\beta$ ratio of 7.5$\pm$0.5. Using the relation by \cite{2012A&A...537A.132B} (their Equation 8), and $A_V/A_{H\alpha}$=1.22 \citep{1989ApJ...345..245C}, we derived an additional host extinction of $A_{V,host}$=3.0$\pm$0.2 mag, which assuming $R_V=$3.1 gives $E(B-V)\sim1.0$, double what we obtained from the Na I D EW and the value assumed by \citetalias{2020ApJ...895...31B}.
The internal extinction we derived from the Balmer decrement of nearby H II regions close to the SN~explosion site is much higher than that adopted by \citetalias{2020ApJ...895...31B}, but it is also similar to the one necessary to obtain a match between the colour curves of SN~1996al and SN~2018ivc. This makes the internal reddening very uncertain, and might be variable across the region.
Therefore, we will not prefer one estimation or the other but evaluate both scenarios equally. We will consider the case of an internal extinction of $A_{V,host}=$1.5 mag (from the Na I doublet) as the `low reddening scenario', while the case of $A_{V,host}=$3.2~mag (from the $B-V$ colour comparison with SN~1996al and the Balmer decrement of close-by H II regions) as the `high reddening scenario'.

In their recent work, \cite{Maeda2023ApJ...942...17M} estimated an upper limit of $\lesssim$0.015 \Msun,on the ejected \Ni\, mass of SN~2018ivc from observations with ALMA, assuming the low reddening scenario and that the \Ni\,heating is not a main power source for the optical luminosity. This value is a factor of $\sim4-5$ smaller than that determined for the canonical Type IIb SN~1993J ($\sim$0.07 \Msun; \citealt{nomoto1993Natur.364..507N, woosley1994ApJ...429..300W}). Instead, in the high reddening case, the upper limit would be larger, at $\sim0.07-0.08$~\Msun. % and would be more consistent with the expectation of a SN~IIb

     % Left: % Right: The spectra of the three H II regions close to the position of SN~2018ivc, labelled as in the left panel.

\subsection{Bolometric light curve} \label{bol}
We constructed the pseudo-bolometric light curve of SN~2018ivc from the contribution from $u-$ to $z-$ bands. For epochs without observations in some bands, we interpolated the available data using the $r$-band light curve as reference and assuming a constant colour index.
For the phases between $\sim$35 and $\sim$70 days, when the NIR magnitudes are available, we calculated the pseudo-bolometric luminosity including also the contributions from $JHK$ bands. In the high-reddening scenario, the NIR bands add a 15\% contribution to the bolometric flux, while this increases to 50\% in the low-reddening one.
However, as the NIR measurements are too few, we did not try to extrapolate the $JHK$ light curves with the assumption of a constant colour.

\begin{figure}
    \includegraphics[width=1.03\columnwidth]{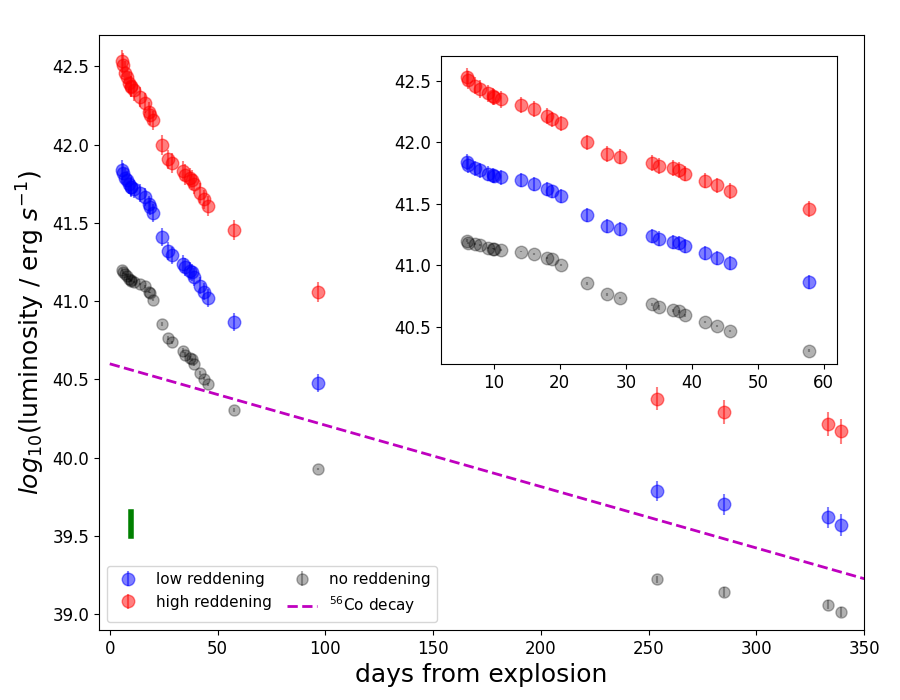}
    \caption{Pseudo-bolometric $uBgVrRiIz$ light curves of SN~2018ivc, for both the low and high reddening scenarios. The inset shows a blow-up of the first two months after the explosion. For reference, we plot also the pseudo-bolometric light curve without any correction for reddening in black; the short plateau feature is more pronounced. The decay slope of $^{56}$Co (0.98~mag~(100~d)$^{-1}$) is reported with a dashed purple line for comparison. At late phases (after +100 days) the evolution of SN~2018ivc is close to being powered by the $^{56}$Co decay. The systematic error bar due to the uncertainty on the distance is reported in the bottom-left corner.}
    \label{fig:bolom}
    \end{figure}

The entire pseudo-bolometric light curves (for the low and high reddening scenarios) are shown in Fig. \ref{fig:bolom}, with a blow-up on the first two months in the inset inbox.
In early phases, we see a steep decline, which is slightly slower between +10 and +20 days, in correspondence with the short-duration plateau observed in the redder bands.
Instead, at phases later than +90~d, the rate of decline is shallower, at 0.92$\pm$0.07~mag~(100 d)$^{-1}$, only slightly slower than the decay of $^{56}$Co (0.98~mag~(100 d)$^{-1}$), though within the error bar can also be compatible with it.

We estimated the ejected \Ni\, mass of SN~2018ivc from the ratio of its bolometric luminosity at the last three epochs (+285 d, +333 d and +339 d) and that of SN~1987A at the same epochs. For SN~1987A, the pseudo-bolometric luminosity was calculated accounting for the contribution of the $UBVRI$ bands. In the low reddening scenario, the average ratio found is $L_{bol}($1987A$)/L_{bol}($2018ivc$)=7.0^{+3.0}_{-2.2}$; assuming a $M$(\Ni) of 0.075 \Msun, for SN~1987A \citep{1989ApJ...346..395W}, the \Ni\, mass derived for SN~2018ivc is 1.1$^{+0.5}_{-0.4}\times10^{-2} M_{\odot}$.
Instead, in the high reddening case, the ratio obtained is $L_{bol}($1987A$)/L_{bol}($2018ivc$)=1.8^{+1.0}_{-0.6}$, hence a mass of \Ni\, for SN~2018ivc of 4.2$^{+2.1}_{-1.5}\times10^{-2} M_{\odot}$.
The estimated \Ni\, mass, even in the high reddening case, is in the lower part of the $M$(\Ni) distribution of SNe IIb: from the analysis of 45 objects of this type, \cite{Rodriguez2023ApJ...955...71R} found that the average \Ni\, mass produced by SNe IIb is $0.066\pm0.006$ \Msun. %\citep[see the samples of][]{2014MNRAS.445.1647M,2018MNRAS.476.3611G}.
The $M$(\Ni) of SN~2018ivc, low for a SN~IIb \citep[see also \citealt{2014MNRAS.445.1647M,2018MNRAS.476.3611G} and the samples of][]{Anderson2019A&A...628A...7A, Meza2020A&A...641A.177M, afs2021ApJ...918...89A}, may explain the missing second \Ni-heating powered peak in the light curves, substituted instead by an interaction-powered plateau \citep{Maeda2023ApJ...942...17M}.
We remark that the above estimates of the $M$(\Ni) are likely only upper limits due to the powering contribution of the interaction to the light curve, even at late phases, as demonstrated by our modelling (see Sect. \ref{Sect:modelling}) and by the late-time rate of decline of the bolometric light curve, which is shallower than that expected from $^{56}$Co decay.

\section{Discussion} \label{discussion}

\subsection{A transitional SN IIL/IIb?}
%It now seems that such a CSM is a fundamental
%feature of RSG stars and may impact, at various levels, all Type
%II SNe (Yaron et al. 2017; Morozova et al. 2017; Dessart et al.
%2017; Moriya et al. 2017; Förster et al. 2018).  Da Hillier & Dessart 2019

Light curves of SNe IIb are typically characterised by a first luminous peak a few days after the explosion, a local minimum and then a second maximum about three weeks later. In some objects (SNe 1993J and 2011fu, more evident in the bluer filters) this second peak is fainter than the first one.
% SN~2013df \citep{2014MNRAS.445.1647M} showed a shallow second peak, that in $B$ even resembles a plateau, and its photometric evolution is remarkably similar to SN~2007fz, which itself is the closest comparison object to SN~2018ivc.

% This second max is powered by 56ni heating, according to Maeda 2022, the lack of the local minimum in the light curves of SN~2018ivc is explained by the ejecta-CSM interaction, whose \textit{radiative contribution} fill up the minimum towards the second maximum, powering the short plateau feature.

% Soderberg+12 pointed out that the progenitor of type IIb SN~2011dh was compact, also 2008ax (AP 08, Chevalier \& Soderberg 2010) e quindi non va bene.
    
% una IIb era senza secondo max, ed era associata a progen esteso
% noi vogliamo esteso perchè lo dice Keichi
    
Between phases +5 and +20 d, the light curves of SN~2018ivc in the red filters show a plateau, while in a few Type IIb SNe (1993J, 2011fu and 2013df; figure 4 of \citealt{2015MNRAS.454...95M}) a secondary maximum is observed at about the same phase.
The absence of a local minimum in the light curve of SN~2018ivc, at the time when the short-duration plateau is visible instead, might be explained by the additional radiation from the interaction between the fast ejecta and a CSM. This is supported by the similar photometric behaviour presented by the SN~IIn 1996al, even though narrow emission lines are not present in SN~2018ivc, and by the results of our hydrodynamical modelling (see Sect. \ref{Sect:modelling}).

\begin{figure*}\centering
    \includegraphics[width=1.8\columnwidth]{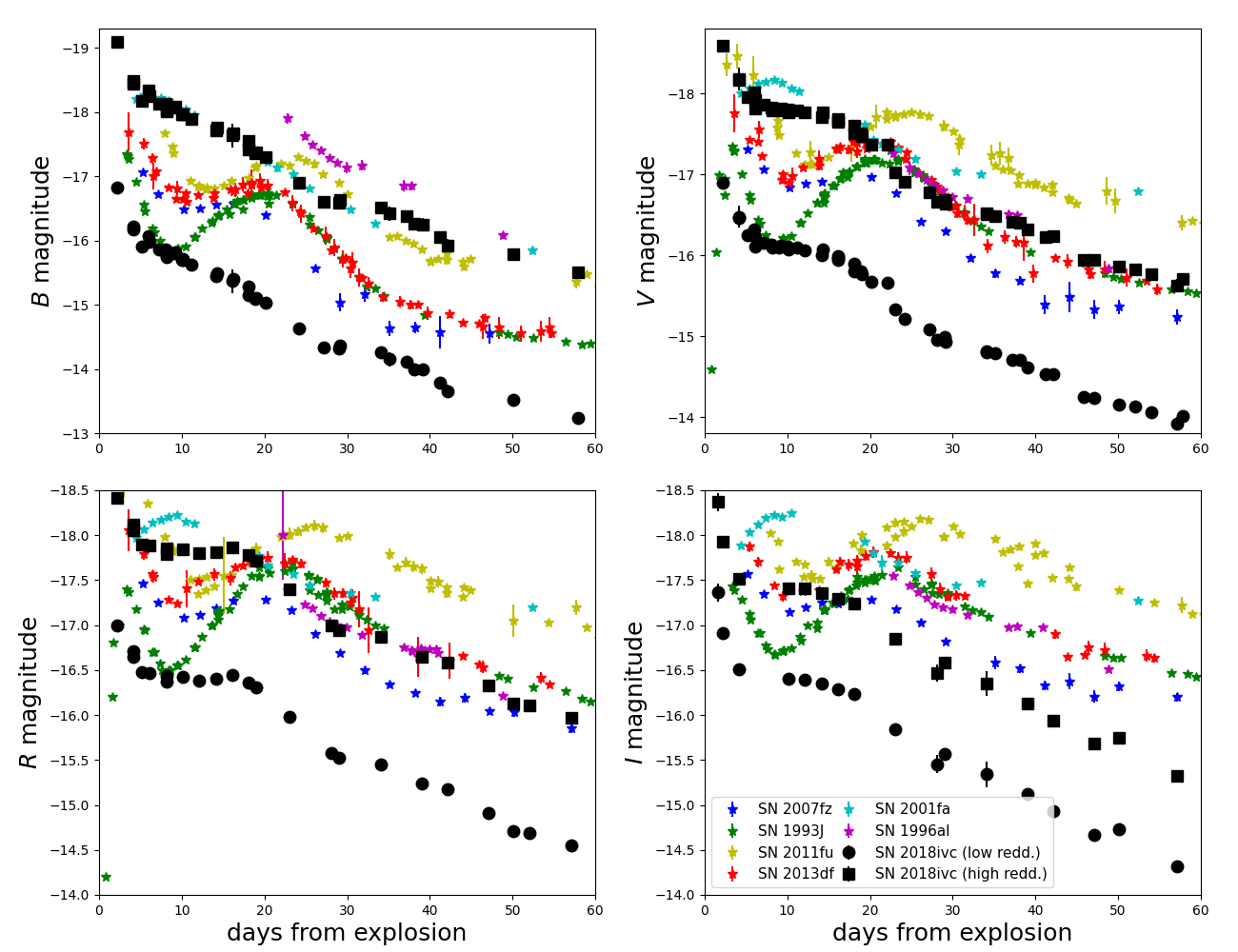}
    \caption{Comparison between the $BVRI$ absolute light curves of SNe~1993J, 2011fu and 2013df (IIb), 1996al (IIn), 2001fa and 2007fz (IIL), and 2018ivc up to 60 days after maximum. For SN~2018ivc, the absolute light curves for both the low and high host reddening scenarios are plotted with circles and squares, respectively. SN~2007fz presents a well-defined secondary peak between one and two weeks after maximum, while SN~2018ivc shows a short plateau. The two objects have quite similar photometric evolution.}
    \label{fig:bvri}
    \end{figure*}

SNe~2007fz and 2001fa \citep{2014MNRAS.445..554F} are two spectroscopically confirmed Type IIL SNe that show a mild increase in brightness between one and two weeks after maximum, similar to that observed in some Type~IIb SNe.
In Fig. \ref{fig:bvri}, we present a comparison of the $BVRI$ absolute light curves of SN~2007fz (together with SNe 1993J and 2001fa) to those of SN~2018ivc in the first 60 days after maximum. We note a good similarity between the two light curves, concerning both the secondary peak and the decline slope at later phases. %We shifted the light curves of SN~2007fz by a constant to have the same absolute magnitude value at phase +20 d after maximum from both objects. 

\begin{figure}
    \includegraphics[width=1.03\columnwidth]{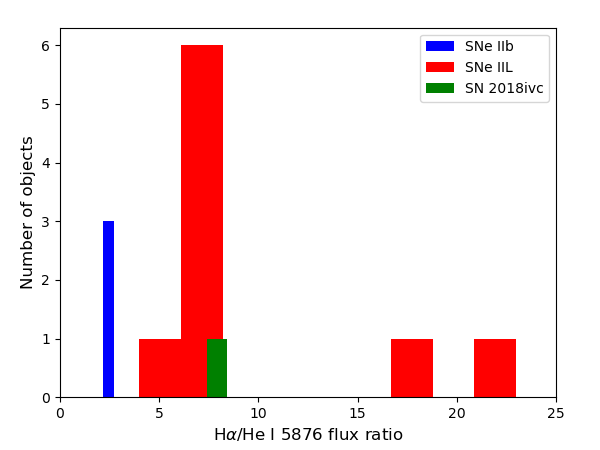}
    \caption{Histogram of the distribution of the average flux ratio H$\alpha$/He I $\lambda5876$ in three SNe IIb (SNe 1993J, 2011fu and 2013df, in blue), 9 SNe IIL from \citealt{2014MNRAS.445..554F} (in red) and SN 2018ivc (in green) between one and three months after the explosion. The majority of SNe IIL have a H$\alpha$/He I $\lambda5876$ ratio of around 7, with some outliers, while SNe IIb have a ratio between 2 and 3. With a ratio of 8.4, SN 2018ivc would be a Type IIL SN. } % da impostor
    \label{fig:istogramma}
    \end{figure}

The second, long linear decline and the late spectrum appearance would make SN~2018ivc a Type IIL SN; however, the presence of He lines and some resemblance of the early spectra with those of SNe~1993J and 2011fu are more coherent with a Type IIb classification. 
As a comparison, in the three late time spectra of SN~2018ivc, we measure a mean intensity ratio H$\alpha$/He~I $\lambda5876\simeq 8.4\pm0.7$. In the spectra of Type IIL SNe from the sample of \cite{2014MNRAS.445..554F}, at phases in the range of +23 to +90~d, the measured ratios are distributed between 4 and 21, but with a majority around 7 (see Fig. \ref{fig:istogramma}).
Differently, in the spectra of SNe~IIb 1993J (from \citealt{1995A&AS..110..513B}), 2011fu and 2013df between +42 and +62 days, the measured ratios are between 2 and 3, clearly separated from the distribution of values for SNe IIL.
According to these measurements, SN 2018ivc belongs to the family of SNe IIL rather than SNe IIb, though it has to be noted that interaction may change the ratio.

%    \begin{figure*}
%    \includegraphics[width=2.1\columnwidth]{Images/sp}
%    \caption{Comparison of the late averaged spectrum of SN~2018ivc (+52 d) with that of SN~2007fz and SN~2001fa, two Type IIL SNe \citep{2014MNRAS.445..554F}, taken at the closest phases. The position of the strongest He I lines are marked with black vertical lines.}
%    \label{fig:spec1}
%    \end{figure*}

\begin{figure*}\centering
    \includegraphics[width=1.9\columnwidth]{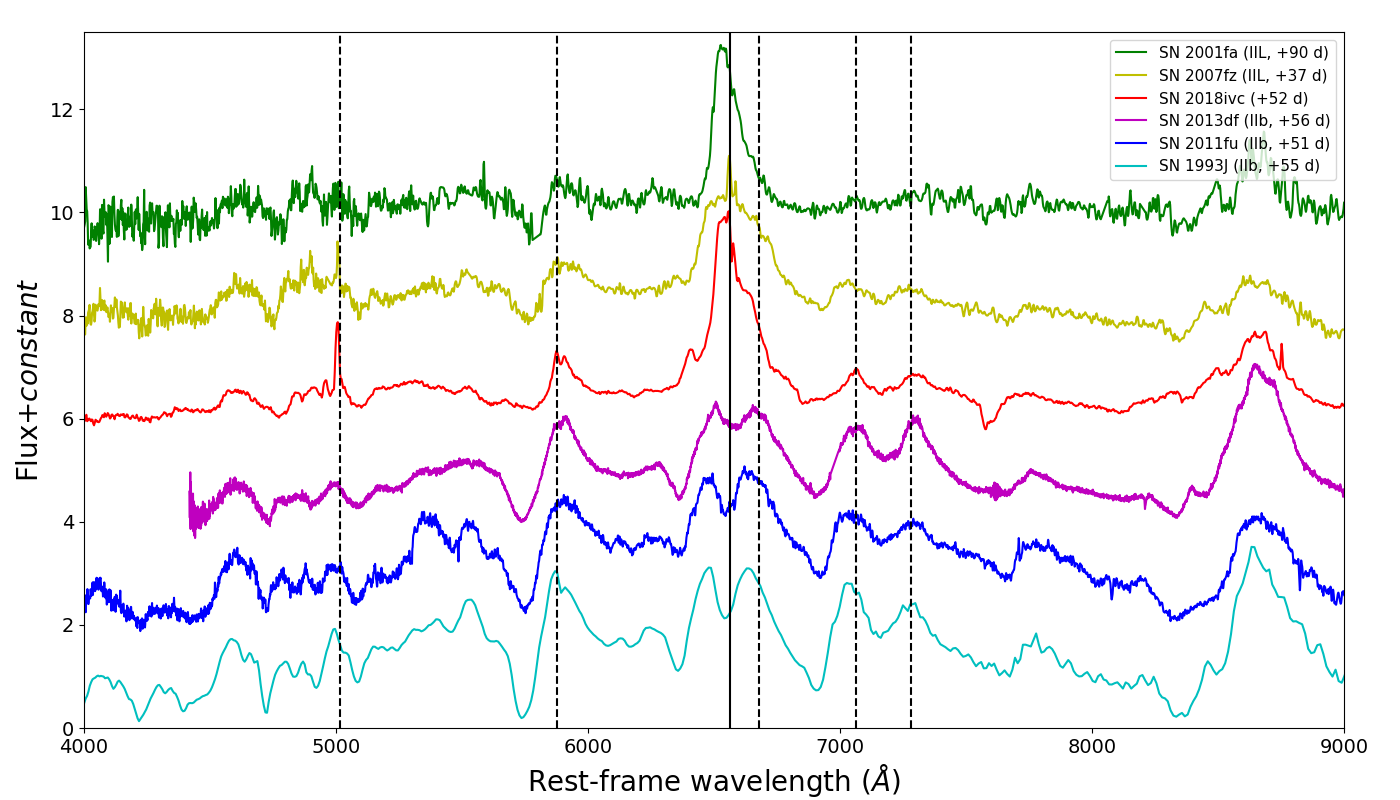}
    \caption{Comparison of the late mean spectrum of SN~2018ivc (+52 d) with those of some reference Types IIL and IIb SNe, taken at late phases. The sample includes: SNe 1993J, 2011fu and 2013df \citep{Shivvers2019MNRAS.482.1545S} for Type IIb, SNe 2001fa and 2007fz \citep{2014MNRAS.445..554F} for Type IIL. SNe IIL have strong H$\alpha$ and faint or no He I lines, the opposite is true for SNe IIb. H$\alpha$ and the principal He I lines are marked by the vertical continuous and dashed lines, respectively.}
    \label{fig:late_spectra} % and SN~2013by \citep{2015MNRAS.448.2608V}
    \end{figure*}

Finally, in Fig. \ref{fig:late_spectra} we compare the late average spectrum of SN~2018ivc with those of a sub-sample of both Type IIL and Type IIb SNe, to see which category the object studied in this work may belong to. SNe IIL are characterised by a strong H$\alpha$, with faint or absent He~I emission lines. On the contrary, SNe IIb show prominent lines from He I, while H$\alpha$ is not visible. By this spectroscopic comparison, SN~2018ivc seems more related to SNe IIL, but with a peculiar light curve. In this sense, probably the best comparison object to SN~2018ivc is SN~2007fz: they have similar H$\alpha$ profiles, He I lines, Ca II NIR triplet and the blue bump at 4600-4700 \AA. Instead, SN~2001fa has a spectrum closer to that of a normal Type II SN, lacking the He I emission lines.

Nonetheless, in their recent paper, \cite{Maeda2023ApJ...942...17M} suggested the SN~type of SN~2018ivc as IIb, based on various indications. The argument was further confirmed by \cite{Maeda2023ApJ...945L...3M} thanks to late-time ALMA observations at 100 and 250 GHz frequencies, which revealed rebrightening synchrotron emission that is consistent with the history of prolonged mass ejection episodes before the explosion proposed by their SN IIb-IIL transition interpretation.

\subsection{An interacting SN IIL?}\label{Sec:interaction}
The light curve of SN~2018ivc, during the early phases of evolution, presents some flattenings (or changes in the declining slope), which are typically interpreted as an indication of an on-going interaction between the SN ejecta and a pre-existing CSM. This interaction is generally revealed by the presence in the spectra of narrow emission lines, from the Balmer series, on top of a much broader component, produced in the fast-moving ejecta.

However, while we detected variations in the slope of the light curve during the post-maximum declining phase, the spectra never show narrow emissions lines. %, a strong evidence of interaction, or a relatively blue continuum at the early stages\footnote{Though, we emphasise the large and uncertain internal reddening suffered by SN~2018ivc, which can affect the shape of the spectral continuum.}.
In Fig. \ref{fig:early_spectra}, we compare the early spectra of SN 2018ivc to those of different SN subtypes, specifically SNe IIn, from which it is clear that SN 2018ivc does not show the typical features of interacting SNe.
%This contrast probably makes SN~2018ivc also a transitional object between the normal Type II and Type IIn classes.

\begin{figure}\centering
    \includegraphics[width=0.95\columnwidth]{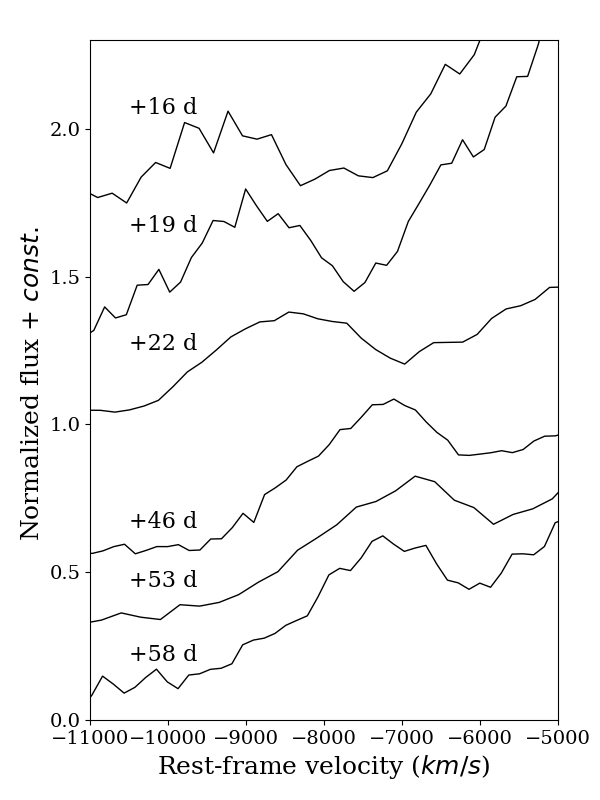}
    \caption{Zoom on the HV feature on the blue side of H$\alpha$ in the rest-frame velocity. The velocity of the feature clearly slows down with the phase.}
    \label{fig:HV}
    \end{figure}

We see the velocity of the HV feature described in Sect. \ref{subsec:line_identification} to slow down with time, as already noted by \citetalias{2020ApJ...895...31B}. The highest velocity, derived by measuring the position of its peak in the +16 d spectrum, is $-$9500 \kms, while the lowest is measured in the +53 d spectrum, at just $-$7000 \kms.
%We performed a linear fit on the velocities of the HV feature with the phase in the six spectra between +16 d and +58 d, obtaining a deceleration of $-$66 \kms d$^{-1}$ and an initial velocity (at phase 0 hence at the explosion) of 10400 \kms.
in Fig. \ref{fig:HV} we plot a zoom on the H$\alpha$ HV feature in the velocity space, where it is evident the movement with the phase towards the red, indicating its slowing down.
The feature seems to be present also in our +46 d spectrum at $\sim$8330 \AA. If it is associated to the Ca~II $\lambda$8542 transition, its recessional velocity would be $-$7400 \kms, consistent with that of the H$\alpha$ HV feature in the same spectrum.
A HV feature may indicate the presence of an asymmetric structure, with which the SN~ejecta later interacted. The CSM may have had a bipolar shape, while the ejecta are more spherically symmetric, or vice-versa. Two HV features (one approaching and one receding) were observed in the H$\alpha$ profile in the spectra of Type IIn SNe~2010jp \citep{2012MNRAS.420.1135S} and 2014G \citep{2016MNRAS.462..137T}, and they offered this interpretation to explain them.
When the fast ejecta interact with a slow-moving CSM, part of its kinetic energy is converted into radiation. This can explain why the velocity slows down from the spectral features of SN 2018ivc at around +1 month, also compared to the average velocity of SNe~IIP/IIL \citep{Gutierrez2017ApJ...850...89G}. %The short plateau observed at around the same phase was likely powered by interaction rather than \Ni\, heating.

%Up to now, it was thought that the progenitors of type IIn SNe were very massive ($>20$ $M_{\odot}$) stars (cit.). 
If indeed SN~2018ivc is also a transitional object between normal Type~II and interacting SNe, it could point in favour of a continuous distribution in properties from SNe II towards SNe IIn, as observed in the luminous low-expansion velocities (LLEV) objects reported by \cite{osmar2020MNRAS.494.5882R}. % be an additional hint of the existence of
LLEV SNe show signs of ejecta-CSM interaction in their light curves during the initial months (4–11 weeks). After that, their spectra evolve as normal Type II SNe.
It is worth noting that LLEV SNe are much more luminous than SN~2018ivc at the characteristic phase of 50 days ($\sim$2 mags even in the high reddening scenario), indicating a stronger interaction, which is also supported by their early appearance as SNe IIn. If this interpretation holds, the different spectral evolution is mainly related to the density profile of the CSM around the progenitor star, which reflects its mass-loss history.  

%the differences being in the initial mass of the progenitor and the density of the CSM around it. %For example, now it is known there is a continuum between Type IIL and Type IIP SNe \citep{2014ApJ...786...67A}.

%It is worth noting that \citetalias{2020ApJ...895...31B}, from $HST$ pre-explosion observations, found a possible progenitor of SN~2018ivc, with an absolute magnitude $M_V\sim-7.5$. Such a luminosity is high even for a massive star in a quiescence state, but it is consistent with an LBV eruption. % Siccome la sorgente è a 22 volte il rms dell'astrometria, forse è un cluster. in v2 mettono solo upper limit anche di -6
On frequent occasions (see, for example, \citealt{Ofek2014ApJ...789..104O, Strotjohann2021ApJ...907...99S, Reguitti2024}), luminous outbursts during a long-lasting eruptive phase were observed a few years before the explosion of some SNe IIn. The mass-loss rate is highly enhanced during these phenomena, leading to the formation of a massive CSM \citep{Smith2014ARA&A..52..487S}. %The detection of such pre-explosion activity would prove the hypothesis of a high massive star as the progenitor of SN~2018ivc.
We searched for pre-explosion images in the archives of major astronomical observatories and all-sky surveys for transients, looking for pre-SN~outbursts from the progenitor of SN~2018ivc in the years prior (between 2000 and 2018), but found no significant variability.

As stated in the previous Section, there is a possible similarity (found by the \texttt{GELATO} algorithm) between the early spectra of SN~2018ivc and those of the Type IIn SN~1998S. This may support the idea of the presence of a CSM close to the progenitor star.
Indeed, looking again at Fig. \ref{fig:late_spectra}, in SN 2018ivc all the broad P Cygni features from H, He, and Ca are shallower than in the comparison objects. This could be an effect due to  interaction with a CSM \citep{Dessart2022A&A...660L...9D}.
Yet, \cite{2015MNRAS.449.1876S} pointed out that, after the initial phases, SN~1998S too shared some properties of Type IIL-like SNe. 
\citetalias{2020ApJ...895...31B} found another Type IIn event, the well-studied SN~1996al \citep{2016MNRAS.456.3296B}, as the object with the most similar light curve to SN~2018ivc, because of a change in slope in the former object during the earliest phases (they even state that SN~1996al can be considered a IIn/IIL SN~transition).

Searching in the literature for other transitional Type IIL/IIn objects, we found one more event with behaviour not dissimilar to SN 2018ivc.
PTF11iqb \citep{2015MNRAS.449.1876S} is a changing-look SN: the spectra at very early phases (a few days after the explosion) show SNe IIn-like features, but then the spectra change and evolve towards a Type II SN, while the light curve presents a plateau, typical of Type IIP SNe. Finally, during the nebular phase, the spectrum returns to be more akin to that of a Type IIn SN. 
PTF11iqb is less luminous than SN~1998S, and its emission lines are fainter, but overall the spectra of the two objects are similar. \cite{2015MNRAS.449.1876S} suggest that PTF11iqb can be explained as a normal core-collapse SN with a weaker CSM interaction than in SN~1998S.
Finally, it has been argued that \textsl{all} SNe IIL experience at least moderate interaction in the initial phase \citep{2015MNRAS.448.2608V, 2015ApJ...806..160B} (though they would not necessarily appear as SNe IIn, with narrow emission lines, if the CSM is not dense enough; \citealt{Dessart2022A&A...660L...9D}). After the first days, the spectra evolved as normal Type IIL or IIP SNe, like it was for PTF11iqb.
For instance, \cite{2017ApJ...838...28M} demonstrated that signs of interaction were present in the light curves of the Type IIL SNe 2013ej and 2013fs, even though the typical features of Type IIn SNe are not revealed in their spectra (see also \citealt{Bullivant2018MNRAS.476.1497B}). %, 2013by

%Moreover, according to \cite{Hillier2019A&A...631A...8H}, the difference between Type IIL and IIP SNe (or fast- and slow-decliners) may actually be explained by a larger or smaller amount of CSM in the proximity of the progenitor. e pendenza declino dip da quanto forte interaz

%SN~2018ivc has revealed to share commonalities of many different Types of SNe, hence a coherent and decisive classification is tricky.

%A change in the spectral appearance may suggest a connection between Type II SNe and interaction with a CSM. 

\subsection{Modelling of the early light curve} \label{Sect:modelling}
The early light curve of SN~2018ivc is suggested to be powered by the interaction between the ejecta and the surrounding CSM in both \citetalias{2020ApJ...895...31B} and \cite{Maeda2023ApJ...942...17M}. \citetalias{2020ApJ...895...31B} derived an upper limit for the ZAMS mass of a `single star’ progenitor, estimated to be less than 11 \Msun, based on pre-supernova $HST$ images captured at the SN's location. In \cite{Maeda2023ApJ...942...17M}, the optical bolometric light curve of SN~2018ivc has been simulated with the SN-CSM interaction model proposed in \cite{Maeda2022}. To explain the unique behaviour of the light curve, characterised by a nearly flat evolution up to 17 days post-explosion followed by a transition phase between 20 and 30 days and, thereafter, a relatively steeper decline, \cite{Maeda2023ApJ...942...17M} proposed a two-component density profile for the CSM enveloping SN~2018ivc. This profile consists of a flatter density configuration ($\rho$~$\propto$ r$^{-1.6}$) spanning from 5 $\times$ 10$^{14}$ cm ($\sim$7200 \Rsun) to 2 $\times$ 10$^{15}$ cm ($\sim$29000~\Rsun) from the progenitor to simulate the flat evolution until 17 days. A steeper density configuration ($\rho$ $\propto$ r$^{-2.5}$) until 1.5 $\times$ 10$^{16}$ cm from the progenitor has been used to reproduce the steeper decline after 30 days. %The initial flat evolution ($\lesssim$ 17 days) and the latter steep decline ($\gtrsim$ 30 days) were computed separately in their work. 
Furthermore, their analysis estimated the synthesised \Ni\ mass resulting from the explosion to be below 0.015 \Msun, and their modelling approach did not consider any contribution from the radiation produced by the decay of $^{56}$Ni. We employed the Supernova Explosion Code (\texttt{SNEC}; \citealt{Morozova2015}), an open-source Lagrangian 1D radiation hydrodynamic code, to simulate the intriguing behaviour of the fast rise followed by a sustained plateau for $\lesssim$ 17 days post-explosion in the multi-band optical LCs of SN~2018ivc. %This simulation aims to establish constraints on the progenitor and CSM parameters of SN~2018ivc. 

In the case of SN~2018ivc, we were able to record a clear and steep rise to maximum brightness, but this was only observed in the Clear band. The fast rise holds substantial importance, particularly given the limited availability of early-time data for events of such peculiar nature. It plays a pivotal role in effectively constraining the characteristics of both the CSM and the progenitor. Thus, our primary focus is centred on conducting multi-band light curve modelling using \texttt{SNEC}, with a specific emphasis placed on the Clear band. The \texttt{SNEC} simulation framework relies on the assumptions of diffusive radiation transport and local thermodynamic equilibrium. Regarding the progenitor model, we explore two scenarios: 
\begin{itemize}
    \item a typical extended progenitor of Type IIb SNe with an H envelope mass (M$_H$) of 0.02 \Msun,
    \item stripped progenitor models with three different M$_H$ values: 0.38, 0.74 and 1.61 \Msun.
\end{itemize}
The best-fitting progenitor model is determined by finding the minimal $\chi^2$, computed as
\begin{equation}
\chi^2 = \sum\limits_{\lambda} \frac{1}{N_\lambda}\sum\limits_{t \textless 20d} \left(\frac{m^\mathrm{obs}_\lambda(t) - m^\mathrm{calc}_\lambda(t)}{\Delta m^\mathrm{obs}_\lambda(t)}\right)^2,
    \label{eq:chi2}
\end{equation}
where $m^\mathrm{obs}_\lambda$(t) and $\Delta m^\mathrm{obs}_\lambda$(t) are the magnitudes and their corresponding errors at time t, $m^\mathrm{calc}_\lambda$(t) are the computed magnitudes at time t, $\lambda$ denotes the various filters, and $N_\lambda$ is the total number of observed data points in filter $\lambda$. We compute $m^\mathrm{calc}_\lambda$ by setting $R_V=3.1$ and varying $E\left(B-V\right)$ in the range of low and high reddening values.

\subsubsection{Light curve modelling with SNe IIb progenitor model}
In this case, the SNe IIb progenitor models are calculated by the \cite{Nomoto1988PhR...163...13N} prescription from the ZAMS to pre-explosion conditions and correspond to H-free structures. Later, an extended H envelope is attached to the He core, assuming thermal and hydrostatic equilibrium. The mixing length parameter used is 2.5, and the stellar models are calculated assuming solar metallicity \citep[see][]{Bersten2012ApJ...757...31B}.
The final progenitor model used corresponds to the stellar evolution of a single star with a ZAMS mass of 15~\Msun, a radius equivalent to 350~\Rsun\footnote{We tried changing the progenitor radius between 350 and 500 \Rsun, but the modelled light curve did not change remarkably.}, a helium core mass (M$_{He}$) of 4~\Msun, and M$_{H}$ of 0.02\,\Msun.
These parameter values are typical for the extended progenitors of SNe IIb \citep[see][]{Chevalier2010ApJ...711L..40C, Ouchi2017ApJ...840...90O, Yoon2017ApJ...840...10Y, Sravan2019ApJ...885..130S}.
The H-envelope mass falls in the range 0.01-0.5 \Msun\ proposed by \cite{Sravan2019ApJ...885..130S} for the progenitors of SNe IIb (see also Table 5 of \citealt{Balakina2021MNRAS.501.5797B}), and agrees with the finding of \cite{Yoon2017ApJ...840...10Y} that SN IIb progenitors have a H-envelope mass $\lesssim$0.15 \Msun, but is lower than the minimum value of 0.033 \Msun\ found by \cite{Gilkis2022MNRAS.511..691G}. However, according to \cite{Dessart2011MNRAS.414.2985D}, a H-envelope mass of just 0.001 \Msun\ is enough to reproduce a normal SN IIb.
%For the H-envelope mass, \cite{Maeda2023ApJ...942...17M} adopted a value of 0.5-1 \Msun, similar to that of SN 1993J (see also Nakar & Piro 14?).

For the explosion, \texttt{SNEC} takes the progenitor and explosion parameters as input and generates a range of outputs including multi-band and bolometric light curves, and photosphere velocity evolution. However, due to the unique characteristics of SN~2018ivc, determining suitable parameter ranges for modelling posed challenges. The common two-step approach adopted in previous studies involving \texttt{SNEC} modelling \citep{Morozova2018}, which entails initially fitting the late-time light curve to constrain mass and explosion energy, followed by fitting the early light curve to determine CSM properties, was not as effective in this case. This is because the evolution of SN~2018ivc is suggested to be predominantly governed by the interaction between the ejected material and the CSM even during later phases \citep{Maeda2023ApJ...942...17M}.

\begin{figure}
%first figure 
\begin{minipage}[h]{1.0\linewidth}
\centering
    \includegraphics[width=1.05\columnwidth, clip, trim={0 0 0 2.9cm}]{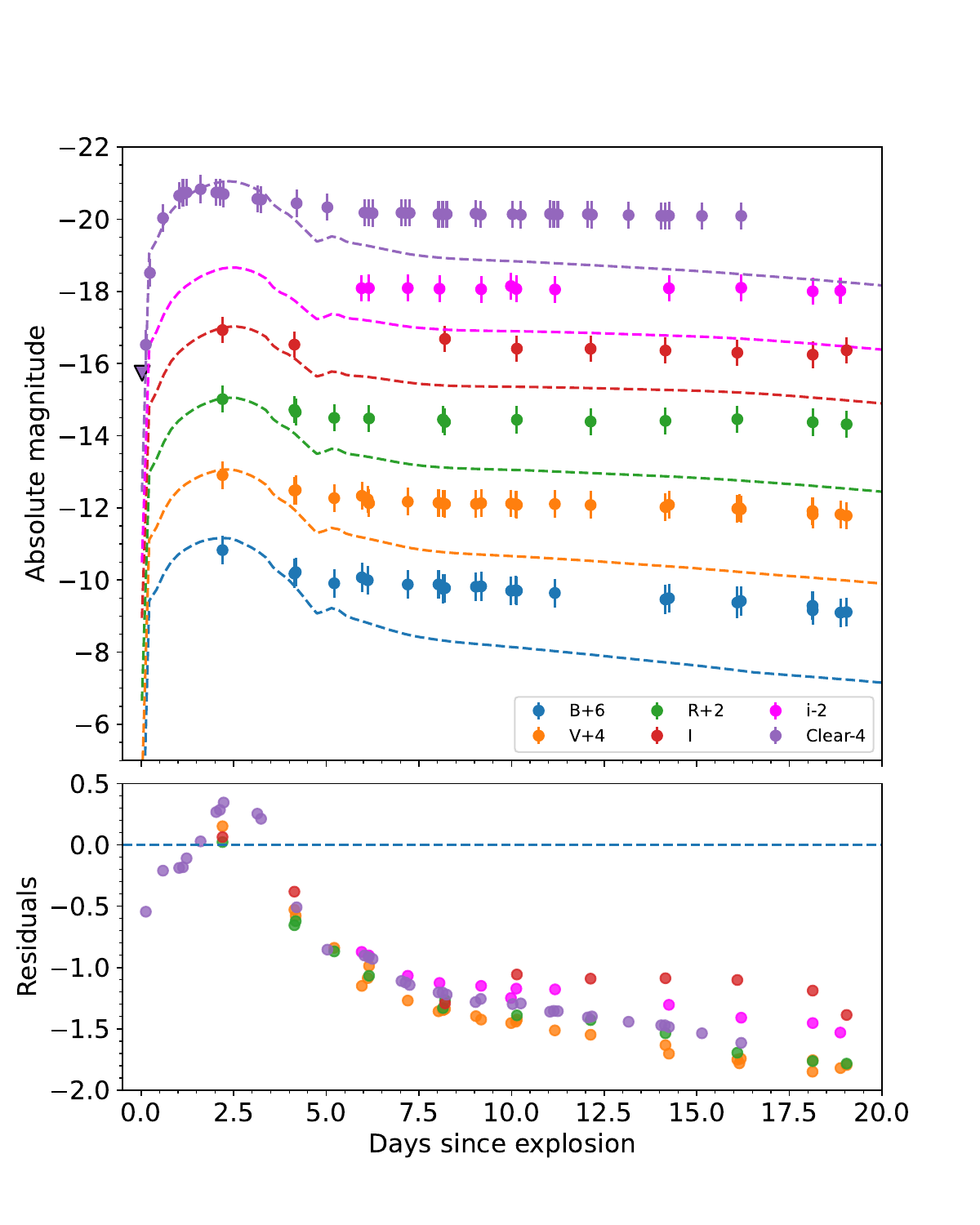}
\end{minipage}

\vspace{-0.50cm} 

%second figure 
\begin{minipage}[h]{1.0\linewidth}
\centering
    \includegraphics[width=1.0\columnwidth, clip, trim={0 0 0 0.cm}]{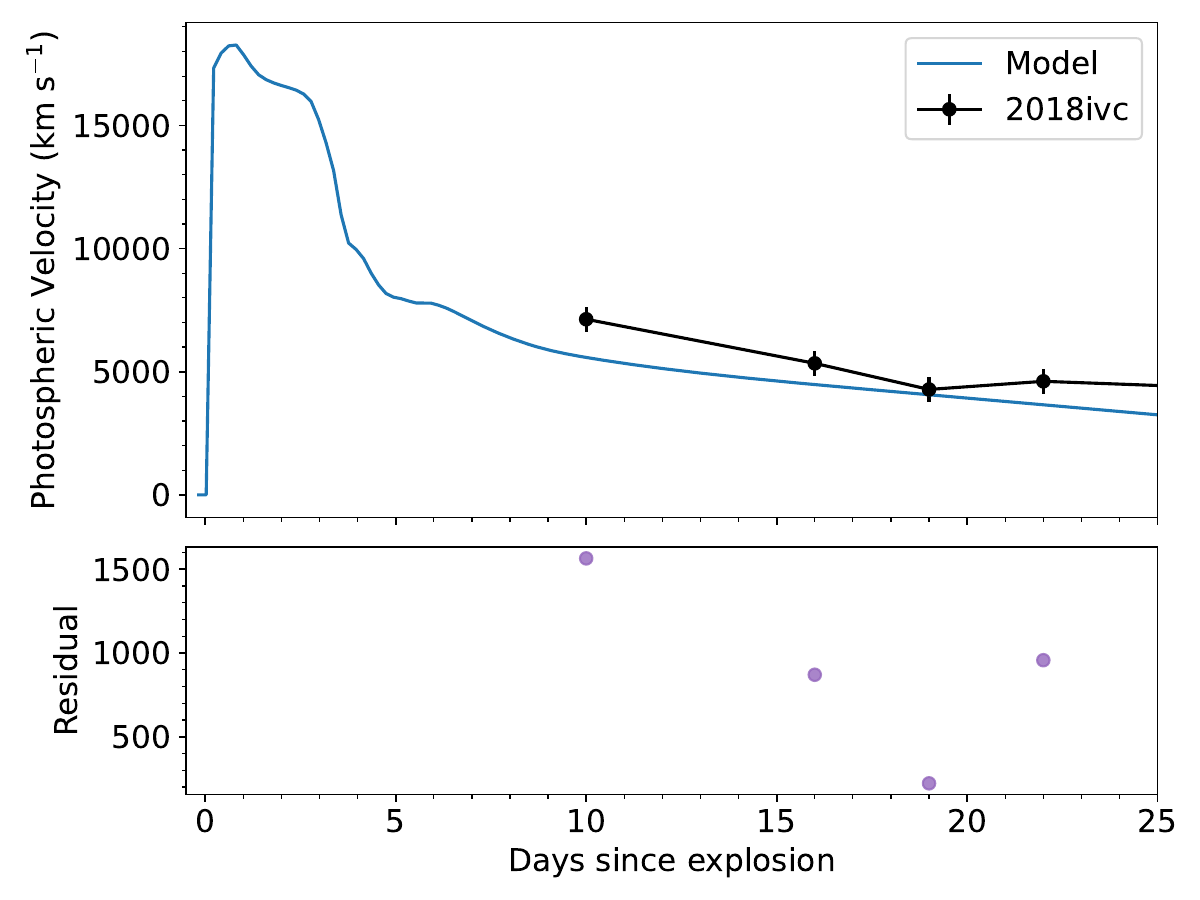}
\end{minipage}   

\caption{Models evolved from the progenitor without adding CSM. The top panel displays the best fit model of the first 20 days of $BVRIi$ and \textit{Clear} LCs of SN~2018ivc. The bottom panel shows the photospheric velocity corresponding to the best fit model along with the observed photospheric velocity.}\label{fig:NoCSM_models}
\end{figure}

To address this, we began by constraining the range of explosion energy through simulations of the progenitor model using \texttt{SNEC}. Multiple simulations were conducted with energies ranging from 0.1 to 1.0 foe (1 foe = $10^{51}$ erg), keeping the \Ni\ mass fixed at 0.015 \Msun. The best fitted model from these simulations corresponds to an A$_V$ of 1.5 mag and explosion energy of 0.7 foe, capable of reproducing the observed multi-band LCs until approximately four days post-explosion. The model photospheric velocities are lower than the expansion velocities estimated from the spectra, as determined by the minima of the \ion{Fe}{ii} $\lambda$5169 line. While increasing the explosion energy can enhance the model photospheric velocity, such models fail to reproduce the rise in the Clear band LC. The best fitted multi-band light curve models without CSM and the photospheric velocities are depicted in Fig.~\ref{fig:NoCSM_models}. 

Next, we added the CSM above the progenitor models with a density profile described by the equation:
\begin{equation}
    {\rm \rho_{\rm CSM}(r) = \frac{\dot M r^{n_{\rm CSM}}}{4\pi \nu_{\rm CSM}}= K_{\rm CSM} r^{n_{\rm CSM}}}
    \label{eqn1}
\end{equation}
Here ${\rm \dot M}$ is the mass-loss rate, ${\rm \nu_{\rm CSM}}$ is the CSM velocity, ${\rm K_{\rm CSM}}$ is the mass-loading parameter, and ${\rm n_{\rm CSM}}$ is the power index of the CSM density radial distribution. Thus, ${\rm \dot M}$ can be estimated from ${\rm K_{\rm CSM}}$ and ${\rm \nu_{\rm CSM}}$. 

The CSM mass can be written as
\begin{equation}
{\rm M_{\rm CSM} = \int_{R_{in}}^{R_{\rm ex}} 4 \pi \rho_{\rm CSM}(r) r^2 dr}
\label{eqn2}
\end{equation}
where ${\rm R_{in}}$ is the inner CSM radius and ${\rm R_{\rm ex}}$ is the outer CSM radius.

Replacing ${\rm \rho_{\rm CSM}(r)}$ in Eqn. \ref{eqn2} with Eqn. \ref{eqn1} and integrating, we get
\begin{equation}
{\rm M_{\rm CSM} = \frac{4\pi K_{\rm CSM}}{n_{\rm CSM}+3} (R_{\rm ex}^{n_{\rm CSM}+3} - R_{in}^{n_{\rm CSM}+3})}
\label{eqn3}
\end{equation}

Thus, we need to derive an expression for ${\rm K_{\rm CSM}}$ to estimate the CSM mass.
To maintain continuity between CSM with varying densities and the progenitor, we introduced an intermediate component bounded by the progenitor radius and the inner CSM radius. While this intermediate component maintains the same density form, it utilises a distinct power value (${\rm n_{INT}}$) for the radius. 

\begin{equation}
    {\rm \rho_{INT}(r) = K_{INT} r^{n_{INT}}}
    \label{eqn4}
\end{equation}

We set the density of the intermediate component at the progenitor radius equal to the progenitor density ${\rm \rho_{prog}}$ at the star's radius (R$_\star$). Thus,

\begin{equation}
    {\rm \rho_{INT}(R_\star) = K_{INT} R_{\star}^{n_{INT}} = \rho_{prog}(R_\star)\\
    \rightarrow K_{INT} = \rho_{prog}(R_\star) R_{\star}^{-n_{INT}}}
    \label{eqn5}
\end{equation}

Also, to maintain continuity between the intermediate shell and the CSM, the density of the CSM at ${\rm R_{in}}$ has to be set to the density of the intermediate shell at ${\rm R_{in}}$. Thus,

\begin{equation}
    {\rm \rho_{\rm CSM}(R_{in}) = K_{\rm CSM} R_{in}^{n_{\rm CSM}} = \rho_{INT}(R_{in})}
    \label{eqn6}
\end{equation}

and, since the density of the intermediate component at ${\rm R_{in}}$ can also be written as (and using ${\rm K_{INT}}$ from Eqn. \ref{eqn5}):

\begin{equation}
{\rm \rho_{INT}(R_{in}) = K_{INT} R_{in}^{n_{INT}} = \rho_{prog}(R_\star) R_{\star}^{-n_{INT}} R_{in}^{n_{INT}}}
\label{eqn7}
\end{equation}

we can finally obtain an expression for ${\rm K_{\rm CSM}}$, by substituting ${\rm \rho_{INT}(R_{in})}$ from Eqn. \ref{eqn7} to Eqn. \ref{eqn6},

\begin{equation}
    {\rm K_{\rm CSM} = \rho_{prog}(R_\star) R_{\star}^{-n_{INT}} R_{in}^{n_{INT}-n_{\rm CSM}}}
    \label{eqn8}
\end{equation}

\begin{figure}
    \includegraphics[width=1.05\columnwidth, clip, trim={0 0.1cm 0 1.0cm}]{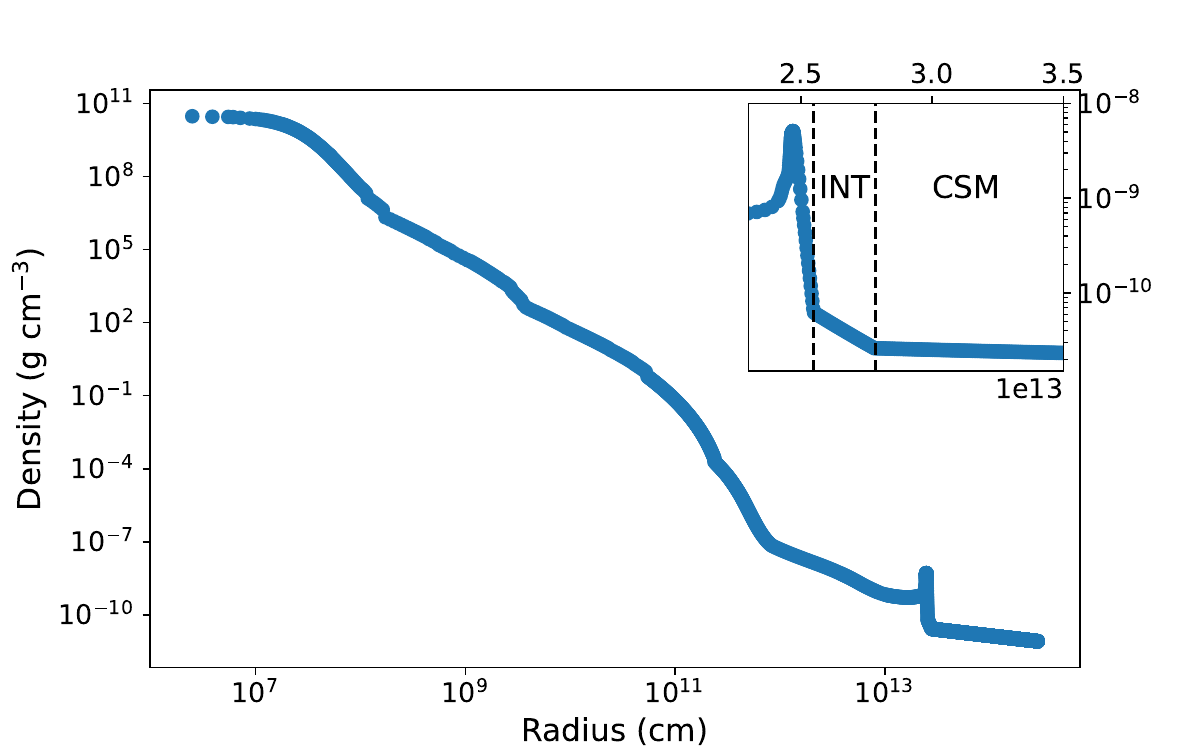}
    \caption{Density profile attached to the progenitor. The inset shows a zoom-in of the intermediate component (`INT') attached between the progenitor and the CSM.}
    \label{density_profile}
\end{figure}

The final CSM structure is shown in Fig.~\ref{density_profile}. The kink in the density at the edge of the progenitor is required to stabilise the hydrostatic structure of the extended envelope.
To accommodate CSM with different densities, we varied ${\rm n_{INT}}$ and chose to keep the inner radius of the CSM fixed at 400 \Rsun. 

\begin{figure}
    \includegraphics[width=1.0\columnwidth, clip, trim={0 0 0 0.0cm}]{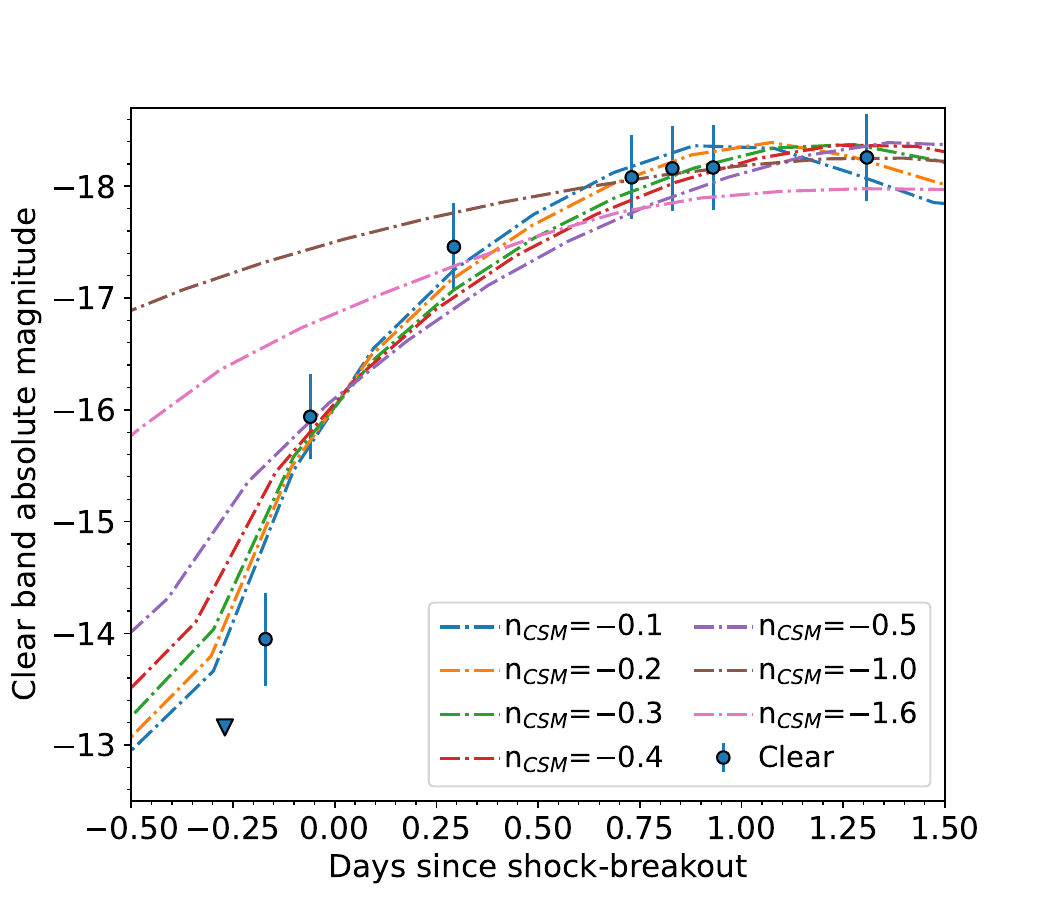} %1.8cm
    \caption{Effect of ${\rm n_{\rm CSM}}$ on the rise of the Clear band LC. Increasing the value of ${\rm n_{\rm CSM}}$ resulted in steeper LCs.}
    \label{fig:models_exp_comp}
\end{figure}

\begin{figure}
%first figure 
\begin{minipage}[h]{1.0\linewidth}
\centering
    \includegraphics[width=1.05\columnwidth, clip, trim={0 0 0 2.9cm}]{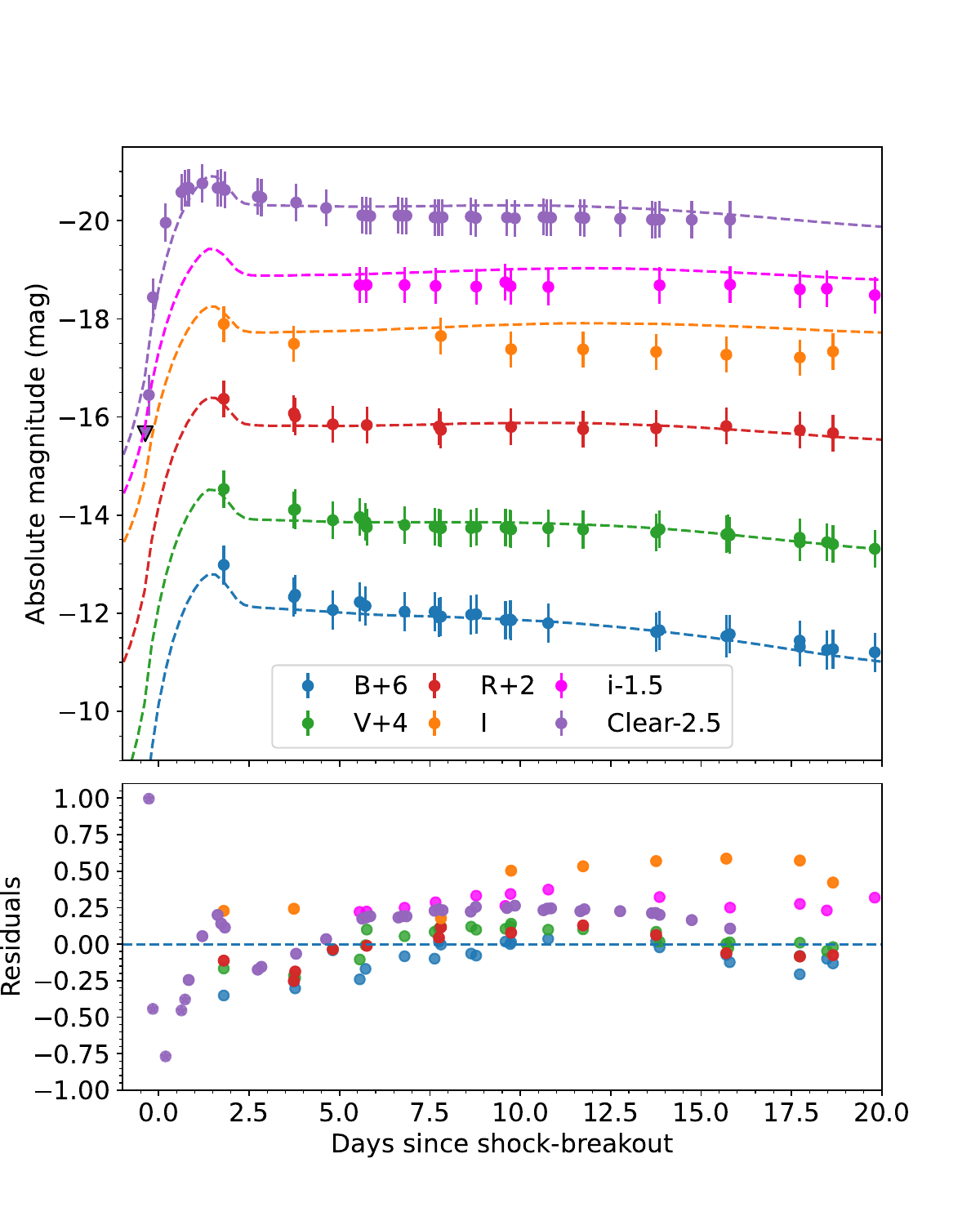}
\end{minipage}

\vspace{-0.60cm} 

%second figure 
\begin{minipage}[h]{1.0\linewidth}
\centering
    \includegraphics[width=1.0\columnwidth, clip, trim={0 0 0 0cm}]{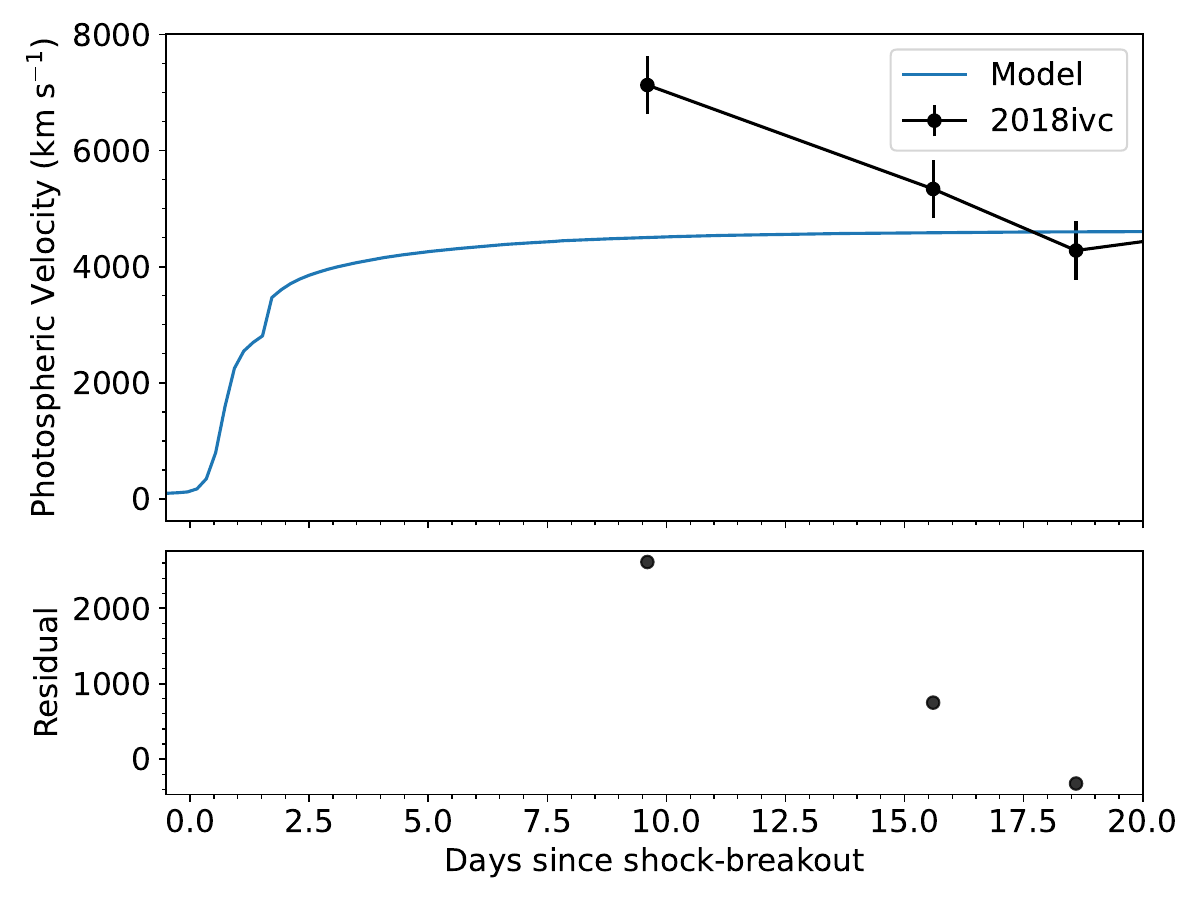}
\end{minipage}   

\caption{Best fit model with the addition of CSM. The top panel features the best fit model of the first 20 days of $BVRIi$ and \textit{Clear} LCs of SN~2018ivc along with the observed multi-band LCs. The bottom panel shows the photospheric velocity evolution corresponding to the best fit model along with the observed photospheric velocity.}\label{fig:models}
\end{figure}

Initially, we used ${\rm n_{\rm CSM}}$ = $-$1.6 in the CSM density profile, as estimated in \cite{Maeda2023ApJ...942...17M}, for the inner CSM responsible for the evolution in the first 17 days. However, the inner CSM's radius in \cite{Maeda2023ApJ...942...17M} was situated at a significantly larger distance compared to our study. Moreover, this choice resulted in a shallower rise, which did not match the steep rise observed in the Clear band light curve of SN~2018ivc. Since reproducing this steep rise was a crucial objective, we investigated how altering ${\rm n_{\rm CSM}}$ impacted the light curve shape. We found that increasing the value of ${\rm n_{\rm CSM}}$ resulted in a steeper rise in the LC. In Fig.~\ref{fig:models_exp_comp}, the effect of varying ${\rm n_{\rm CSM}}$ on the modelled Clear band LCs up to 1.5 d post-explosion for a specific set of explosion energy, ${\rm R_{\rm ex}}$, and ${\rm n_{INT}}$ values (0.3~foe, 4000~\Rsun, and $-$10, respectively) are shown.

%\begin{figure}
%first figure 
%\begin{minipage}[h]{1.0\linewidth}
%\centering
%    \includegraphics[width=1.05\columnwidth, clip, trim={0 0.8cm 0 2.9cm}]{Models/Takashi_model_R400_lowred.pdf}
%\end{minipage}

%\vspace{-0.2cm} 

%second figure 
%\begin{minipage}[h]{1.0\linewidth}
%\centering
%    \includegraphics[width=1.0\columnwidth, clip, trim={0 0 0 0.cm}]{Models/vel_18ivc_lowred.pdf}
%\end{minipage}   

%\caption{Best fit model with the addition of CSM for the low-reddening case. The top panel features the best fit model of the first 20 days of $BVRIi$ and \textit{Clear} LCs of SN~2018ivc along with the observed multi-band LCs. The bottom panel shows the photospheric velocity evolution corresponding to the best fit model along with the observed photospheric velocity.}\label{fig:models_lowred}
%\end{figure}

The slope of the rise in the model LCs is also affected by the outer CSM radius (${\rm R_{\rm ex}}$). A higher ${\rm R_{\rm ex}}$ leads to a shallower rise and a broader peak. We also explored the influence of the \Ni\ mass on the early light curve and found that the \Ni\ mass had a negligible impact on the early LC. Thus, we fixed the \Ni\ mass at 0.015 \Msun. The \Ni\ mixing parameter, which signifies the mass coordinate up to which \Ni\ is distributed outward in the ejecta, was fixed at 4 \Msun. Nonetheless, this parameter had no discernible impact on the early LC.

Considering these factors, we are left with four modelling parameters: ${\rm n_{INT}}$, ${\rm n_{\rm CSM}}$, ${\rm R_{\rm ex}}$, and explosion energy. We proceeded by generating models with explosion energies in the range 0.1 to 0.7~foe, incrementing in steps of 0.1 foe. For the CSM extent parameter, we systematically explored a range spanning from 1500 to 5000 $R_\odot$, incrementing in steps of 100 $R_\odot$. For ${\rm n_{\rm CSM}}$, we explored a range from $-$0.5 to $-$0.1, with increments of 0.1. For ${\rm n_{INT}}$, we explored the range between $-$8 and $-$15. From these simulations, we identified the best fit parameters. %for both the low-reddening and high-reddening LCs. When the low-reddening value is adopted, the explosion energy that best reproduces the observed multiband LCs is 0.1 foe. The estimations of the other parameters are as follows: ${\rm R_{\rm ex}}$~=~2200 $R_\odot$, ${\rm n_{\rm CSM}}$ = $-$0.2, ${\rm n_{INT}}$ = $-$10. The best fit LCs and the corresponding photospheric velocities are shown in Fig.~\ref{fig:models_lowred}. However, we note that due to the low explosion energy, the model photospheric velocities are underestimated.

The best-fitted model correspond to ${\rm R_{\rm ex}}$~=~4100 $R_\odot$, explosion energy = 0.3 foe, ${\rm n_{\rm CSM}}$ = $-$0.5, ${\rm n_{INT}}$ = $-$10. The $A_V$ value for this model fit is 3.2 mag. These parameter choices yielded a model that achieved an optimal alignment with the observed multiband LCs. The best fit model and the corresponding photospheric velocity evolution are shown in Fig.~\ref{fig:models_new}, along with the observed multiband LCs and expansion velocities, respectively. While it is feasible to achieve a better fit for the rising portion by decreasing ${\rm R_{\rm ex}}$ and increasing n$_{\rm CSM}$, such an adjustment leads to a deterioration in the fit for the light curve following the initial rise.
Using these parameter values in Eqn. \ref{eqn3} and \ref{eqn8}, we estimate a CSM mass of 0.47 \Msun. % in 96al M_CSM entro 3x10^16 cm è 0.13 Msol.
%With this CSM mass, the electron scattering optical depth will be very high ($>>100$). Hence, the mean free path of recombination photons responsible for the formation of the narrow features, typically observed in the spectra of SNe IIn, will be much smaller than ${\rm R_{\rm CSM}}$. This explains why we did not observe narrow lines in the spectra of SN 2018ivc. The narrow features can only form in case of radiation leakage from the shock into the CSM above, and this cannot occur when the CSM density is high \citep[on the order of $10^{-10}-10^{-9}$ g cm$^{-3}$,][]{Dessart2017A&A...605A..83D}. For instance, at $3\times10^{13}$ cm (${\rm R_{in}^{\rm CSM}}$), the mean free path of the photons will be 1/$\kappa\rho$. In our model, $\rho\sim 2\times10^{-11}$ g cm$^{-3}$, assuming $\kappa = 0.3$ cm$^{2}$ g$^{-1}$, the mean free path is just $1.7\times10^{11}$ cm.  

%Our best fit model is unable to replicate the initial data point in the Clear band acquired on MJD 58445.03.  

%\textit{work done by Takashi on early light curve, his findings, if true that interacts. Fig. \ref{fig:models}.}

%\begin{figure}
%    \includegraphics[width=1.03\columnwidth]{Models/sn2018ivc_nocsm.eps}
%    \includegraphics[width=1.03\columnwidth]{Models/sn2018ivc_nocsm_breakout.eps}
%    \caption{Takashi Moriya models without CSM. Top: first 20 days. Bottom: Zoom on the first day.}
%    \label{fig:models_nocsm}
%\end{figure}

%qui discussion modelli nakar plottati in Fig 5, modelli RSG e WR indicano progen +compatto, in linea con SN~IIb.

\subsubsection{Light curve modelling with stripped progenitor model}
In this case, we used the stripped-mSGB series progenitor models from \cite{Morozova2015}, where a ZAMS star with an initial mass of 15 \Msun\ is evolved using Modules for Experiments in Stellar Astrophysics (\texttt{MESA}, \citealt{Paxton2019, Jermyn2023}, and references therein) until the middle of the subgiant branch phase (mSGB). At this stage, a portion of the star's mass is instantaneously removed, after which the evolution continues until the onset of core collapse. We considered progenitor models with 5, 6, and 7\,\Msun\ of stripping, resulting in residual M$_H$ values of 0.38, 0.74, and 1.61 \Msun, respectively. The 5\,\Msun\ stripped progenitor has a pre-supernova mass of 6.82\,\Msun, a He core mass ofF 5.21\,\Msun\ and a radius of 828\,\Rsun. The 6\,\Msun\ stripped progenitor has a pre-supernova mass of 5.94\,\Msun, a He core mass of 5.20\,\Msun\ and a radius of 663\,\Rsun, while for the 7\,\Msun\ stripped progenitor, the pre-supernova mass is 5.59\,\Msun, the He core mass is 5.21\,\Msun\ and the progenitor radius is 555\,\Rsun.

\begin{figure}
    \includegraphics[width=1.05\columnwidth, clip, trim={0 0.1cm 0 0.8cm}]{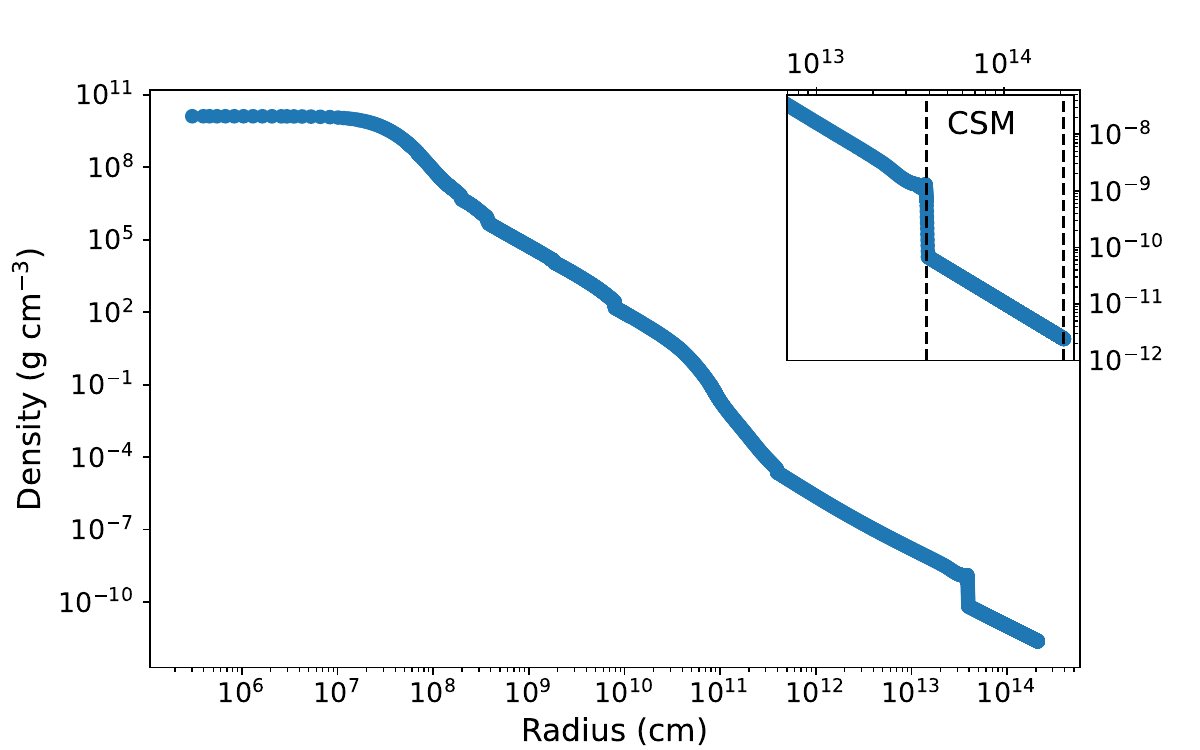}
    \caption{Density profile attached to the stripped progenitor. The inset shows a zoom-in of the CSM profile attached.}
    \label{density_profile_stripped}
\end{figure}

\begin{figure}
%first figure 
\begin{minipage}[h]{1.0\linewidth}
\centering
    \includegraphics[width=1.05\columnwidth, clip, trim={0 0 0 2.9cm}]{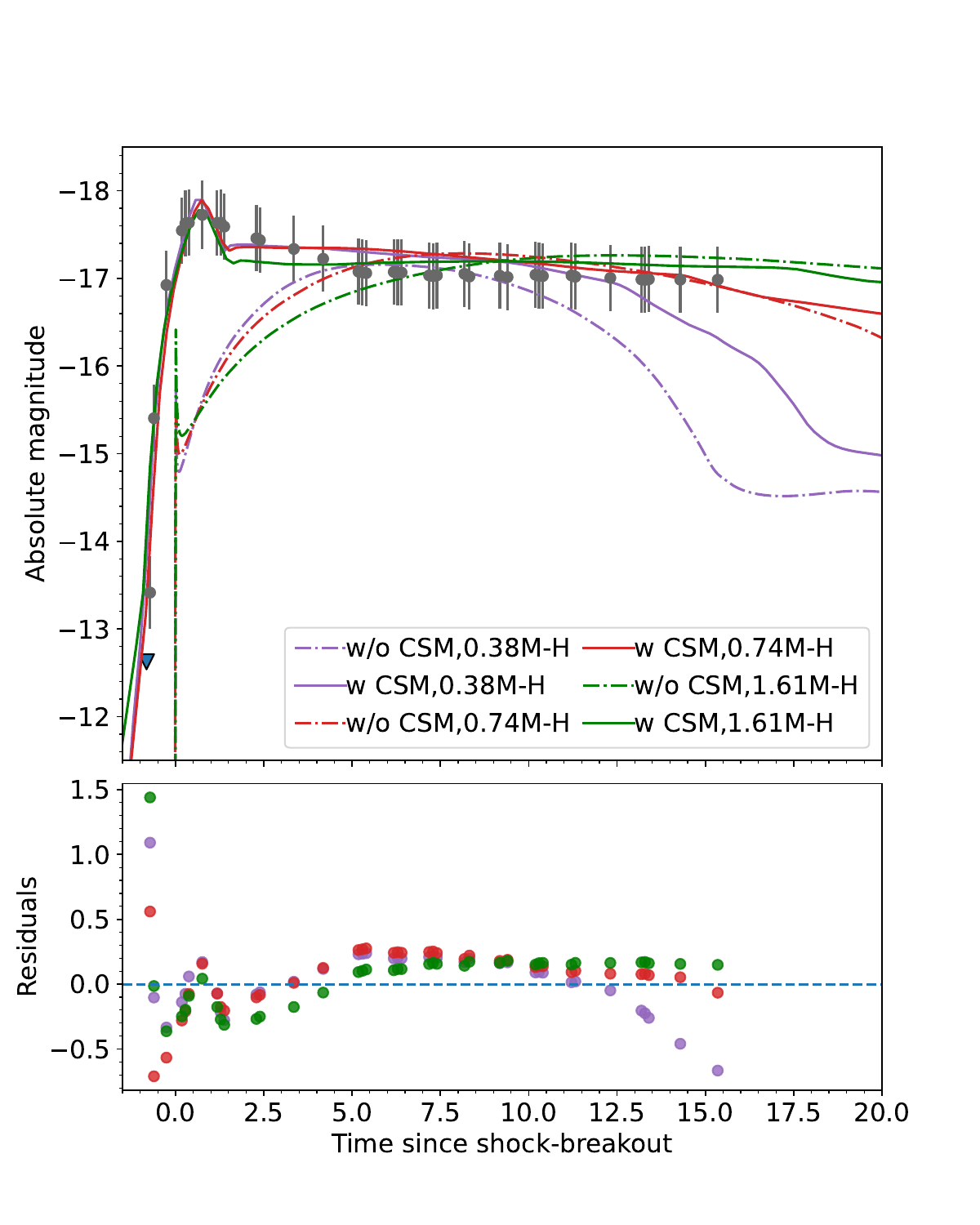}
\end{minipage}

\vspace{-0.60cm} 

%second figure 
\begin{minipage}[h]{1.0\linewidth}
\centering
    \includegraphics[width=1.0\columnwidth, clip, trim={0 0 0 0cm}]{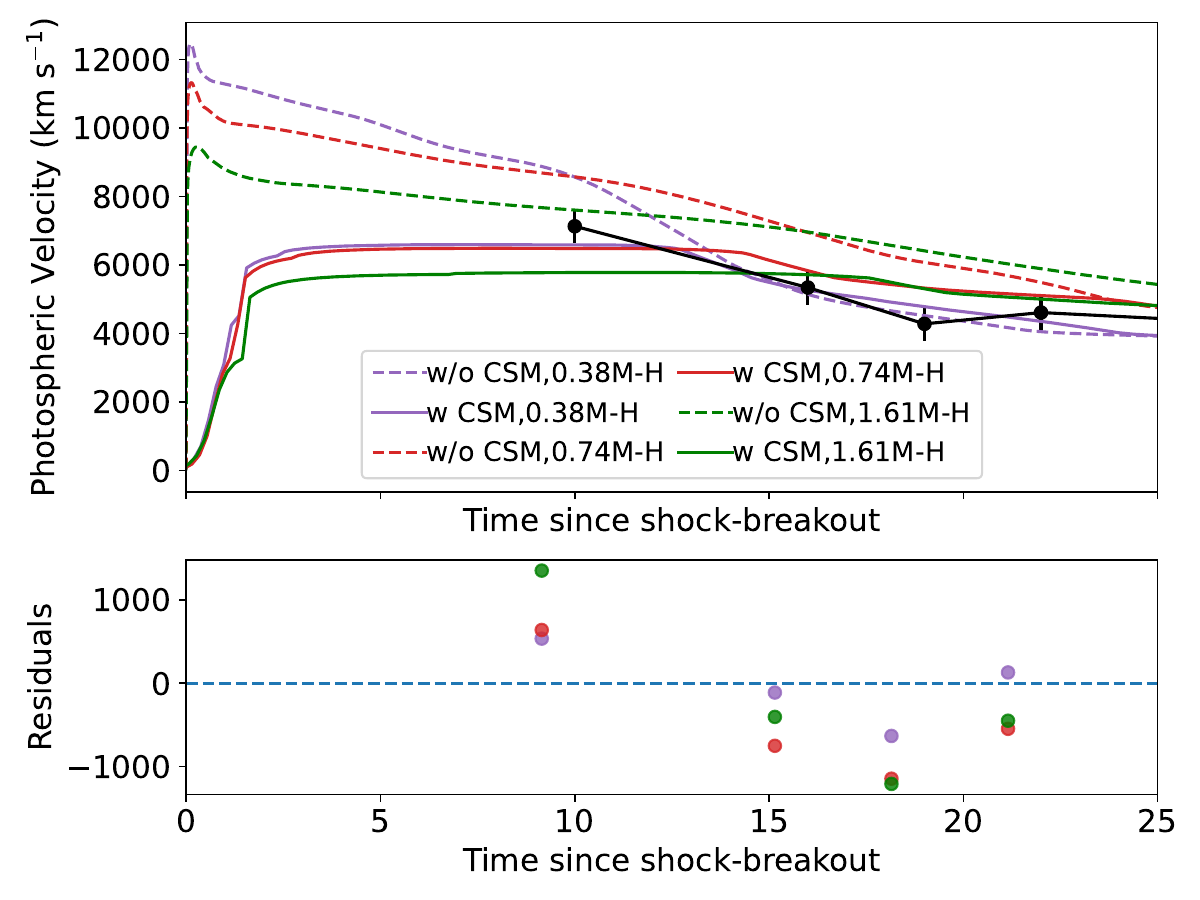}
\end{minipage}   

\caption{Top panel features the model LCs corresponding to the three stripped models of the first 20 days of \textit{Clear} band of SN~2018ivc along with the observed Clear band LCs. The bottom panel shows the photospheric velocity evolution corresponding to these models along with the observed photospheric velocity.}\label{fig:models_clear}
\end{figure}

To these progenitor models, we attached a wind-like density profile for the CSM (see Fig.~\ref{density_profile_stripped}) and exploded the models with \texttt{SNEC}. In this case, we have three free parameters: explosion energy, CSM extent (R$_{\rm ex}$) and mass-loading parameter (K$_{\rm CSM}$). We explored a range of explosion energies from 0.1 to 0.7 foe, R$_{\rm ex}$ from 2000 to 4000 \Rsun\ in steps of 100 \Rsun, and K$_{\rm CSM}$ ranging from 0.5$\times$10$^{17}$ g cm$^{-1}$ to 1.5$\times$10$^{17}$ g cm$^{-1}$ in steps of 0.1$\times$10$^{17}$ g\,cm$^{-1}$. The \Ni\ mass and \Ni\ mixing parameters are kept fixed at 0.015 \Msun\ and 3 \Msun, respectively. To observe the impact of increasing the progenitor's H envelope mass on the LCs, we depict the Clear band model LCs for the three progenitor models in Fig.~\ref{fig:models_clear}, employing the same K$_{\rm CSM}$ of 1.0$\times$10$^{17}$ g cm$^{-1}$ and R$_{\rm ex}$ of 3000 \Rsun, which corresponds to an approximate CSM mass of 0.1 \Msun. It is evident that as the H envelope mass increases, the models better fit the LCs after 12 days. However, lower H envelope masses tend to yield higher velocities, aligning more closely with the observed velocities. Comparing the three progenitor models, the $\chi^2$ attains its minimum value for the 6 \Msun\ stripped model, corresponding to a H-envelope mass of 0.74 \Msun. Nevertheless, the model velocities associated with H-envelope mass of 0.38 \Msun\ better match the observed velocities.

Finally, the best-fitted model to the multiband LCs corresponds to the 6 \Msun\ stripped model with an explosion energy of 0.4 foe, R$_{\rm ex}$ of 3200 \Rsun\ and K$_{\rm CSM}$ of 0.8$\times$10$^{17}$ g cm$^{-1}$. This would correspond to a CSM mass of 0.09 \Msun. The A$_V$ corresponding to this fit is 2.6 mag.

\begin{figure}
%first figure 
\begin{minipage}[h]{1.0\linewidth}
\centering
    \includegraphics[width=1.05\columnwidth, clip, trim={0 0 0 1.3cm}]{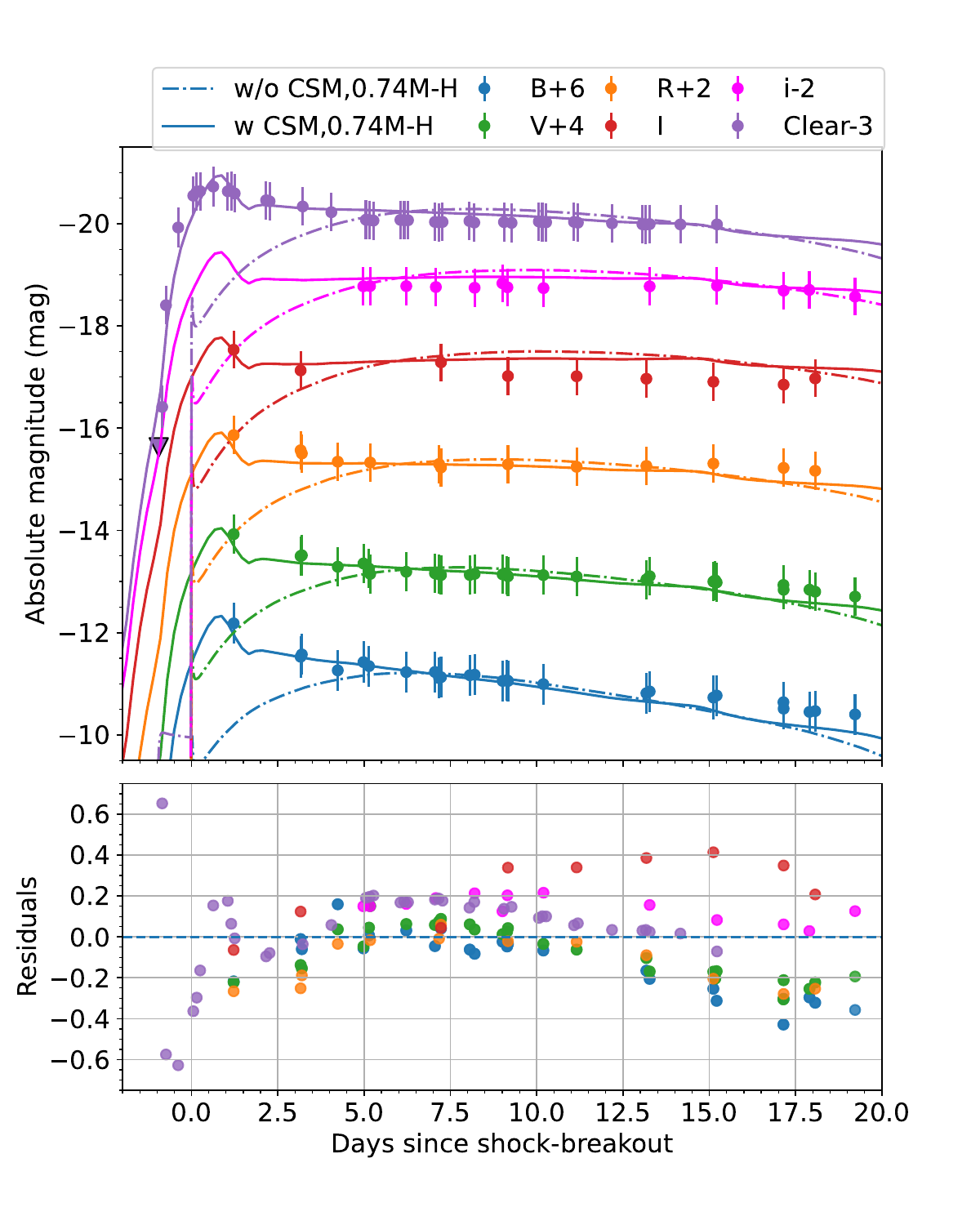}
\end{minipage}

\vspace{-0.60cm} 

%second figure 
\begin{minipage}[h]{1.0\linewidth}
\centering
    \includegraphics[width=1.0\columnwidth, clip, trim={0 0 0 0cm}]{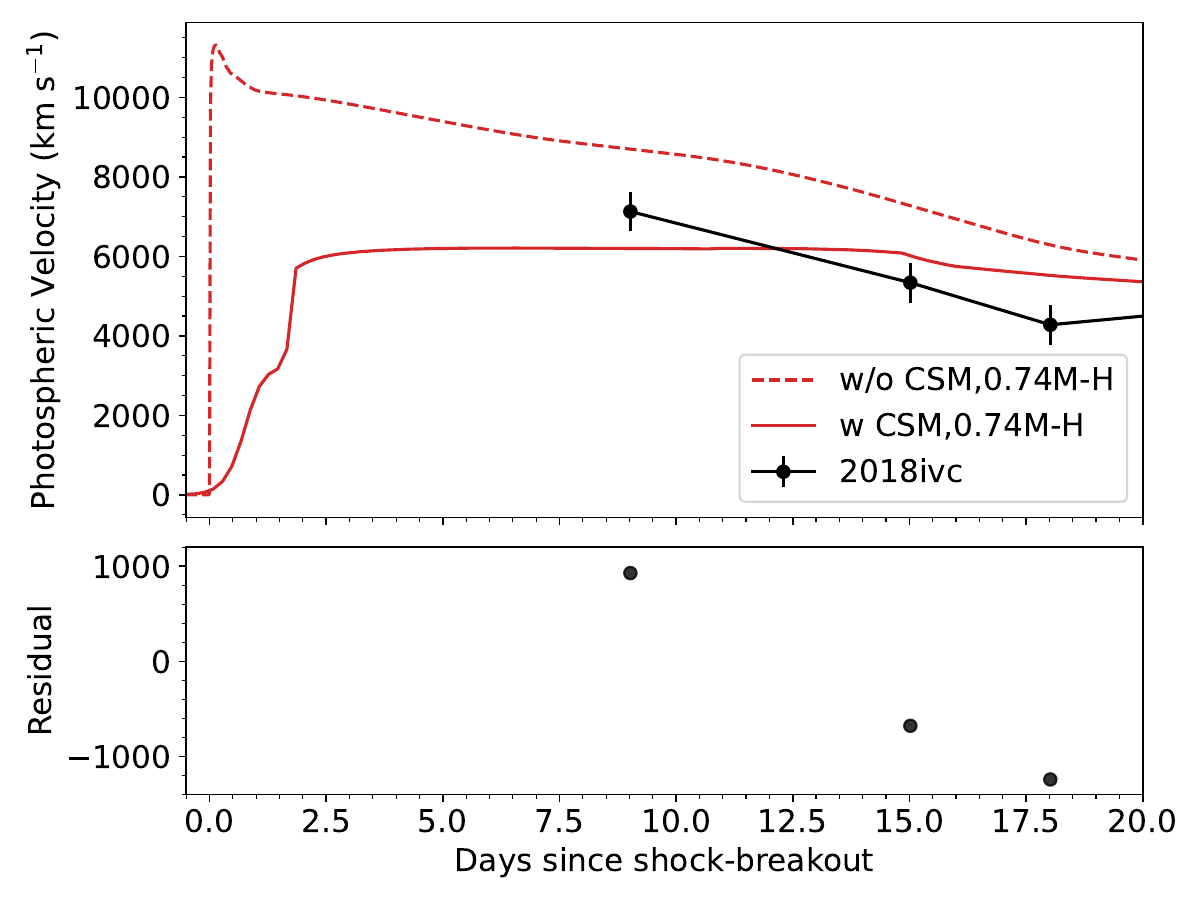}
\end{minipage} 
\caption{Top panel features the model light curves of the first 20 days of $BVRIi$ and \textit{Clear} bands of SN~2018ivc along with the observed multi-band LCs. The bottom panel shows the photospheric velocity evolution corresponding to these models along with the observed photospheric velocity.}\label{fig:models_new}
\end{figure}

\begin{table*}
\centering
\caption{Pre-supernova structure summary and CSM parameters.} \label{comp_model}
\renewcommand{\arraystretch}{1.1}
\footnotesize
\begin{tabular}{llllllll}
\hline
Progenitor Model  & R$_\star$ (R$_\odot$)  & M$_{H}$ (M$_\odot$)  & M$_{He}$ (M$_\odot$) & R$_{ex}$ (R$_\odot$) & A$_{\rm V,tot}$ (mag)      & M$_{\rm CSM}$ (M$_\odot$) & $\chi^2$\\
\hline
IIb           &   350   &      0.02 & 4   &    4100   & 3.2 & 0.47  &   4.8\\ %+0.4
stripped\_6M  &   663   &      0.74 & 5.2 &    3200   & 2.6 & 0.09 &    2.2\\ %-0.1
\hline
\end{tabular}
\end{table*}

An H-envelope mass of 0.74 \Msun, falls in between the values estimated for the typical progenitors of SNe IIb (up to 0.5 \Msun, \citealt{Sravan2019ApJ...885..130S}) and SNe IIL ($\sim$1 \Msun,for \citealt{Swartz1991ApJ...374..266S}; 1-2 \Msun,for \citealt{Blinnikov1993A&A...273..106B}).
This is consistent with SN 2018ivc being a transitional IIL/IIb object, as confirmed by our other indicators.
While an $M_H$ of 0.38 \Msun,is within the proposed range of SNe IIb progenitors, it still remains in the upper side of the range. 
Considering the large progenitor radius of 350~\Rsun, we adopted for the SN IIb progenitor model, and 663~\Rsun,for the `stripped\_6M model' (see Table~\ref{comp_model}), a higher $M_H$ is also consistent with the fact that more extended SNe~IIb progenitors should be also more H-rich \citep{Prentice2017MNRAS.469.2672P}.

The pre-supernova configuration and the optimal CSM parameters, along with the corresponding $\chi^2$ values for the two progenitor scenarios, are summarised in Table~\ref{comp_model}. The stripped progenitor model yields the minimum $\chi^2$, corresponding to an M$_H$ of 0.74 \Msun, which is intermediate to those of IIb and IIL progenitors. Additionally, the model photospheric velocities corresponding to the stripped progenitor model are a closer match to the observed velocities than those of the Type IIb progenitor model. We note that in \cite{Maeda2023ApJ...942...17M}, a low-density CSM located at a greater radial distance has been favoured. Hence, the CSM we are examining here differs from the component proposed by \cite{Maeda2023ApJ...942...17M}. However, a dense CSM (as used in our models) in close proximity to the progenitor would decelerate the shock, delaying the peak in radio emission, which contradicts observed radio data. Introducing such low densities near the progenitor in \texttt{SNEC} modelling produces LCs resembling those of "no-CSM" models, capable of replicating the early rise in the Clear band, seemingly unaffected by the low-density CSM. While \cite{Maeda2023ApJ...942...17M} suggested that their CSM would negligibly affect the rising part of the LC, placing the low-density CSM at a higher radius, as suggested by \cite{Maeda2023ApJ...942...17M}, results in model light curves that appear unphysical. Indeed, all \texttt{SNEC} modelling based studies have demonstrated the necessity of a confined CSM to accurately replicate observed LCs. The disparity may be reconciled with a non-spherical CSM featuring both low and high dense regions.

Additionally, the dense CSM close to the progenitor could explain the absence of narrow emission lines in SN~2018ivc, typically observed in the spectra of Type IIn SNe. The dense CSM would result in a high electron scattering optical depth ($>>100$). Hence, the mean free path of recombination photons responsible for the formation of the narrow features will be much smaller than ${\rm R_{\rm CSM}}$. The narrow features can only form in case of radiation leakage from the shock into the CSM above, and this cannot occur when the CSM density is high \citep[on the order of $10^{-10}-10^{-9}$ g cm$^{-3}$,
][]{Dessart2017A&A...605A..83D}. For instance, at $4\times10^{13}$ cm (${\rm R_{in}^{\rm CSM}}$), the mean free path of the photons will be 1/$\kappa\rho$. In our model, $\rho\sim 5\times10^{-11}$ g cm$^{-3}$, assuming $\kappa = 0.3$ cm$^{2}$ g$^{-1}$, the mean free path is $7.7\times10^{10}$ cm, which is much smaller than ${\rm R_{in}^{\rm CSM}}$.

\section{Conclusions} \label{conclusion}
Thanks to the high cadence of the CHASE survey, we observed a very fast rise to maximum light during the first day of evolution of the peculiar Type II SN~2018ivc. % we detected the Type II SN~2018ivc very early, just a few hours after the explosion, and observed a very fast rise to maximum light during the first day of evolution.
This rise, with a slope of about 18 mag/day, is unusual, and we know only the case of SN~2016gkg \citep{2018Natur.554..497B} as a faster rise from a Type II SN, at $43\pm6$ mag/day.
The $r$-band light curve can be decomposed in four distinct phases: the fast rise to maximum, which is reached in just 1.6 days, a first steep linear decline, a short duration plateau and a second, long and slower linear decline, which makes SN~2018ivc a Type IIL SN. The 10-day plateau is also observed in the redder bands ($i$ and $z$), resembling a secondary peak, as observed in some Type~IIb~SNe, while the light curve in the bluer filters is steadily declining. The linear decline continued for one year.
In Autumn 2022, after the announcement of a rebrightening in radio \citep{Maeda2023ApJ...945L...3M}, we detected SN~2018ivc at $r\sim21$ mag, four years after the explosion.

The early spectra of SN~2018ivc show a blue featureless continuum and then evolve, revealing a strong H$\alpha$, a broad Ca II NIR triplet and prominent He I lines. On top of the He I $\lambda$5876 line, a deep absorption from the Na ID doublet indicates the presence of a conspicuous additional extinction from the host galaxy. From its EW, we estimate $A_{V,host}$ to be 1.5$\pm$0.2 mag, but this measure can be uncertain. Indeed, by matching the $B-V$ colour of SN 2018ivc to that of SN 1996al at early phases, we obtain a higher $A_{V,host}$ of $\sim$3 mag, which seems to be favoured also by our hydrodynamical modelling with CSM. This value is supported by our measurement of the Balmer decrement in the spectra of three H II regions near the SN explosion site. 
%The contribution from the host galaxy is evident also from the spectra, with the presence of contamination in the form of narrow emissions from H$\alpha$ and H$\beta$, and forbidden lines like [O~III] $\lambda\lambda$4959,5007 and [N II] $\lambda$6584. This last one is always present over the broad component of H$\alpha$.

From the comparison of the light curves and the late spectrum of SN~2018ivc with those of objects from both Type IIL and IIb SNe, we found some aspects in common with both typologies, thus we propose that SN~2018ivc may be a transitional object between the two families of events, a Type IIL/IIb SN. In this regard, SN~2007fz is a good comparison object. However, we found a few common aspects with Type IIn SNe, such as SN 1996al, especially for the signs of interaction with a CSM.
%If SN~2018ivc is indeed a Type IIb SN, no one has shown such a steep rise to maximum after the explosion.
The peculiar \Ha profile, with a red shoulder and a flat-topped peak blue-shifted by $-$2000 \kms, together with the presence of an HV feature, point towards an asymmetric ejecta, despite a relatively weak explosion energy.

We modelled the early multi-band light curve evolution of SN~2018ivc to constrain the properties of the explosion and the surrounding CSM.
For the high-reddening scenario, the best fit model is found for an explosion energy of 0.3 foe, an external radius of the CSM of 4100 \Rsun, and a CSM mass of 0.47 \Msun. The values for the low-reddening case are 0.1 foe (a rather weak explosion), 2200 \Rsun, and 0.15~\Msun, respectively, but fails to reproduce the photospheric velocity. Therefore, our hydrodynamical modelling tend to favour the high reddening case.

%Given the share of observational properties of SN 2018ivc with both SNe IIb and IIL, the progenitor was likely also in an evolutionary stage in between those typical of the two classes of SNe.
Our hydrodynamical modelling reproduces better the observables of SN 2018ivc with a stripped progenitor, with a final mass of 6 \Msun~and a radius of $\sim$660 \Rsun.
These parameters are consistent with a yellow supergiant/hypergiant progenitor, whose H envelope was partly stripped during the Blue Loop stage as a single star, or through some mass-loss mechanism within a binary system.
This explains the He I lines stronger than in a normal SN IIL and the presence of a CSM surrounding the progenitor star.

SN 2018ivc turned out to be a peculiar and interesting object, sharing properties of Type IIb, IIL and IIn~SNe.
SN 2018ivc highlighted the importance of a high cadence survey for transients, able to catch the fast evolution of a SN in the very first hours after the explosion.
The ULTRASAT satellite \citep{ULTRASAT2024ApJ...964...74S} is expected to be launched in 2026. It will conduct a wide-field survey of transients in the ultraviolet, with the goal to discover rapidly evolving ones, including fast-rising SNe. Thus, it will detect and allow to study numerous SBO events, improving our knowledge on what happens in the first instants of a newly-born SN.

%-------------------------------------------------------------------

\begin{acknowledgements}
\begin{small}
We thank the anonymous referee for their insightful comments and suggestions, which helped to improve the paper.
% Io
AR acknowledges financial support from the GRAWITA Large Program Grant (PI P. D’Avanzo) and the PRIN-INAF 2022 \textit{"Shedding light on the nature of gap transients: from the observations to the models"}.
% RD
RD acknowledges funds by ANID grant FONDECYT Postdoctorado Nº 3220449.
% Giuliano
GP acknowledges support by FONDECYT Regular 1201793.
% Maeda
KM acknowledges support from the Japan Society for the Promotion of Science (JSPS) KAKENHI grant (JP20H00174).
% HK
HK was funded by the Research Council of Finland projects 324504, 328898, and 353019.
% JP Anderson
JP acknowledges ANID, Millennium Science Initiative, ICN12\_009.
% Panos
PC acknowledges support via Research Council of Finland (grant 340613).
% MF
MF is supported by a Royal Society - Science Foundation Ireland University Research Fellowship.
% YC
Y.-Z. Cai is supported by the National Natural Science Foundation of China (NSFC, Grant No. 12303054), the Yunnan Fundamental Research Projects (Grant No. 202401AU070063) and the International Centre of Supernovae, Yunnan Key Laboratory (No. 202302AN360001).
% Fondi Maeda?
The work is partly supported by the JSPS Open Partnership Bilateral Joint Research Project between Japan and Chile (JPJSBP120239901), as well as that between Japan and Finland (JPJSBP120229923).
% NER
NER also acknowledges support from the PRIN-INAF 2022 `Shedding light on the nature of gap transients: from the observations to the models'.
% LG
LG acknowledges financial support from the Spanish Ministerio de Ciencia e Innovaci\'on (MCIN) and the Agencia Estatal de Investigaci\'on (AEI) 10.13039/501100011033 under the PID2020-115253GA-I00 HOSTFLOWS project, from Centro Superior de Investigaciones Cient\'ificas (CSIC) under the PIE project 20215AT016 and the program Unidad de Excelencia Mar\'ia de Maeztu CEX2020-001058-M, and from the Departament de Recerca i Universitats de la Generalitat de Catalunya through the 2021-SGR-01270 grant.
% CPG
CPG acknowledges financial support from the Secretary of Universities and Research (Government of Catalonia), the Horizon 2020 Research and Innovation Programme of the European Union under the Marie Sk\l{}odowska-Curie, the Beatriu de Pin\'os 2021 BP 00168 programme from the Spanish Ministerio de Ciencia e Innovaci\'on (MCIN), the Agencia Estatal de Investigaci\'on (AEI) 10.13039/501100011033 under the PID2020-115253GA-I00 HOSTFLOWS project, and the program Unidad de Excelencia Mar\'ia de Maeztu CEX2020-001058-M.
% MR
MR acknowledge support from National Agency for Research and Development (ANID) grants ANID-PFCHA/Doctorado Nacional/2020-21202606.
% CHASE
The CHASE project is founded by the Millennium Institute for Astrophysics.
% NOT and LT
This work is based in part on observations made with the Nordic Optical Telescope, operated by the Nordic Optical Telescope Scientific Association, and with the Liverpool Telescope, operated on the island of La Palma by Liverpool John Moores University, with financial support from the UK Science and Technology Facilities Council. Both telescopes are located at the Observatorio del Roque de los Muchachos, La Palma, Spain, of the Instituto de Astrofisica de Canarias. 
% NUTS
NUTS is funded in part by the IDA (Instrument Centre for Danish Astronomy). 
% ESO PESSTO
Based in part on observations collected at the European Organisation for Astronomical Research in the Southern Hemisphere, Chile, under ESO programs 105.20DF.002, 0103.D-0338(A), 0110.A-9012(A), and as part of ePESSTO project (the extended Public ESO Spectroscopic Survey for Transient Objects Survey), under ESO programs ID 199.D-0143, 1103.D-0328.
% NOAO
Based in part on observations at Cerro Tololo Inter-American Observatory, National Optical Astronomy Observatory (NOAO), which is operated by the Association of Universities for Research in Astronomy (AURA), Inc. under a cooperative agreement with the National Science Foundation.
% ZTF
ZTF is supported by the National Science Foundation.
% INAF Schmidt
Based in part on observations collected at 67/92 Schmidt telescope (Asiago, Italy) of the INAF~-~Osservatorio Astronomico di Padova.
% SUBARU
Based in part on data collected at the Subaru Telescope, operated by the National Astronomical Observatory of Japan, under S19B-055.
% CFHT MegaPrime
Based in part on observations obtained with MegaPrime/MegaCam, a joint project of CFHT and CEA/IRFU, at the Canada-France-Hawaii Telescope (CFHT) which is operated by the National Research Council (NRC) of Canada, the Institut National des Science de l'Univers of the Centre National de la Recherche Scientifique (CNRS) of France, and the University of Hawaii.
% ATLAS
Support to ATLAS was provided by NASA grant NN12AR55G.
% NED
This research has made use of the NASA/IPAC Extragalactic Database (NED) which is operated by the Jet Propulsion Laboratory, California Institute of Technology, under contract with the National Aeronautics and Space Administration.
% ADS
This research has made use of NASA's Astrophysics Data System Bibliographic Services.
\end{small}
\end{acknowledgements}

\bibliographystyle{aa} % style aa.bst
\bibliography{bib} % your references file.bib
%-------------------------------------------------------------------

\begin{appendix} %First appendix
\section{Discrepancy in the g-band magnitude}\label{g-discrepancy} % 1st Appendix

We noted a minor discrepancy, although with a big scatter, between our photometric measurements and those presented by \citetalias{2020ApJ...895...31B}. The difference in the Sloan-$g$ filter reaches almost half a magnitude (with our photometry, also conducted on template-subtracted images, being always brighter), and it remains almost constant among their entire two months-long light curve.
In particular, the short duration plateau around +10 d, visible in the red filters light curves in both \citetalias{2020ApJ...895...31B} and our work, is noticeable also in our $g$-band light curve, while it is less evident in theirs.

\begin{figure}
    \includegraphics[width=1.05\columnwidth]{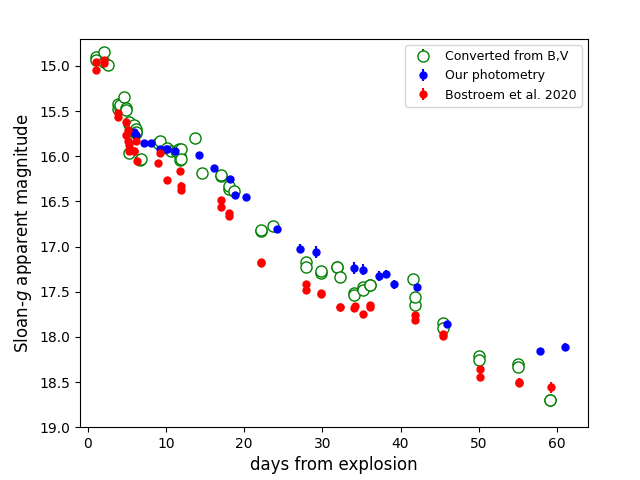}
    \caption{Comparison of the $g$-band light curves from this work (blue points), \citetalias{2020ApJ...895...31B} (red points) and \citetalias{2020ApJ...895...31B} converted from their $BV$ magnitudes using the formula from \citet{2005AJ....130..873J} (green points).}
    \label{Fig:g-Bostroem}
    \end{figure}

We performed a sanity check by calculating the $g$-band magnitudes from the photometry in the JC $B$ and $V$ filters, applying the conversion formula of \cite{2005AJ....130..873J} and comparing them with the observed $g$-band magnitudes.
The calculated magnitudes in the $g$-band from our $BV$ photometry are consistent (within 0.1 mag) with our measured ones, while those derived from the $BV$ photometry published by \citetalias{2020ApJ...895...31B} do not. Instead, between 5 and 20 days after maximum, they are even closer to our $g$-band measurements than theirs.
Our $g$-band light curve, the one published by \citetalias{2020ApJ...895...31B}, and the one calculated with the conversion formula using the $BV$ photometry from their work are plotted in Fig. \ref{Fig:g-Bostroem}.

\section{Determination of the explosion epoch from the light curve modelling}\label{Appendix_B} % 2nd Appendix

\begin{figure}
    \includegraphics[width=0.95\columnwidth]{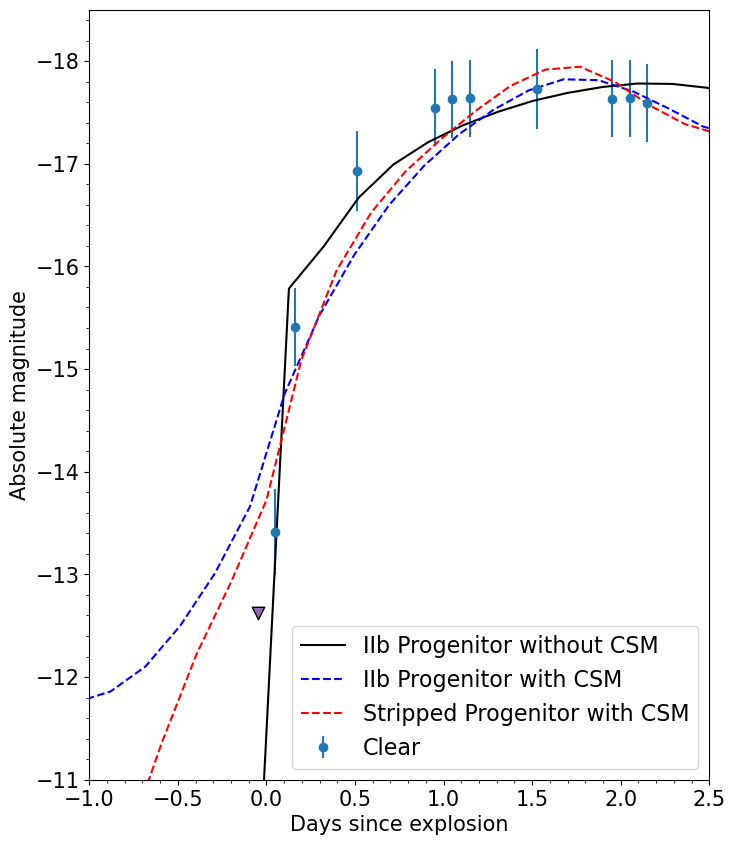}
    \caption{Comparison of the observed early light curve rise with those of the best-fit models: IIb progenitor model without CSM, and models with CSM for both IIb and stripped progenitors.
    }
    \label{Fig:explosion_epoch_from_modeling}
    \end{figure}

Allowing $t_0$ to vary as a free parameter in the modelling (within a range of $\pm$1 day), we identified the best fit by considering only the first 2.5 days after our first detection, as these data points provide the strongest constraint on the explosion epoch. The resulting time offsets from the assumed explosion epoch (MJD 58845.08) were 0.06 days, $-$0.3 days, and $-$0.9 days for the `IIb progenitor without CSM,' `IIb progenitor with CSM,' and `stripped progenitor with CSM' models, respectively. 
The rising portions of the shifted model light curves, along with the observed \textit{Clear} band light curve, are shown in Fig. \ref{Fig:explosion_epoch_from_modeling}. As discussed in Sect. \ref{discussion}, the model without CSM best reproduces the rising part of the light curve, which is also evident from Fig. \ref{Fig:explosion_epoch_from_modeling}. 
For the other two scenarios, a trade-off was necessary between fitting the early rising part and the later flattened part, as improving the fit for one part negatively impacted the fit for the other. 
Consequently, we were unable to obtain a good fit for the rising part of the light curve in these two cases. Given that the model without CSM provides the best fit to the rising part of the light curve, the explosion time derived from this model is likely to be more accurate than the other two. This explosion time is also consistent with all other methods discussed in the paper.

%\newpage
\section{Photometry tables} % 3rd Appendix

\begin{table*}\centering
\caption{Optical Sloan magnitudes of SN~2018ivc.}
\label{tab:sloan}
\begin{threeparttable}[b]
\begin{tabular}{cccccccl}
\hline\hline
Date & MJD & $u$ & $g$ & $r$ & $i$ & $z$ & Instrument\tnote{a} \\
\hline
2018-11-28 & 58450.95 & 16.89 0.05 & 15.73 0.02 & 15.16 0.02 & 15.02 0.02 & - & SCH\\
2018-11-29 & 58451.15 & - & 15.76 0.04 & 15.17 0.01 & 15.01 0.02 & 14.90 0.02 & PROM\\ 
2018-11-30 & 58452.20 & - & 15.86 0.02 & 15.16 0.03 & 15.01 0.02 & 14.87 0.04 & PROM\\ 
2018-12-01 & 58453.05 & - & 15.85 0.02 & 15.15 0.03 & 15.03 0.03 & 14.90 0.03 & PROM\\ 
2018-12-02 & 58454.18 & - & 15.92 0.03 & 15.17 0.02 & 15.05 0.03 & 14.92 0.03 & PROM\\ 
2018-12-02 & 58454.98 & 18.02 0.04 & 15.92 0.01 & 15.15 0.03 & 14.96 0.01 & 14.94 0.01 & ALFOSC\\
2018-12-03 & 58455.13 & - & 15.92 0.03 & 15.16 0.02 & 15.04 0.02 & 14.91 0.02 & PROM\\ 
2018-12-04 & 58456.17 & - & 15.94 0.02 & 15.19 0.01 & 15.05 0.03 & 14.89 0.02 & PROM\\ 
2018-12-07 & 58459.25 & - & 15.99 0.02 & 15.17 0.02 & 15.02 0.02 & 14.88 0.02 & PROM\\ 
2018-12-09 & 58461.20 & - & 16.13 0.01 & 15.17 0.02 & 15.01 0.02 & 14.90 0.02 & PROM\\ 
2018-12-11 & 58463.13 & - & 16.25 0.02 & 15.22 0.02 & 15.10 0.01 & 14.90 0.02 & PROM\\ 
2018-12-11 & 58463.87 & 19.81 0.04 & 16.43 0.01 & 15.34 0.01 & 15.09 0.01 & 14.82 0.01 & ALFOSC\\
2018-12-13 & 58465.20 & - & 16.45 0.03 & 15.37 0.02 & 15.22 0.02 & 15.07 0.03 & PROM\\ 
2018-12-17 & 58469.17 & - & 16.81 0.04 & 15.77 0.04 & 15.59 0.03 & 15.36 0.04 & PROM\\ 
2018-12-20 & 58472.20 & - & 17.03 0.05 & 16.00 0.04 & 15.91 0.02 & 15.52 0.04 & PROM\\ 
2018-12-22 & 58474.14 & - & 17.06 0.06 & 16.01 0.03 & 15.98 0.03 & 15.67 0.03 & PROM\\ 
2018-12-27 & 58479.07 & - & 17.24 0.06 & 16.14 0.03 & 16.20 0.03 & 15.74 0.03 & PROM\\ 
2018-12-28 & 58480.17 & - & 17.26 0.06 & 16.23 0.03 & 16.22 0.04 & 15.85 0.06 & PROM\\ 
2018-12-30 & 58482.18 & - & 17.33 0.06 & 16.21 0.03 & 16.30 0.03 & 15.90 0.03 & PROM\\ 
2018-12-31 & 58483.16 & - & 17.30 0.04 & 16.28 0.03 & 16.35 0.03 & 15.82 0.04 & PROM\\ 
2019-01-01 & 58484.11 & - & 17.42 0.05 & 16.35 0.05 & 16.35 0.04 & 16.02 0.03 & PROM\\ 
2019-01-04 & 58487.09 & - & 17.45 0.04 & 16.57 0.04 & 16.53 0.04 & 16.17 0.03 & PROM\\ 
2019-01-05 & 58488.90 & - & - & 16.59 0.01 & - & - & LT\\
2019-01-07 & 58490.89 & - & 17.85 0.01 & 16.64 0.01 & 16.61 0.01 & 16.31 0.02 & ALFOSC\\
2019-01-19 & 58502.86 & - & 18.15 0.02 & 17.19 0.12 & 16.97 0.01 & 16.82 0.02 & ALFOSC\\
2019-01-23 & 58506.05 & - & 18.11 0.04 & - & - & - & PROM\\ 
2019-02-06 & 58520.24 & - & - & - & 17.74 0.01 & - & CFHT\\
2019-02-20 & 58534.18 & - & 18.52 0.20 & - & - & - & ZTF\\
2019-02-25 & 58539.13 & - & 18.83 0.10 & - & - & - & ZTF\\
2019-02-27 & 58541.84 & - & 19.08 0.02 & 18.23 0.02 & 18.03 0.02 & 17.84 0.03 & ALFOSC\\
2019-03-04 & 58546.23 & - & - & - & 18.37 0.01 & - & CFHT\\
2019-03-09 & 58551.23 & - & - & - & 18.40 0.03 & - & CFHT\\
2019-03-12 & 58554.23 & - & - & - & 18.45 0.02 & - & CFHT\\
2019-03-14 & 58556.24 & - & - & - & 18.49 0.01 & - & CFHT\\
2019-08-01 & 58696.39 & - & - & 20.06 0.09 & - & - & FORS \\
2019-08-04 & 58699.15 & - & 22.16 0.13 & 20.03 0.07 & 19.74 0.04 & 19.73 0.08 & ALFOSC\\
2019-08-12 & 58707.60 & - & - & - & 19.83 0.04 & - & CFHT\\
2019-09-04 & 58730.30 & - & - & 20.12 0.10 & 19.72 0.15 & - & EFOSC\\
2019-10-22 & 58778.18 & - & - & 20.24 0.07 & 19.89 0.04 & - & EFOSC\\
2019-10-28 & 58784.26 & - & - & 20.37 0.08 & - & - & EFOSC\\
2019-12-28 & 58845.05 & - & - & $>$21.3 & $>$20.8 & - & EFOSC\\
%2022-07-24 & 59784.57 & - & $>20.9$ & - & - & - & CFHT\\
%2022-07-26 & 59786.55 & - & $>20.9$ & - & - & - & CFHT\\
2022-09-18 & 59840.29 & - & - & $>$20.8 & - & - & FORS \\
2022-09-23 & 59845.33 & - & - & 21.11 0.07 & - & - & FORS \\
2022-09-27 & 59849.21 & - & - & 21.08 0.05 & - & - & FORS \\
\hline
\end{tabular}
\begin{tablenotes}
\footnotesize
\item[a] SCH = 0.92-m Asiago Schmidt+Moravian, PROM = 0.4-m Prompt5+Apogee, ALFOSC = 2.56-m NOT+ALFOSC, LT = 2.0-m Liverpool+IO:O, CFHT = 3.58-m CFHT+MegaPrime, ZTF = 1.2-m `S. Oschin' Schmidt+ZTF survey MOSAIC, FORS = 8.2-m VLT+FORS2, EFOSC = 3.58-m NTT+EFOSC2 
\end{tablenotes}
\end{threeparttable}
\end{table*}

\begin{table*}
\caption{Optical Johnson magnitudes of SN~2018ivc.}
\label{tab:johnson}
\begin{threeparttable}[b]
\begin{tabular}{ccccccl}
\hline\hline
Date & MJD & $B$ & $V$ & $R$ & $I$ & Instrument\tnote{a} \\
\hline
2018-11-25 & 58447.20 & 15.32 0.06 & 14.72 0.06 & 14.35 0.01 & 14.05 0.01 & ANDICAM\\
2018-11-27 & 58449.14 & 15.97 0.09 & 15.15 0.02 & 14.64 0.02 & 14.45 0.06 & ANDICAM\\
2018-11-27 & 58449.17 & 15.93 0.05 & 15.13 0.14 & 14.70 0.06 & - & PROM6\\
2018-11-28 & 58450.21 & 16.24 0.04 & 15.36 0.06 & 14.86 0.05 & - & PROM6\\
2018-11-28 & 58450.96 & 16.08 0.07 & 15.29 0.03 & - & - & SCH\\
2018-11-29 & 58451.12 & 16.15 0.02 & 15.39 0.03 & - & - & PROM5\\
2018-11-29 & 58451.16 & - & 15.50 0.03 & 14.88 0.01 & - & ANDICAM\\
2018-11-30 & 58452.20 & 16.27 0.02 & 15.45 0.01 & - & - & PROM5\\ 
2018-12-01 & 58453.03 & 16.27 0.02 & 15.48 0.03 & - & - & PROM5\\ 
2018-12-01 & 58453.15 & 16.39 0.07 & 15.51 0.02 & 14.91 0.02 & - & ANDICAM\\
2018-12-01 & 58453.20 & 16.37 0.03 & 15.52 0.02 & 14.98 0.02 & 14.29 0.02 & PROM8\\
2018-12-02 & 58454.04 & 16.33 0.02 & 15.52 0.01 & - & - & TRAP\\ 
2018-12-02 & 58454.18 & 16.32 0.04 & 15.50 0.02 & - & - & PROM5\\ 
2018-12-02 & 58454.99 & 16.45 0.01 & 15.51 0.03 & - & - & ALFOSC\\
2018-12-03 & 58455.12 & 16.43 0.03 & 15.53 0.02 & - & - & PROM5\\ 
2018-12-03 & 58455.15 & 16.45 0.07 & 15.54 0.03 & 14.92 0.02 & 14.56 0.01 & ANDICAM\\
2018-12-04 & 58456.17 & 16.51 0.03 & 15.52 0.02 & - & - & PROM5\\ 
2018-12-05 & 58457.13 & - & 15.55 0.04 & 14.97 0.01 & 14.57 0.01 & ANDICAM\\
2018-12-07 & 58459.16 & 16.69 0.03 & 15.60 0.04 & 14.95 0.02 & 14.61 0.02 & ANDICAM\\
2018-12-07 & 58459.25 & 16.65 0.03 & 15.54 0.02 & - & - & PROM5\\ 
2018-12-09 & 58461.10 & 16.77 0.18 & 15.65 0.04 & 14.90 0.04 & 14.67 0.03 & ANDICAM\\
2018-12-09 & 58461.15 & - & 15.62 0.04 & - & - & EFOSC\\ 
2018-12-09 & 58461.19 & 16.73 0.03 & 15.66 0.02 & - & - & PROM5\\ 
2018-12-11 & 58463.12 & 16.86 0.03 & 15.71 0.02 & - & - & PROM5\\ 
2018-12-11 & 58463.13 & 16.99 0.10 & 15.81 0.05 & 14.99 0.01 & 14.73 0.03 & ANDICAM\\
2018-12-11 & 58463.88 & 17.05 0.01 & 15.81 0.01 & - & - & ALFOSC\\
2018-12-12 & 58464.04 & 17.04 0.03 & 15.84 0.02 & 15.04 0.01 & 14.61 0.03 & TRAP\\ 
2018-12-13 & 58465.20 & 17.10 0.05 & 15.94 0.02 & - & - & PROM5\\ 
2018-12-15 & 58467.12 & - & 15.95 0.02 & - & - & EFOSC\\ 
2018-12-16 & 58468.08 & - & 16.28 0.03 & 15.37 0.02 & 15.12 0.01 & ANDICAM\\
2018-12-17 & 58469.17 & 17.51 0.06 & 16.40 0.04 & - & - & PROM5\\ 
2018-12-20 & 58472.20 & 17.80 0.05 & 16.53 0.06 & - & - & PROM5\\ 
2018-12-21 & 58473.12 & - & 16.65 0.04 & 15.77 0.02 & 15.50 0.10 & ANDICAM\\
2018-12-22 & 58474.05 & 17.82 0.10 & 16.62 0.03 & 15.82 0.01 & 15.39 0.02 & TRAP\\
2018-12-22 & 58474.14 & 17.78 0.09 & 16.67 0.06 & - & - & PROM5\\ 
2018-12-27 & 58479.07 & 17.88 0.07 & 16.79 0.04 & - & - & PROM5\\
2018-12-27 & 58479.10 & - & 16.80 0.04 & 15.90 0.03 & 15.62 0.14 & ANDICAM\\
2018-12-28 & 58480.17 & 17.99 0.12 & 16.83 0.05 & - & - & PROM5\\ 
2018-12-30 & 58482.17 & 18.03 0.10 & 16.90 0.05 & - & - & PROM5\\ 
2018-12-31 & 58483.16 & 18.14 0.09 & 16.91 0.04 & - & - & PROM5\\
2019-01-01 & 58484.06 & - & - & 16.11 0.03 & 15.84 0.03 & ANDICAM\\
2019-01-01 & 58484.10 & 18.15 0.09 & 16.99 0.03 & - & - & PROM5\\ 
2019-01-03 & 58486.18 & 18.36 0.09 & 17.08 0.04 & - & - & PROM5\\ 
2019-01-04 & 58487.11 & 18.48 0.10 & 17.08 0.03 & 16.18 0.03 & 16.03 0.03 & ANDICAM\\
2019-01-07 & 58490.89 & - & 17.37 0.01 & - & - & ALFOSC\\
2019-01-09 & 58492.05 & - & 17.37 0.03 & 16.44 0.03 & 16.29 0.03 & ANDICAM\\
2019-01-12 & 58495.05 & 18.62 0.09 & 17.45 0.06 & 16.64 0.03 & 16.23 0.03 & TRAP\\
2019-01-14 & 58497.05 & - & 17.48 0.07 & 16.66 0.05 & - & ANDICAM\\
2019-01-16 & 58499.05 & - & 17.55 0.04 & - & - & PROM5\\
2019-01-19 & 58502.07 & - & 17.69 0.08 & 16.79 0.04 & 16.65 0.05 & ANDICAM\\
2019-01-19 & 58502.87 & 18.90 0.02 & 17.60 0.02 & - & - & ALFOSC\\
2019-01-24 & 58507.05 & 18.95 0.04 & 17.84 0.05 & 16.87 0.02 & 16.61 0.02 & ANDICAM\\
2019-02-01 & 58515.03 & 19.07 0.06 & 18.16 0.03 & 17.17 0.03 & 16.91 0.02 & ANDICAM\\
2019-02-27 & 58541.85 & 20.07 0.06 & 18.86 0.02 & - & - & ALFOSC\\
2019-07-14 & 58678.40 & - & - & - & 19.32 0.08 & ANDICAM\\
2019-08-06 & 58701.40 & 21.10 0.11 & 20.02 0.08 & 19.42 0.07 & 19.30 0.05 & EFOSC\\
2019-09-04 & 58730.30 & - & 20.21 0.12 & - & - & EFOSC\\ 
2019-10-22 & 58778.17 & - & 20.78 0.09 & - & - & EFOSC\\ 
2019-12-08 & 58825.48 & - & 21.47 0.13 & 20.80 0.31 & - & SUBARU\\ 
2019-12-28 & 58845.04 & - & $>$20.9 & - & - & EFOSC\\
2022-10-21 & 59873.17 & - & - & 20.87 0.15 & - & WFI \\
\hline
\end{tabular}
\begin{tablenotes}
\footnotesize
\item[a] PROM5 = 0.4-m Prompt5+Apogee, PROM6 = 0.4-m Prompt6+Apogee, PROM8 = 0.6-m Prompt8+Apogee, SCH = 0.92-m Asiago Schmidt+Moravian, EFOSC = 3.58-m NTT+EFOSC2, TRAP = 0.6-m TRAPPIST+Fairchild, ALFOSC = 2.56-m NOT+ALFOSC, ANDICAM = 1.3-m SMARTS+ANDICAM, SUBARU = 8.2-m SUBARU+FOCAS, WFI = 2.2-m MPG/ESO+WFI
\end{tablenotes}
\end{threeparttable}
\end{table*}

\begin{table}
\caption{Open filter magnitudes of SN~2018ivc, calibrated as Sloan-$r$.}
\label{tab:clear}
\begin{threeparttable}[b]
\begin{tabular}{cccl}
\hline\hline
Date & MJD & $Open$ & Instrument\tnote{a} \\
\hline
2018-11-22 & 58444.18 & $>$18.3 & PROM \\
2018-11-23 & 58445.03 & $>$19.7 & PROM \\
2018-11-23 & 58445.13 & 18.91 0.18 & PROM \\
2018-11-23 & 58445.24 & 16.92 0.05 & PROM \\
2018-11-23 & 58445.59 & 15.40 0.10 & Itagaki\tnote{b} \\
2018-11-24 & 58446.03 & 14.78 0.02 & PROM \\
2018-11-24 & 58446.13 & 14.70 0.02 & PROM \\
2018-11-24 & 58446.23 & 14.69 0.02 & PROM \\
2018-11-24 & 58446.61 & 14.60 0.10 & Tanaka\tnote{b} \\
2018-11-25 & 58447.03 & 14.69 0.01 & PROM \\
2018-11-25 & 58447.13 & 14.69 0.01 & PROM \\
2018-11-25 & 58447.23 & 14.73 0.01 & PROM \\
2018-11-26 & 58448.14 & 14.87 0.01 & PROM \\
2018-11-26 & 58448.24 & 14.89 0.01 & PROM \\
2018-11-27 & 58449.20 & 14.99 0.01 & PROM \\
2018-11-28 & 58450.03 & 15.10 0.01 & PROM \\
2018-11-29 & 58451.03 & 15.25 0.02 & PROM \\
2018-11-29 & 58451.14 & 15.26 0.01 & PROM \\
2018-11-29 & 58451.25 & 15.27 0.01 & PROM \\
2018-11-30 & 58452.03 & 15.25 0.01 & PROM \\
2018-11-30 & 58452.14 & 15.26 0.01 & PROM \\
2018-11-30 & 58452.25 & 15.26 0.01 & PROM \\
2018-12-01 & 58453.03 & 15.29 0.01 & PROM \\
2018-12-01 & 58453.14 & 15.30 0.01 & PROM \\
2018-12-01 & 58453.25 & 15.29 0.01 & PROM \\
2018-12-02 & 58454.03 & 15.27 0.01 & PROM \\
2018-12-02 & 58454.17 & 15.30 0.02 & PROM \\
2018-12-03 & 58455.03 & 15.29 0.02 & PROM \\
2018-12-03 & 58455.25 & 15.31 0.01 & PROM \\
2018-12-04 & 58456.04 & 15.28 0.01 & PROM \\
2018-12-04 & 58456.14 & 15.29 0.01 & PROM \\
2018-12-04 & 58456.25 & 15.30 0.01 & PROM \\
2018-12-05 & 58457.06 & 15.29 0.01 & PROM \\
2018-12-05 & 58457.17 & 15.31 0.01 & PROM \\
2018-12-06 & 58458.16 & 15.32 0.01 & PROM \\
2018-12-06 & 58458.27 & 15.33 0.01 & PROM \\
2018-12-07 & 58459.04 & 15.34 0.02 & PROM \\
2018-12-07 & 58459.14 & 15.34 0.01 & PROM \\
2018-12-07 & 58459.25 & 15.33 0.01 & PROM \\
2018-12-08 & 58460.14 & 15.34 0.02 & PROM \\
2018-12-09 & 58461.20 & 15.34 0.01 & PROM \\
2018-12-10 & 58462.04 & 15.35 0.01 & PROM \\
\hline
\end{tabular}
\begin{tablenotes}
\footnotesize
\item[a] PROM = 0.4-m Prompt1+Apogee \item[b] https://www.rochesterastronomy.org/sn2018/sn2018ivc.html
\end{tablenotes}
\end{threeparttable}
\end{table}

\begin{table}
\caption{NIR $JHKs$ Vega magnitudes of SN~2018ivc.}
\label{tab:nir}
\begin{tabular}{cccccl}
\hline\hline
Date & MJD & $J$ & $H$ & $Ks$ & Instrument \\
\hline
2018-12-31 & 58483.04 & 15.02 0.02 & 13.96 0.04 & 12.71 0.03 & SOFI\\
2019-01-14 & 58497.06 & 15.36 0.05 & 14.17 0.06 & 12.70 0.01 & SOFI\\
2019-01-24 & 58507.00 & 15.82 0.17 & 14.47 0.07 & - & ANDICAM-IR\\ 
2019-02-01 & 58515.02 & 16.03 0.17 & 14.42 0.07 & - & ANDICAM-IR\\
\hline
\end{tabular}
\end{table}

\begin{table}
\caption{ATLAS $orange$ and $cyan$ magnitudes of SN~2018ivc.}
\label{tab:atlas}
\begin{tabular}{cccccl}
\hline\hline
Date & MJD & $orange$ & $cyan$ \\
\hline
2018-11-26	&	58448.45	&	14.76	0.09	&-	\\
2018-11-28	&	58450.45	&	15.07	0.03	&-	\\
2018-11-30	&	58452.44	&	15.06	0.02	&-	\\
2018-12-02  & 58454.44 &- & 15.49 0.07 \\
2018-12-04	&	58456.40	&	15.19	0.14	&-	\\
2018-12-06  & 58458.40 &- & 15.53 0.08 \\
2018-12-08	&	58460.39	&	15.13	0.09	&-	\\
2018-12-10  & 58462.40 &- & 15.77 0.05 \\
2018-12-12	&	58464.38	&	15.46	0.06	&-	\\
2018-12-16	&	58468.34	&	15.59	0.04	&-	\\
2018-12-18	&	58470.35	&	15.82	0.07	&-	\\
2018-12-20	&	58472.36	&	15.88	0.28	&-	\\
2018-12-22	&	58474.43	&	16.08	0.04	&-	\\
2018-12-24	&	58476.35	&	16.06	0.29	&-	\\
2018-12-26	&	58478.35	&	16.05	0.12	&-	\\
2018-12-28	&	58480.35	&	16.18	0.15	&-	\\
2019-01-03  & 58486.35 &- & 17.13 0.18 \\
2019-01-05	&	58488.34	&	16.45	0.46	&-	\\
2019-01-07  & 58490.34 &- & 17.10 0.28 \\
2019-01-09	&	58492.33	&	16.57	0.11	&-	\\
2019-01-13	&	58496.31	&	16.59	0.28	&-	\\
2019-01-21	&	58504.31	&	16.97	0.41	&-	\\
2019-01-27	&	58510.29	&	17.09	0.28	&-	\\
\hline
\end{tabular}
\end{table}

\end{appendix}

\end{document}